\PassOptionsToPackage{pdfpagelabels=false}{hyperref}
\documentclass[fleqn,usenatbib]{mnras}
\usepackage[T1]{fontenc}
\usepackage{ae,aecompl}
\usepackage{graphicx}	
\usepackage{amsmath}	
\usepackage{amssymb}	
\usepackage{txfonts}
\usepackage[symbol]{footmisc}
\usepackage{url}
\usepackage{color} 

\title[Morphology dependent $M_{\rm bh}$--$\sigma_0$ relations]{Galaxy morphology-dependent (black hole mass)-(velocity dispersion) relations: implications for gravitational-wave forecasts and cosmological simulations}

\author[Graham]
{Alister W.\ Graham$^1$\thanks{E-mail: AGraham@swin.edu.au}
\\
$^1$ Centre for Astrophysics and Supercomputing, Swinburne University of Technology, Hawthorn, VIC 3122, Australia
}

\date{Accepted XXX. Received YYY; in original form ZZZ}
\pubyear{2026}

\begin{document}
\label{firstpage}
\pagerange{\pageref{firstpage}--\pageref{lastpage}}
\maketitle

\begin{abstract}

The correlation between black hole mass, $M_{\rm bh}$, and stellar velocity dispersion, $\sigma_0$, is revisited using 137 galaxies with quantitative bar strengths and enhanced morphological awareness. Interpreted within the proposed `Triangal' evolutionary framework, gas-rich and gas-poor assembly pathways emerge in the $M_{\rm bh}$--$\sigma_0$ diagram. To robustly quantify these scaling relations, a symmetric  Bayesian hierarchical regression code, dubbed the Symmetric COvariance Population Estimator (\textsc{scope}), is introduced. Unlike conditional estimators (e.g., \textsc{linmix}), \textsc{scope} derives the intrinsic population covariance, natively accommodating asymmetric measurement errors while guaranteeing directional invariance between axes.
Applying \textsc{scope} reveals that the current sample of primeval, dust-poor S0 galaxies 
(including dwarf early-type galaxies with $R_{\rm e,gal}\approx1$~kpc) 
follow a shallow relation ($M_{\rm bh}\propto\sigma_0^{2.5\text{--}3.1}$). Explained via the virial theorem, this flattening reframes expectations for intermediate-mass black holes. In contrast, tracing the `Disc Down-sizing' sequence --- where dry mergers erase discs --- yields a steep relation for massive elliptical and ellicular galaxies ($M_{\rm bh}\propto\sigma_0^{7.8\pm1.3}$). 
Applying a single, monolithic scaling relation across all morphologies inadvertently averages over these distinct galaxy formation histories, 
potentially influencing AGN virial $f$-factor calibrations and systematically under-predicting estimates of 
the ultra-massive black hole population thought to contribute to the nanohertz gravitational wave background. 
Furthermore, although lost in the complexity of spiral galaxies, strongly barred, dust-poor S0 galaxies appear offset to higher velocity dispersions, qualitatively consistent with 
theoretical expectations of kinematic heating by stellar bars. Ultimately, these morphology-dependent relations provide physically-motivated benchmarks for cosmological simulations and a framework for disentangling regimes driven by AGN feedback from those driven by mergers.

\end{abstract}


\section{Introduction}
\label{Sec_Intro}

The correlation between supermassive black hole mass, $M_{\rm bh}$, and the stellar velocity dispersion, $\sigma$, of the host spheroid \citep{2000ApJ...539L...9F, 2000ApJ...539L..13G} is often heralded as the most fundamental of the black hole scaling relations. 
Most, if not all, massive spheroidal stellar systems are believed to host a central supermassive black hole 
\citep[e.g.,][]{1969Natur.223..690L, 1982MNRAS.200..115S, Rees1984, 1998AJ....115.2285M} 
whose growth proceeds through gas accretion and, during mergers, coalescence with other black holes.
The `active' accretion phase --- which is distinct from black hole merger events --- is thought to be coupled to the host-galaxy evolution, as the energetic output from the active galactic nucleus (AGN) can heat or expel the surrounding interstellar medium. 
Initial reports of the tightness of the $M_{\rm bh}$--$\sigma_0$ relation were regarded as evidence for self-regulating feedback from these AGN \citep[e.g.,][]{1998MNRAS.300..817H, 1998A&A...331L...1S, 2024ApJ...961L..39S}. 
However, as sample sizes have grown, so too have the observed scatter and the complexity of the relation. 

The physical mechanisms governing the $M_{\rm bh}$--$\sigma_0$ relation, and the origin of its scatter, remain active areas of research.
Early explorations of the relation's morphological dependence reported distinct behaviors for barred galaxies and `pseudobulges' \citep{2008ApJ...680..143G, 2008MNRAS.386.2242H}, 
an idea later pursued by other investigations 
\citep[e.g.,][]{2009ApJ...698..812G, 2009ApJ...698..198G, 2013ARA&A..51..511K, 2017MNRAS.471.2187D}. 
In parallel, additional distinct relations were identified in the $M_{\rm bh}$--$\sigma_0$ plane by 
\citet{2013ApJ...764..151G}, which revealed that early-type galaxies (ETGs) built from major (dry) mergers and displaying partially depleted stellar cores (and thus having core-S{\'e}rsic light profiles: \citealt{2003AJ....125.2951G}) follow a distinct sequence compared to galaxies with non-power-law \citet{1963BAAA....6...41S} light profiles.\footnote{\citet{2005PASA...22..118G} provides an overview and expanded formulation of the \citet{1963BAAA....6...41S} model in terms of effective half-light parameters, while \citet{graham2026rrmerrme} expresses the equations in terms of the more practical $R_5$ parameters.} 
A non-causal interpretation has been 
proposed, in which a correlation arises statistically rather than through
direct physical coupling between the growth of the black hole and the host galaxy's stellar mass, $M_\star$, such that repeated
galaxy mergers drive initially uncorrelated $M_{\rm bh}/M_\star$ ratios toward
a common value via the central-limit theorem, without requiring AGN feedback
\citep{2007ApJ...671.1098P, 2011ApJ...734...92J}. 
However, this statistical-convergence picture predicts
an asymptotic linear ($M_{\rm bh}\propto M_\star$) relation, and the observed 
quadratic-like $M_{\rm bh}$--$M_{\star,\rm gal}$ relation
for the ensemble of major-merger-built galaxies 
\citep[e.g.,][]{2023MNRAS.522.3588G} 
is difficult to reconcile with such a purely stochastic,
merger-driven origin, favouring instead some form of physical coupling
between black hole and spheroid growth. 
The above ideas have been reviewed in 
\citet{2013ARA&A..51..511K} and \citet[][and references therein]{2016ASSL..418..263G}. 

\begin{figure}
\begin{center}
\includegraphics[width=1.0\columnwidth]{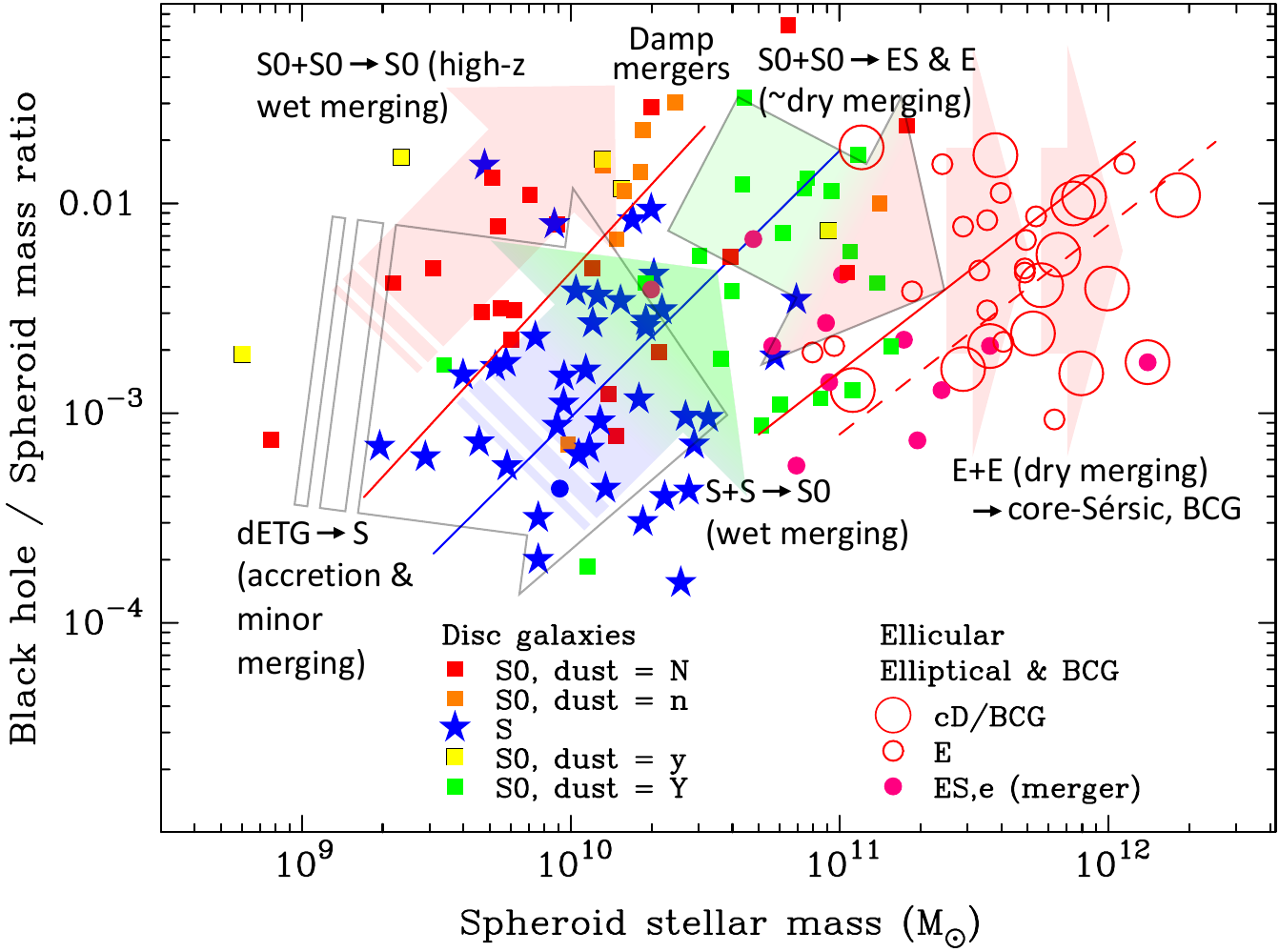}
\caption{
Adaptation and extension of the $M_{\rm bh}$--$M_{\rm \star,sph}$ diagram from \citet[][figure~3]{2026arXiv260424084G}, shown here as the $M_{\rm bh}/M_{\rm \star,sph}$ ratio versus $M_{\rm \star,sph}$. The left-most line represents dust-poor (dust $=$ N) S0 galaxies, and is likely offset to the right of the trend set by primeval S0 and dwarf ETGs due to (i) a few faded S galaxies that are now dust-poor S0/a and S0 galaxies \citep{2026arXiv260424084G} and (ii) a few major-merger-built S0 galaxies that have lost their dust.  The blue line tracks the S galaxies, some of which merge to become dust-rich S0 galaxies, and the non-(cD/BCG) E and ES,e galaxies are represented by the right-most solid line. The dashed line denotes the BCGs. Lines are taken from \citet{2023MNRAS.522.3588G}.  These sequences illustrate the morphology-dependent trends that gave rise to the triangular-like distribution discussed there and briefly in  Section~\ref{Sec_Intro}. The arrows indicate the broad evolutionary pathways linking the different populations. 
}
\label{bh-ratio-sph}
\end{center}
\end{figure}

In this study, new lessons garnered from the $M_{\rm bh}$--$M_{\rm \star}$ diagram are examined in the $M_{\rm bh}$--$\sigma_0$ diagram.
When galaxy morphology and `dust bin' type \citep[][see Section~\ref{Sec_Data} for details]{2023MNRAS.521.1023G} are considered together, the black hole scaling relations 
can be interpreted as tracing 
multiple evolutionary pathways through the $M_{\rm bh}$--$M_{\rm \star}$ diagram, 
as discussed within the framework proposed by 
\citet{2023MNRAS.522.3588G}.  This 
behaviour 
is shown in Figure~\ref{bh-ratio-sph} in terms of the $M_{\rm bh}/M_{\rm \star}$ ratio. 
Gas-rich systems --- including spiral (S) galaxies and dust-rich lenticular (S0) galaxies ---  
follow steep growth tracks in this diagram, 
plausibly 
associated with dissipative (minor and major) mergers and rapid black hole accretion. In contrast, the most massive elliptical (E) and ellicular (ES,e)\footnote{Following \citet{Graham:Sahu:22b}, ES,e galaxies are ES galaxies \citep{1966ApJ...146...28L} with fully-embedded intermediate-scale discs and {\it large} extended spheroids like E galaxies, rather than the {\it small} compact (bulge-like) spheroids that are characteristic of the ES,b population.\label{foot_one}} galaxies, 
that likely 
assembled through largely dry mergers, evolve along an approximately constant $M_{\rm bh}/M_{\rm \star}$ track, 
as major mergers roughly double both $M_{\rm bh}$ and $M_{\rm \star}$, somewhat preserving the $M_{\rm bh}\propto M_{\rm \star}^2$ scaling seen among the gas-rich galaxies. 

Appreciating such refinements is critical for several pressing astrophysical domains. First, they provide a deeper understanding of galaxy-black hole coevolution across cosmic time. Second, accurate local baseline relations 
serve as direct demographic inputs 
for population-synthesis models used to predict and interpret the
stochastic gravitational-wave background 
(see Section~\ref{sec:gravitational_waves}) probed by
the nanohertz signal seen by Pulsar Timing Arrays \citep[PTAs; e.g.,][]{2023ApJ...951L...9A} and the individual merger events targeted by the upcoming Laser Interferometer Space Antenna \citep[LISA; e.g.,][]{2017arXiv170200786A}. 
Such gravitational-wave detections, in turn, inform about galaxy/black hole coevolution. Refined $M_{\rm bh}$--$M_{\rm \star}$ and $M_{\rm bh}$--$\sigma$ relations also enable precise calibrations of the virial $f$-factor --- required to convert AGN virial products into true black hole masses \citep[][]{1972ApJ...171..467B, 1982ApJ...255..419B, 2004ApJ...615..645O, 2009ApJ...694L.166B, 2011MNRAS.412.2211G} --- and anchor the array of secondary relations used to infer black hole masses in the distant Universe. 
Furthermore, many cosmological simulations and semi-analytic models have
calibrated sub-grid prescriptions for black-hole seeding, accretion, and AGN
feedback so that their emergent z=0 galaxy populations reproduce a single,
monolithic $M_{\rm bh}$--$\sigma$ (or $M_{\rm bh}$--$M_\star$) relation, valued
for its simplicity as a single calibration target. However, as modern
simulations increasingly resolve the kinematic and morphological diversity of
galaxies through distinct merger trees, these prescriptions can instead be
tested against, and calibrated to reproduce, galaxy-morphology-dependent relations
(Section~\ref{sec:simulations}), 
offering a more discriminating benchmark than matching a single blended relation alone. 

With galaxy morphologies and `dust bins' \citep{2023MNRAS.521.1023G} in hand, reflecting the galaxies' 
formation histories, the demographics in the $M_{\rm bh}/M_{\rm \star,sph}$--$M_{\rm \star,sph}$ diagram (Figure~\ref{bh-ratio-sph}) 
can be viewed as an 
informative map in which $M_{\rm bh}$ and $M_{\rm \star,sph}$, both measures of entropy, serve as broad proxies of this history. 
Under this interpretation, Figure~\ref{bh-ratio-sph} can be used to trace potential pathways such as 
accretion and the emergence of spiral patterns (primeval S0 $\rightarrow$ S) and hierarchical merging (S $\rightarrow$ dust-rich S0 $\rightarrow$ E), mapping out a possible sequence of galaxy speciation.  This has been used to create the `Triangal' evolutionary scheme \citep{2023MNRAS.522.3588G}, building on the `Tuning Fork' diagram that is absent of evolutionary pathways. 

With so much diversity now apparent in the $M_{\rm bh}/M_{\rm \star,sph}$--$M_{\rm \star,sph}$ and $M_{\rm bh}$--$M_{\rm \star,sph}$ diagrams, it is pertinent to re-examine the other black hole scaling diagrams. Here, this refined morphological framework is applied to the $M_{\rm bh}$--$\sigma_0$ diagram. 

The steepening distribution of galaxies at the high-mass end of the $M_{\rm bh}$--$\sigma_0$ diagram noted by \citet{2013ApJ...764..151G} and \citet{2018ApJ...852..131B} was explained and quantified by \citet{2023MNRAS.518.6293G} using the virial theorem. 
In this work, the low-mass end of the $M_{\rm bh}$--$\sigma_0$ diagram is also explored through  
updated $M_{\rm bh}$--$\sigma_0$ scaling relations derived for separate morphological sub-populations, including E (and ES,e) galaxies -- systems formed by dry (and perhaps damp) mergers --- and low-mass, dust-poor S0 galaxies, 
to investigate potential variations in their scaling behavior.
Specifically, we test whether the low-mass dust-poor S0 sequence exhibits a shallower slope. 
This study explores whether these distinct regimes can be explained by 
a single physical principle --- dubbed the 'Virial Mirror' --- in which the scalar virial theorem, coupled with the curved size–mass relation of ETGs \citep{2019PASA...36...35G}, 
is hypothesised to 
simultaneously drive both a low-mass flattening and a high-mass steepening, perhaps with an (AGN-driven) intermediate sequence bridging the two extremes.

It is also investigated whether the `hidden' signal of bar heating can be isolated. Studies suggest that bars should increase vertical (out of the disc plane) velocity dispersion and drive radial motions, elevating $\sigma_0$ \citep{2013ApJ...765...23D, 2014MNRAS.441.1243H}. Yet, this effect was not clearly evident in a recent large compilation \citep{2019ApJ...887...10S}, warranting closer scrutiny. 

The sample of 137 galaxies from \citet{2026arXiv260424084G} is used here 
to investigate potential morphology-dependent scaling relations in the $M_{\rm bh}$--$\sigma_0$ diagram. 
This sample is briefly described in Section~\ref{Sec_Data} along with a discussion of an alleged sample selection bias.  
A novel hierarchical Bayesian regression is introduced in this work and used to establish $M_{\rm bh}$--$\sigma_0$ scaling relations for various galaxy morphological types. 
Section~\ref{Sec_Methodology} provides an overview of this statistical methodology. 
Section~\ref{Sec_RandI} presents the key observations and results from these data,  
which is seen to lend itself to four primary divisions: 
E and ES,e galaxies built from relatively dry major mergers define a steep $M_{\rm bh}$--$\sigma_0$ relation with a slope approaching 8 to 9 (Section~\ref{Sec_top}); 
mid-range S0 galaxies define a relation having the more commonly recognised slope around 4 to 5 \citep{2000ApJ...539L...9F, 2000ApJ...539L..13G}, as described in Section~\ref{Sec_central}; 
and  
the low-mass dwarf dETGs and dust-poor S0 galaxies, regarded as primeval systems 
appear to define a shallow relation with a slope around $\sim$2.5 to 3 (Section~\ref{Sec_flat}).  Tentative evidence for an offset population of barred dust-poor S0 galaxies is also presented in Section~\ref{Sec_flat}, while Section~\ref{Sec_flatter} explores extensions toward the realm of intermediate-mass black holes (IMBHs: $10^2 \lesssim M_{\rm bh}/M_{\odot} < 10^5$). The fourth division pertains to a relatively broad distribution of S galaxies, as shown in Section~\ref{Sec_spiral}, while no preferred location is seen in the $M_{\rm bh}$--$\sigma_0$ diagram for either S0/a galaxies that host weak spiral patterns or spiral galaxies with particularly strong arms. Section~\ref{Sec_unknown} offers a simple $\sigma_0$-based prescription for estimating $M_{\rm bh}$ if the precise ETG morphology is unknown. 

Section~\ref{Sec_disc} provides a detailed discussion covering such topics as 
(i) the departures at low- and high-masses from the classical monolithic $M_{\rm bh}$--$\sigma_0$ relation applied to past smaller galaxy ensembles with less regard for galaxy morphology (Section~\ref{sec:feedback_and_mergers}), 
(ii) the signature of an offset barred population in the less complex, dust-poor S0 galaxies (Section~\ref{Sec_bar_offset}), 
(iii) the scatter seen among the S galaxies (Section~\ref{Sec_scat}), 
(iv) the need to re-calibrate AGN virial products used to estimate distant black hole masses (Section~\ref{sec:f_factors}), 
(v) IMBHs in star clusters (Section~\ref{Sec_IMBH}), 
(vi) implications for ultra-massive black holes (UMBHs: $>10^{10}$~M$_\odot$) and the expected gravitational wave signal from their merging (Sections~\ref{Sec_UMBH} and \ref{sec:gravitational_waves}), 
(vii) connections with the `Triangal' galaxy evolution framework (Section~\ref{Sec_connect}), and 
(viii) the suggested use of these refined $M_{\rm bh}$--$\sigma_0$ relations (Table~\ref{Table_rel}) to calibrate galaxy simulations and semi-analytical code (Section~\ref{sec:simulations}).

Finally, Appendix~\ref{Sec_Appdx} details the Bayesian hierarchical regression code called \textsc{scope}, including a comparison with the popular \textsc{linmix} code \citep{2007ApJ...665.1489K}. The new code \textsc{scope} treats the data symmetrically, while \textsc{linmix} assumes a dependent and an independent variable, resulting in \textsc{linmix} generating different slopes if the ordering of the data is switched.

\subsection{Further background/context}
\label{Sec_Intro2}

Some additional background is provided here for readers accustomed only to monolithic scaling relations derived from single power-law fits to heterogeneous galaxy types.

Over a decade ago, two studies updated the local black hole mass scaling relations using expanded samples of galaxies with dynamically measured black hole masses \citep{2013ApJ...764..151G,2013ApJ...764..184M}. While both studies reported a single $M_{\rm bh}$--$\sigma_0$ power-law relation across their full samples,  \citet{2013ApJ...764..151G} noted that this relation may be fundamentally `broken', reflecting underlying differences in galaxy morphology and thus formation history. Building on the identification of a bend in the $M_{\rm bh}$--$M_{\rm dyn}$ diagram \citep{2012ApJ...746..113G}, subsequent work showed that separating galaxies into S\'ersic and core-S\'ersic populations reveals distinct scaling relations involving a near-quadratic (or `super-quadratic')\footnote{This term is used here to describe a log-log relation with a slope slightly greater than 2, and up to $\sim$2.5.} $M_{\rm bh}$--$M_{\rm \star,sph}$ relation for S\'ersic galaxies \citep{2013ApJ...764..151G,2013ApJ...768...76S}, and a steep $M_{\rm bh}$--$\sigma_0$ relation (slope $\sim$7) for core-S\'ersic galaxies \citep{2013ApJ...764..151G}.

Follow-up work demonstrated that this apparent dichotomy represented only part of a broader, morphology-dependent landscape. Late-type galaxies (aka S galaxies) follow a steep $M_{\rm bh}$--$M_{\rm \star,sph}$ relation \citep{2016ApJ...817...21S,2019ApJ...873...85D}, while mixed samples of ETGs (E, ES, and S0) display a near-linear (or `super-linear')\footnote{This term is used here to describe a log-log relation with a slope slightly greater than 1, and up to $\sim$1.5.} relation whose slope is sensitive to sample composition \citep{2020ApJ...903...97S}. In particular, the observed slope depends on the relative fractions of E and S0 galaxies that are included, rendering single power-law fits problematic and potentially biasing expectations in both the low- and high-mass regime.

To better capture this diversity, an interpretive galaxy evolution
scheme --- the `Triangal' framework 
\citep{2023MNRAS.522.3588G} 
--- was
proposed, which infers, from observed morphology and dust content, an
evolutionary separation of S0 galaxies into three physically distinct types:
primordial systems, faded S galaxies that have lost their spiral pattern, and dust-rich merger remnants.
This development has resulted in significant modifications to evolutionary pathways through the colour-magnitude/mass diagram \citep{2024MNRAS.531..230G} and the mass-(star formation rate) diagram \citep{2024MNRAS.52710059G}. Indeed, the augmentation of the latter diagram with refined galaxy morphological types has revealed that morphologies, rather than bulge mass, bulge-to-total stellar mass ratio, or feedback from black holes (as measured by the BH-to-galaxy-stellar-mass ratio), better trace a galaxy’s specific star formation rate, as shown in \citet{2024MNRAS.52710059G}. 
This morphology-aware approach has also led to revised interpretations of galaxy evolution for both systems with inactive and active galactic nuclei \citep[AGN;][]{2025PASA...42...68G} and has highlighted the importance of formation history --- rather than galaxy parameters alone --- in shaping the black hole--galaxy relations.

In this context, the $M_{\rm bh}/M_{\rm \star,sph}$--$M_{\rm \star,sph}$ diagram (Figure~\ref{bh-ratio-sph}) can be understood as tracing distinct evolutionary channels. For instance, 
gradual accretion and minor mergers may build up discs and bulges at a similar or faster fractional pace than the central black hole mass, leading to constancy or a decline in the $M_{\rm bh}/M_{\rm \star,sph}$ ratio with increasing $M_{\rm \star,sph}$ when transitioning from primeval dETGs to S galaxies. 
Major S galaxy collisions erase spiral patterns and fold considerable S galaxy disc mass into the merger remnant to increase $M_{\rm \star,sph}$, while a greater fractional growth arises in the black hole mass --- from gas (and stars) losing orbital angular momentum and making their way inward, resulting in an increase in the $M_{\rm bh}/M_{\rm \star,sph}$ ratio with increasing $M_{\rm \star,sph}$. Indeed, if this dramatic increase in $M_{\rm bh}$ occurs quickly, one may expect a quasar phenomenon.  At high-redshift, when today's dust-poor (dust $=$ N, see later) S0 galaxies still contained cold gas, their major merging also likely powered quasars and a somewhat parallel steep ascent in the $M_{\rm bh}/M_{\rm \star,sph}$--$M_{\rm \star,sph}$ diagram. This is the quadratic growth in the $M_{\rm bh}$--$M_{\rm \star,dyn}$ and $M_{\rm bh}$--$M_{\rm \star,sph}$ diagrams noted by  \citet{2012ApJ...746..113G}, \citet{2013ApJ...764..151G}, and \citet{2013ApJ...768...76S}. 
The subsequent relatively dry merging of S0 galaxies initially results in a decrease in the $M_{\rm bh}/M_{\rm \star,sph}$ ratio as the pre-merged S0 galaxy discs are folded into the spheroidal component of either the resultant ES galaxy with an intermediate-scale disc \citep[see][]{2024MNRAS.535..299G} or E galaxy, possibly with a nuclear dust disc or ring.\footnote{The S0$+$S0 $\rightarrow$ ES path tends to occur below, i.e., at lower $M_{\rm bh}/M_{\rm \star,sph}$ ratios (and lower $M_{\rm \star,sph}$ values), than the S0$+$S0 $\rightarrow$ E pathway.}  
The largely dry merging of these ES and E galaxies preserves the $M_{\rm bh}/M_{\rm \star,sph}$ ratio as $M_{\rm \star,sph}$ increases.

If some primeval ETGs instead undergo major mergers rather than disc-building accretion and minor mergers, they will follow a different evolutionary path towards the more massive S0 galaxies, ES,b galaxies, E galaxies with nuclear discs, and finally E galaxies with depleted cores.  This divergent history creates an evolutionary fork, yielding the triangular-like pattern noticed in the $M_{\rm bh}$--$M_{\rm \star,sph}$ diagram \citep{2023MNRAS.522.3588G} --- and shown here in the $M_{\rm bh}/M_{\rm \star,sph}$--$M_{\rm \star,sph}$ diagram (Figure~\ref{bh-ratio-sph}) --- with pathways reuniting when galaxies become dynamically-hot pressure-supported ETGs after major mergers.  Presumably, those galaxies that spent some time on the spiral sequence will have somewhat younger mean stellar ages than those primarily built through the always-ETG path, i.e., via the summation of primeval ETGs.

Extending this framework, \citet{2026arXiv260424084G} examined the role of bar and spiral strength in the $M_{\rm bh}$--$M_{\rm \star,sph}$ diagram and found that bars do not drive galaxies to preferred locations in that plane. The present work builds on this by deriving morphology-specific $M_{\rm bh}$--$\sigma_0$ scaling relations and investigating whether such structural features (bars and arms) influence these relations.

\section{Data: Galaxy Sample} 
\label{Sec_Data}

\subsection{Sample construction and measurement compilation}
\label{Sec_sample_construction}

The initial  sample comprises 137 galaxies with directly measured black hole masses and multi-component decompositions, as detailed in \citet{2026arXiv260424084G}. 
However, 
IC~1481 has no available velocity dispersion, 
NGC~6926 is both bulgeless and lacks a velocity dispersion, 
and the bulgeless galaxies NGC~2748 and NGC~4395 are similarly (initially) excluded since their dispersions reflect disc rather than bulge kinematics.
This reduced (137$-$4) sample of 133 galaxies includes 53 S0 and S0/a galaxies (including four ES,b galaxies), 40 S galaxies, and 40 ETGs (E, ES,e, and brightest cluster galaxies, aka BCGs).

The S galaxy NGC~5495 is added here\footnote{NGC~5495 was excluded from \citet{2026arXiv260424084G} due to contamination from a bright star impacting the luminosity.}, as is the 
gravitationally-stripped, dust $=$ N, S0 galaxy NGC~4342 \citep{2014MNRAS.439.2420B}, 
and the so-called `compact elliptical' (cE) galaxies NGC~4486B and, in Section~\ref{Sec_flat}, NGC~221 (M32), 
thereby giving a sample count of 137 galaxies. 

Central velocity dispersions are adopted from \citet{2019ApJ...876..155S} and had previously been homogenized to a uniform aperture size of 0.595~kpc within the Hyperleda database\footnote{http://leda.univ-lyon1.fr}  \citep{2003A&A...412...45P, 2014A&A...570A..13M}.  Following \citet{2019ApJ...876..155S}, a ten per cent uncertainty is assigned to these velocity dispersions. 
Five galaxies with particularly low velocity dispersions relative to
their respective morphological classes are subsequently excluded from the
primary regression fits, for reasons of statistical robustness detailed in
Section~\ref{sec:sample_pruning}.

The black hole masses were compiled by
\citet{2019ApJ...887...10S} 
and can be traced back to the source following the reference chain therein.  
All masses were obtained via standard methods based around spatially resolving the black hole's sphere-of-influence, using stellar, gas disc, or maser dynamics 
\citep[e.g.,][]{2005SSRv..116..523F}. 
Some masses have been adjusted according to the updated distances tabulated in 
\citet{2026arXiv260424084G}.

\subsection{Morphological and dust-bin classifications}
\label{Sec_dust_bins}

As noted in \citet{2026arXiv260424084G}, 
refined galaxy classifications were obtained for the S0 galaxies based on the presence, prominence, and distribution of visible dust. In the context of the proposed `Triangal' framework 
\citep{2023MNRAS.522.3588G}, 
the different `dust bins' introduced by
\citet{2023MNRAS.521.1023G} 
serve as observational proxies to investigate potential gas accretion and merger histories. The sub-populations are defined observationally, and interpreted within this evolutionary framework, as follows:
\begin{itemize} 
    \item \textbf{S0 (dust $=$ N):} 
    S0 galaxies displaying no visible signs of dust (dust-poor). Within the Triangal framework, these systems are hypothesized to represent primeval, gas-stripped, or early-quenched systems.
    \item \textbf{S0 (dust $=$ n):} 
    S0 galaxies containing a concentrated, nuclear dust disc or nuclear dust ring. These are interpreted as systems that have retained or accreted a modest reservoir of cold gas in their core, potentially via minor mergers or damp major mergers.
    \item \textbf{S0 (dust $=$ y):} 
    S0 galaxies containing faint, widespread dust that is often centrally concentrated but does not form a highly structured nuclear disc or ring.  
    \item \textbf{S0 (dust $=$ Y):} 
    S0 galaxies displaying prominent, widespread dust. These dust-rich systems with angular momentum are likely to be major wet merger remnants 
    \citep{1983MNRAS.205.1009N, 1992ApJ...393..484B, 1996ApJ...471..115B}, 
    and most have been identified as such 
    \citep[][and references therein]{2023MNRAS.521.1023G}.  
    They also occupy the `Green Valley' transition zone, which may be a `Green Mountain' after correcting for dust 
    \citep{2018MNRAS.481.1183E, 2024MNRAS.531..230G}.
    \item \textbf{S:} 
    Spiral galaxies, defined observationally by the presence of an extended, star-forming stellar disc and spiral arm patterns.  Within the evolutionary model, their disc growth and emergent spiral structure mark an evolutionary stage beyond the primeval ETGs, achieved through gas accretion.
    \item \textbf{E and ES,e:} 
    Elliptical (E) and ellicular (ES,e) galaxies, defined observationally as dynamically hot, pressure-supported spheroidal stellar systems (with the latter possessing fully embedded, intermediate-scale stellar discs), are widely understood to be spheroids built by repeated mergers \citep{2006ApJ...636L..81N}.  
    Within the Triangal framework, they are further distinguished as having formed via relatively dry mergers or damp mergers, which may produce nuclear dust discs 
    \citep{2026arXiv260424084G}.
\end{itemize}

The `dust bin' criteria and their evolutionary relevance are explained further in \citet{2023MNRAS.521.1023G}.

\subsection{Reassigned dust-bin classifications for individual galaxies}
\label{Sec_individual_reassign}

There are three dust $=$ N S0 galaxies that were classified as E galaxies in the \citet{1991rc3..book.....D} RC3 and are likely in the wrong dust bin from an evolutionary perspective.  They are 
NGC~3640 and NGC~4649 (aka M60), recognised and suspected merger products \citep{2014ApJ...783...18D, 2024A&A...691A.104M}, and  
NGC~3091 \citep[in Hickson Compact Group No.~42 and with 
an outer exponential (halo?) having a scalelength of 15~kpc:][]{2024MNRAS.535..299G}. 

NGC~3091 and NGC~4649 appear in the X-ray bright catalogue of \citet{2010MNRAS.404..180D}, likely explaining their absence of dust. It is also possible that these galaxies formed from a dry (S0+S0), rather than wet and dusty\footnote{Perhaps the term `muddy' would capture this combination.} (S+S, S+S0) major merger, as shown by the formation paths in \citet[][figure~1]{2023MNRAS.522.3588G}.  
From an evolutionary point of view, they very likely do not belong in the S0 galaxy dust $=$ N bin full of suspected primeval S0 galaxies, and, as such, these three galaxies have been moved to the S0 galaxy dust $=$ Y bin full of wet major-merger-built S0 galaxies.  This reflects the imperfection (the slightly dirty nature) of the `dust bins' as a proxy for accretion/merger history.  Nonetheless, the `dust bins' have their merit, despite coming slightly unstuck at the high-mass-end where the existence of hot gas halos can strip/deprive galaxies of their dust, and/or some relatively rarer dry-major-merger-built S0 galaxies exist.

\begin{figure}
\begin{center}
\includegraphics[width=1.0\columnwidth]{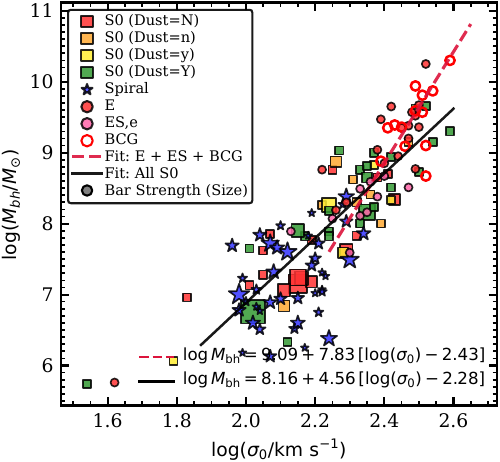}
\caption{
The $M_{\rm bh}$--$\sigma_0$ diagram, categorised by the `Triangal' morphology and dust classes. 
S0 and S0/a galaxies are shown as squares, colour-coded by dust content. 
S galaxies are shown as stars.
E and ES,e galaxies are shown as red- and pink-filled circles, respectively.
BCGs are shown as larger open red circles.
Symbol size for disc galaxies scales with the bar strength parameter $P$.
The dashed red line represents the Bayesian hierarchical regression for the E and ES,e galaxies, including the BCG (Equation~\ref{Eq_ESeE}), while the solid black line is the fit to the S0 galaxies (Equation~\ref{Eq_S0_all}). 
To provide a more robust result, a few data points were excluded from the regression but are shown here; these are noted in Section~\ref{Sec_Data}. 
Note the large scatter in the S galaxy population.
The unbarred S0/a galaxy Mrk~1029 (dust $=$ Y) at the bottom of the S galaxy distribution displays shells and what might be considered tidal debris/arms from (minor or major?) accretion events.
}
\label{Fig-M-sigma-2}
\end{center}
\end{figure}

The dust $=$ N S0 galaxy NGC~1374, which was also classified as an E galaxy in the RC3, may be an old `damp' major merger remnant, perhaps formed by the collision of two S0 galaxies long ago, rather than moulded by a more recent collision involving a gas-rich S galaxy. Although it lacks the proposed \citep{2026arXiv260424084G} signature of a damp merger --- namely a nuclear dust disc (dust $=$ n) --- its bimodal distribution of globular cluster colours \citep{2006MNRAS.367..156B} suggests it was built from a (non-dry) major merger \citep{2009ApJ...691...83S}. After re-assigning `dust bin' types for NGC~3091 and NGC~4649 (and NGC~3640), NGC~1374 is the dust $=$ N S0 galaxy with the highest black hole mass shown in Figure~\ref{Fig-M-sigma-2}.  Its dust bin type is not, however, reassigned.

\subsection{Bar and spiral strength}
\label{Sec_bar_strength}

The galaxy bar strengths used in this paper were quantified using the $P$ parameter (the bar-to-total luminosity ratio) from \citet{2026arXiv260424084G}. Galaxies with $P \ge 0.1$ are considered to have strong bars, while those with $0 < P < 0.1$ have weak bars. 
The spiral strength was also noted by \citet{2026arXiv260424084G}, which reported on seven (possibly nine) S0/a galaxies with weak spiral patterns plus six S galaxies with particularly strong/prominent spiral arms.

\subsection{Sample Representativeness and Selection Bias} 
\label{sec:sample_bias}

Because the present work seeks to infer the intrinsic scaling relations
between supermassive black holes and their associated galaxies, it is important
to note that the adopted galaxy sample is representative of the
underlying population. Any systematic selection bias in the galaxies with
direct black hole mass measurements would propagate into the inferred slope,
normalisation, and intrinsic scatter of the scaling relations. The principal
concern raised in recent years has been whether galaxies with dynamically
measured black holes preferentially have unusually above-average stellar velocity
dispersions for a given stellar mass, 
which would otherwise introduce a systematic selection bias into the observed black hole--galaxy relations.

This issue has been investigated extensively. While
\citet{2016MNRAS.460.3119S} 
suggested that the sample with directly measured black hole masses {\em might} be biased
toward galaxies with unusually large velocity dispersions, this conclusion
was subsequently resolved by
\citet{2023MNRAS.518.1352S}, 
who demonstrated that
the apparent offset arose from comparing those galaxies
with a reference population having an inconsistent stellar mass-to-light ratio. After constructing a matched
comparison sample with consistent morphological classifications and stellar 
mass-to-light ratios, 
\citet{2023MNRAS.518.1352S} 
found no significant evidence that
galaxies with direct black hole mass determinations are systematically offset
from the general galaxy population. This possibility was flagged in 
\citet{2016MNRAS.460.3119S}, 
and the subsequent resolution by 
\citet{2023MNRAS.518.1352S} 
indicate that the currently available dynamical SMBH sample is consistent with being representative in terms of these global host properties. 
While the absence of an identified systematic offset does not definitively prove that the dynamically measured sample is a perfectly representative slice of all local galaxies, it demonstrates that previous claims of severe, order-of-magnitude selection biases were methodological artifacts rather than physical features of the dataset.

More recently, 
\citet{2025MNRAS.541.2070S} 
used 
a subsample of the \textit{Spitzer Survey of Stellar Structure in Galaxies} 
\citep[S$^4$G:][]{2010PASP..122.1397S}\footnote{\url{https://irsa.ipac.caltech.edu/data/SPITZER/S4G/}} 
sample to 
argue that a residual selection bias 
remains. However, that claim rested on an analysis that did not account for a fundamental feature of ETG structure. Specifically, their analysis relied on an assumed single-power-law $\sigma$--(luminosity, $L$) relation that overlooked the well-established broken nature of the $\sigma$--$L$ relation.  In reality, the relation scales as $L \propto \sigma^{2\text{--}2.5}$ at faint magnitudes 
\citep[e.g.,][]{1983ApJ...266...41D, 2005MNRAS.362..289M, 2014ApJS..215...17T} 
and steepens to $L \propto \sigma^{4\text{--}5}$ or steeper at high luminosities 
\citep[e.g.,][]{1976ApJ...204..668F, 1980AJ.....85..801S, 1981ApJ...251..508M, 2019ApJ...887...10S}. 
Fitting a single power law across the full luminosity range, separately to
each sample, the relations of 
\citet[][their Figure~A1]{2025MNRAS.541.2070S}
did not correctly capture the broken $\sigma$--$L$ relation for either
population. Extrapolating the SMBH sample's fit toward its own fainter,
lower-mass members overestimates $\sigma$ relative to the genuinely different
low-luminosity distribution, while the S$^4$G reference fit, spanning both branches
of the broken relation, is pulled down at these same faint magnitudes by the
large number of low-luminosity ETGs following the shallower
$L\propto\sigma^{2\text{--}2.5}$ branch. The combination generates an
(artificial) apparent offset between the two samples, specifically at faint
luminosities ($M_i \gtrsim -22$; see their fig.~A1) rather than a genuine
physical bias. As demonstrated by 
\citet{2019ApJ...887...10S}, 
using an observationally consistent, broken $L$--$\sigma$ relation removes this
apparent discrepancy entirely.

\section{Statistical Methodology}
\label{Sec_Methodology}

\subsection{The Symmetric COvariance Population Estimator (\textsc{scope})}
\label{sec:scope_desc}

For ease of reference, the hierarchical Bayesian regression framework based on the intrinsic covariance of the latent population is denoted as the \emph{Symmetric COvariance Population Estimator}
(\textsc{scope}, hereafter \textsc{scope}). 
Throughout this work, the latent population inferred by \textsc{scope}
is assumed to represent the underlying galaxy population discussed in
Section~\ref{sec:sample_bias}. Consequently, the inferred covariance structure
and regression parameters are interpreted as intrinsic population
properties rather than corrections to an observationally biased sample.
Rather than treating either variable as intrinsically dependent upon the other, \textsc{scope} models the latent population as a bivariate normal distribution with mean vector $(\mu_x,\mu_y)$ and an intrinsic 
covariance matrix. Measurement uncertainties for both variables are incorporated directly into the hierarchical likelihood via latent-variable measurement models, including skewnormal models to report asymmetric uncertainties. 
The intrinsic correlation coefficient $\rho$ 
and dispersions $(\sigma_x,\sigma_y)$ are inferred simultaneously, with the 
regression slope obtained from the covariance structure as
\begin{equation}
\beta = \rho\,\frac{\sigma_y}{\sigma_x}, \nonumber
\end{equation}
rather than being treated as a free directional parameter.
This distinction is important because the scientific objective is to characterise the intrinsic scaling relation linking galaxy properties and black-hole mass. Predictive applications may subsequently be derived from the inferred covariance structure.

This formulation is invariant to exchanging the roles of X and Y: neither variable is treated as intrinsically dependent upon the other. In contrast to ordinary least-squares and other conditional regressions, which minimise residuals in a chosen direction, \textsc{scope} estimates the covariance structure of the underlying galaxy population itself. 
This makes the resulting regression slope and intrinsic scatter mathematically optimal 
not only for comparison with theoretical models of galaxy--black hole coevolution, 
but also for predicting the black hole mass of a new galaxy given its velocity dispersion. 
Because the derived parameters are recovered from the intrinsic population covariance, any subsequent predictions are corrected for the standard attenuation bias (regression dilution), which typically flattens slopes when predictor variables carry measurement uncertainty. Appendix~A presents the complete hierarchical Bayesian 
model, prior specification, and computational implementation. 

For every application presented here, four independent Markov chains were
run using Hamiltonian Monte Carlo with the No-U-Turn Sampler (NUTS) as
implemented in Stan. Convergence was assessed using the potential scale
reduction statistic ($\hat{R}$), effective sample sizes, and the absence of
divergent transitions. All analyses satisfied the standard convergence
criteria ($\hat{R}<1.01$ for every parameter, no divergent transitions, and
effective sample sizes exceeding 900).

The \textsc{scope} 
implementation used in this work is written in \textsc{R} and Stan and is publicly 
available as open-source 
software.\footnote{\url{https://github.com/A-Graham/SCOPE}}

\subsection{Sample Clipping for Regression Robustness}
\label{sec:sample_pruning}

To protect the global bivariate normal regressions from being disproportionately influenced by high-leverage points at the boundaries of the parameter space, five galaxies with particularly low velocity dispersions are pruned from the primary regression samples. 
Specifically, the following five systems are excluded from the main fitting runs:
\begin{itemize}
    \item the E galaxy NGC~5206 ($\log \sigma_0 = 1.62$~dex);
    \item the ES,e galaxy NGC~3377 ($\log \sigma_0 = 2.13$~dex);
    \item the S0 (dust $=$ N) galaxy NGC~7457 ($\log \sigma_0 = 1.83$~dex);
    \item the S0 (dust $=$ y) galaxy NGC~5102 ($\log \sigma_0 = 1.79$~dex); and
    \item the S0 (dust $=$ Y) galaxy NGC~404 ($\log \sigma_0 = 1.54$~dex).
\end{itemize}
NGC~7457 is a somewhat unrelaxed merger product with cylindrical rotation about its major axis \citep{2019MNRAS.488.1012M}. 
The bulgeless S galaxy NGC~4395 is already excluded (Section~\ref{Sec_Data}), but is also noted here as an undesirable high-leverage data point.

Each of these five galaxies stands out as a clear outlier in velocity
dispersion relative to the other members of its respective morphological
class, lying well below the bulk of its population's distribution in
Figure~\ref{Fig-M-sigma-2} 
and thereby acting as a high-leverage point capable
of disproportionately steepening or flattening the fitted slope.  
The
regression analysis assumes the `true' black hole masses and velocity
dispersions follow a Bivariate Normal (Gaussian) distribution
(Appendix~\ref{Sec_Appdx}), 
under which such extreme points can exert
outsized influence on the inferred slope and intercept.

These five galaxies are still displayed in the general scaling diagrams to
show the full distribution of the data, but they are excluded from the
primary \textsc{scope} fits. A quantitative evaluation of the sensitivity of
the derived scaling relations to the inclusion or exclusion of these points
is presented in 
Section~\ref{sec:exclusions_results}.

\subsection{Quantifying the Statistical Significance of Slope Differences}
\label{sec:slope_comparisons}

The uncertainties on the slope and intercept are larger than those obtained 
from classical regression methods such as BCES \citep[][]{1996ApJ...470..706A} or modified FITEXY \citep[][]{2002ApJ...574..740T}.  
This reflects the fact that the present hierarchical model infers the 
full intrinsic covariance structure of the bivariate population and 
propagates uncertainty in the intrinsic dispersions and correlation 
coefficient into the derived slope. In contrast, classical methods 
estimate the regression parameters directly under more restrictive 
assumptions, typically leading to smaller formal uncertainties.
Although {\em symmetric} uncertainties ($1\sigma$ posterior standard deviations) are reported throughout this work for simplicity, the \textsc{scope} code derives the full posterior distributions, seamlessly capturing and allowing for any underlying asymmetries in the inferred parameter uncertainties.

Because \textsc{scope} samples the full posterior probability distribution of every
model parameter, differences between regression slopes can be quantified directly
from the posterior draws rather than inferred from overlapping confidence intervals
or approximate error propagation. This provides an exact Bayesian probability that
one population possesses a steeper scaling relation than another.

The separate morphological subsamples (e.g., elliptical and lenticular galaxies)
are fitted independently, producing independent posterior samples for each regression.
The posterior probability that the slope of population $A$ exceeds that of population
$B$ ($P(\beta_A > \beta_B)$) is then estimated directly from the MCMC draws by 
calculating the proportion of pairwise MCMC draws for which the condition holds:
\begin{equation}
P(\beta_A > \beta_B) \approx \frac{1}{N} \sum_{i=1}^N \mathbb{I}\left(\beta_A^{(i)} > \beta_B^{(i)}\right), \nonumber 
\end{equation}
where $\beta_A^{(i)}$ and $\beta_B^{(i)}$ are the $i$-th posterior draws from the respective 
MCMC chains, $N$ is the total number of post-warmup sampling draws (assumed to be identical), 
and $\mathbb{I}$ is the indicator function equal to 1 if the inequality is true and 0 otherwise.

Using this metric, the statistical significance of the apparent slope differences 
between key populations discussed in 
Section~\ref{Sec_RandI} 
is quantified there.

\section{Results and Interpretation}
\label{Sec_RandI}

\subsection{The high-rise top-end of town: Elliptical galaxies and UMBHs}
\label{Sec_top}

\begin{figure}
\begin{center}
\includegraphics[width=1.0\columnwidth]{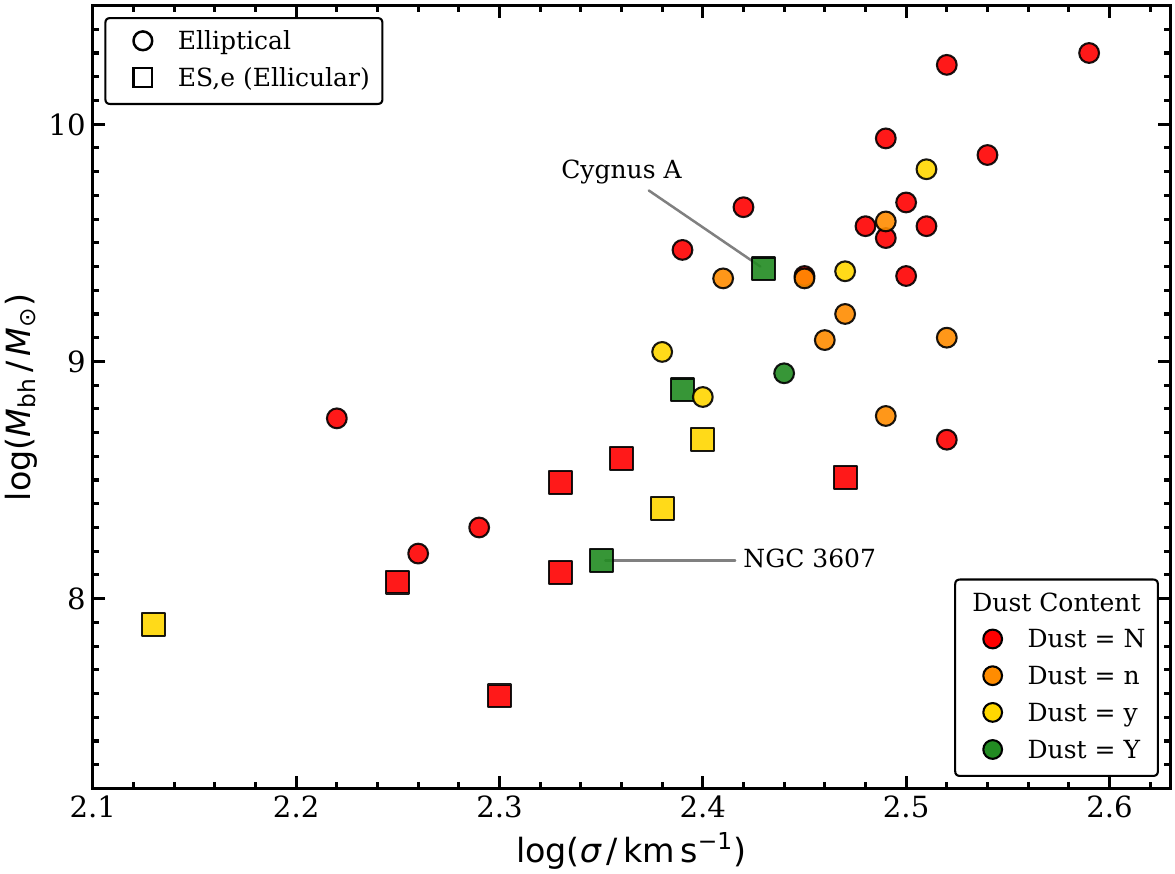}
\caption{
Zoomed-in portion of Figure~\ref{Fig-M-sigma-2} to better display the E and ES,e galaxies. 
The ES,e tend to reside towards the 
lower-left, while the E galaxies with nuclear dust discs tend to bridge the ES,e and the pure (discless) E galaxies in the upper-right.  This behavior was discovered in the $M_{\rm bh}$--$M_{\rm \star,sph}$ diagram by \citet{2026arXiv260424084G}.
}
\label{Fig-M-sigma-dust}
\end{center}
\end{figure}

Figure~\ref{Fig-M-sigma-dust} suggests the E galaxies with dust $=$ n (nuclear dust discs) tend to reside between the ES,e galaxies with intermediate-scale discs and the pure (discless) E galaxies. This structural pattern visually traces the `Disc Down-sizing' sequence (S0 $\rightarrow$ ES $\rightarrow$ E with dust $=$ n $\rightarrow$ E, including E with depleted cores) recently discovered by \citet{2026arXiv260424084G} in the $M_{\rm bh}$--$M_{\rm \star,sph}$ diagram. 

The concept that the Jeans-Reynolds sequence might reflect a merger-driven erosion of discs was foreshadowed by \citet[][their figure~8]{1994A&A...283....1N}. They distinguished `disky ellipticals' (with pointed isophotes) from `boxy ellipticals' (merger-remnants), proposing a physical continuity where `disky Es' are essentially S0s with faint or unfavourably oriented, and thus hard to detect, discs.  While their `disky E' category likely contained the galaxies now classified as ES,e, the distinction drawn here is structural and photometric. The ES,e galaxies in the `Triangal' are not simply S0s with faint large-scale discs; they possess \textit{intermediate-scale} discs that are fully embedded within the spheroid. This refinement, tied to the recognition of ES galaxies by \citet{1966ApJ...146...28L}, allows one to map the `Disc Down-sizing' sequence not just as a fading of light, but as a physical reduction of the disc scale relative to the bulge.

Built from relatively dry major mergers, E galaxies represent a departure from the gas-rich systems to which the theories of \citet{1998MNRAS.300..817H} and \citet{1998A&A...331L...1S} were intended.  

The high-mass end of the $M_{\rm bh}$--$\sigma_0$ diagram is of particular interest for the study of UMBHs. While early linear scaling relations often treated these giants as over-massive outliers \citep[e.g.,][]{2013ARA&A..51..511K}, the steep $M_{\rm bh}$--$\sigma_0$ relations established for core-S\'ersic galaxies \citep{2013ApJ...764..151G, 2019ApJ...887...10S} suggest that UMBHs are the natural, predictable result of hierarchical assembly. In particular, BCGs and some brightest group galaxies (BGGs) have been shown to follow a significantly steeper trajectory than lower-mass ETGs \citep{2018ApJ...852..131B}, yielding black hole masses of $10^{10}$--$10^{11}\,M_{\odot}$ for $337 \le \sigma_0 \le 428$, somewhat in line with the UMBHs originally predicted by \citet{1969Natur.223..690L}.

However, the `Triangal' framework suggests that further refinement is necessary to isolate the true UMBH scaling law. While BGGs are the most luminous members of their local environments, they are morphologically diverse and can include S0 or even S galaxies that follow different evolutionary tracks. In contrast, the most massive BCGs in rich clusters are almost exclusively E galaxies formed through multiple generations of dry mergers. By deriving the $M_{\rm bh}$--$\sigma_0$ relation specifically for these E-type BCGs --- that likely contain cores scoured of stars due to the coalescence of infalling massive black holes \citep{1980Natur.287..307B, 2007ApJ...671...53M, 2015ApJ...804..128M, 2017MNRAS.470..940M} --- and excluding the more heterogeneous BGG population, a more robust and physically consistent predictor for the UMBH regime could be established. 

Consistent with the virial-theorem argument of 
\citet{2023MNRAS.518.6293G}, 
this segregation is not merely taxonomic but reflects the different physical processes governing black hole and spheroid growth in cluster centres versus group environments. As galaxies advance along the `Disc Down-sizing' and `Dust Attrition' sequences into the high-mass E galaxy regime, the virial constraints on dry galaxy merging ensure that black hole mass increases more rapidly than velocity dispersion. Consequently, the steepening of the relation at high $\sigma_0$ is an expected signature of the transition from gas-rich accretion to the merger-dominated assembly of the Universe's largest black holes.

Eleven of the 13 BCGs in the sample are E galaxies; Cygnus~A and NGC~1275 are ES,e.
Combining the 13 BCGs with the other E and ES,e galaxies (excluding the two cE galaxies, NGC~4486B and M32)  
results in a distribution with a correlation strength of $0.87\pm0.08$. 
The associated scaling relation is given by 
\begin{equation}
\log M_{\rm bh} = (9.09 \pm 0.10) + (7.8 \pm 1.3) \left(\log \sigma_0 - 2.43 \right),
\label{Eq_ESeE}
\end{equation}
with a posterior-mean observed root-mean-square (rms) vertical scatter of 
$\Delta_{\rm rms} = 0.42 \pm 0.04$~dex, 
and an intrinsic scatter about the relation of $\sigma_{y|x} = 0.28 \pm 0.08$~dex.
Throughout this work, $\Delta_{\rm rms}$ denotes the posterior mean of the observed RMS vertical scatter evaluated across the posterior draws of the fitted relation, rather than the single-value classical RMS estimator.

\subsubsection{Ultra-massive black holes}

\begin{figure}
\begin{center}
\includegraphics[width=1.0\columnwidth]{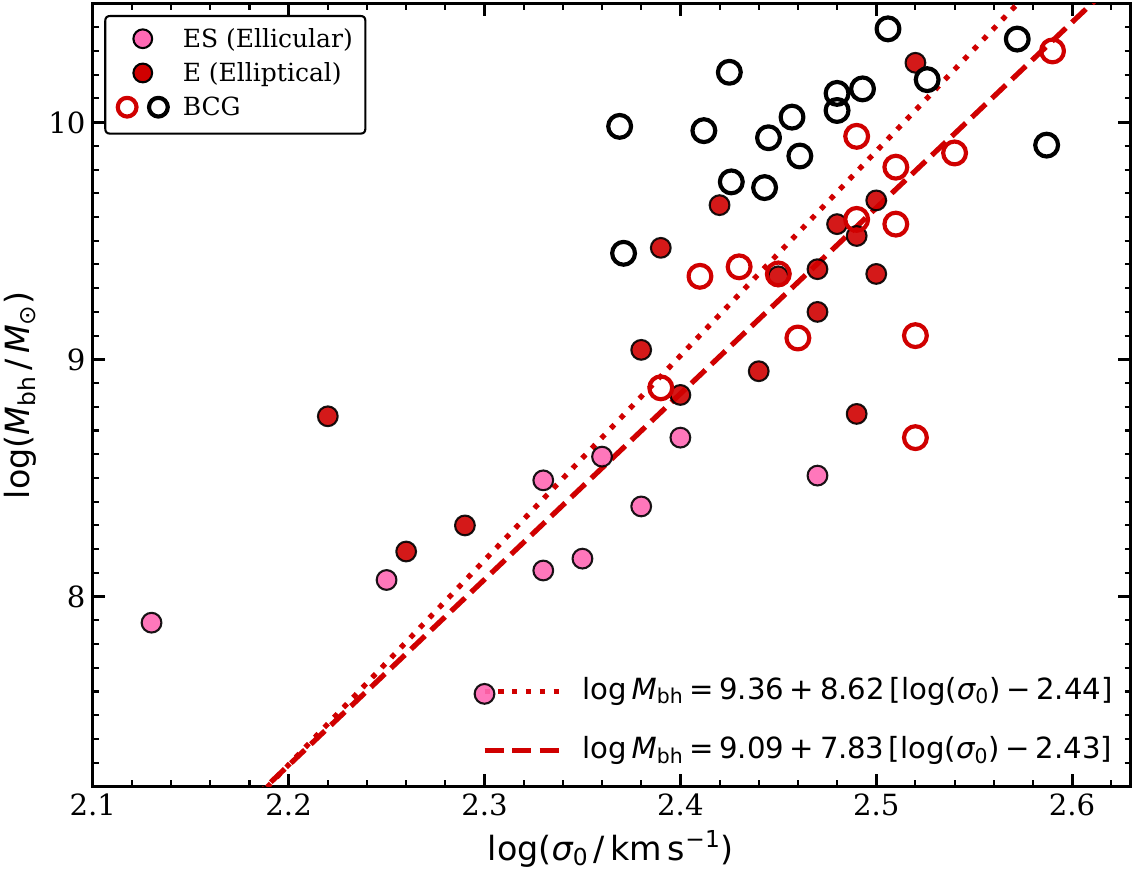}
\caption{
Zoomed-in variant of Figure~\ref{Fig-M-sigma-2} highlighting the high-mass regime dominated by E and ES,e galaxies. The 16 newly measured BCGs from \citet{2025arXiv251204178D} have been added as open black circles. The dashed line represents the scaling relation from Equation~\ref{Eq_ESeE} that was fit to the initial combined ES,e, E, and BCG sample, while the dotted line is the fit including the 16 new BCGs (Equation~\ref{Eq_ESeE_tentative}).
Shown here, but excluded from the fit, are NGC~3377 (the ES galaxy with the lowest $\sigma_0$ value) and NGC~4486B (the cE galaxy with $\log\sigma_0=2.22$~dex). 
}
\label{Fig-M-sigma-UMBH}
\end{center}
\end{figure}

Recently, \citet{2025arXiv251204178D} reported dynamically measured black hole masses and velocity dispersions for a sample of 16 BCGs, eight of which were reported to host UMBHs. Although the details of their triaxial orbit-based dynamical models were deferred to a companion paper (de Nicola et al., in prep.), this data provides a valuable opportunity to test the high-mass end of the scaling relations and has been incorporated into Figure~\ref{Fig-M-sigma-UMBH} (open black circles).\footnote{These results are preliminary pending the companion paper de Nicola et al., in prep.} 

These 16 additional BCGs preferentially populate the upper-mass region of the diagram, reinforcing the expected steepening of the distribution. When combined with the pre-existing sample of 13 BCGs, the new BCG population continues to span a broad range in black hole mass across a very narrow range in velocity dispersion, which precludes a robust regression on the BCGs alone. However, integrating these 16 massive systems into the broader sample of ES,e and E galaxies (including all BCGs) yields a revised fit. 
While the reported formal uncertainties on the velocity dispersion are very small for this additional sample of 16 BCGs \citep{2025arXiv251204178D}, a larger 10 per cent uncertainty has been adopted in \textsc{scope} to better capture additional issues, such as aperture effects, that can influence the value of $\sigma_0$. 
The updated regression (potentially revising Equation~\ref{Eq_ESeE}) is given by
\begin{equation}
\log M_{\rm bh} = (9.36 \pm 0.09) + (8.6 \pm 1.5) \left(\log \sigma_0 - 2.44 \right),
\label{Eq_ESeE_tentative}
\end{equation}
with $\Delta_{\rm rms} = 0.53 \pm 0.04$~dex
and $\sigma_{y|x} = 0.39\pm0.08$~dex. 
This revised slope of 8.6 further solidifies the dry-merger-driven steepening at the high-mass end of the $M_{\rm bh}$--$\sigma_0$ diagram.\footnote{To ensure that the posterior inferences are strictly data-driven, weakly informative priors scaled to the observed variance of the sample were employed. It was verified that adopting significantly broader priors --- including a uniform prior on the intrinsic correlation matrix (LKJ $\eta=1$, see Appendix~\ref{Sec_Appdx}) --- did not meaningfully alter the posterior medians or credible intervals, confirming that the results are, at the $1\sigma$ level, insensitive to the specific choice of regularising priors.  With LKJ($\eta=1$) and 5$\times$ broader population mean and intrinsic scatter priors applied simultaneously, the slope is $9.1\pm1.6$, while the pivots (9.36 and 2.44) are consistent at the 0.01 level.\label{foot_prior}}

It is visually apparent that the new data from \citet{2025arXiv251204178D} tend to reside on the low-velocity-dispersion side of the fitted relation. Rather than representing a physical deviation, this offset is highly likely to be a methodological artefact arising from how the velocity dispersions were extracted.  \citet{2025arXiv251204178D} report luminosity-weighted dispersions integrated over all available radial bins, thereby capturing the extended, dynamically cooler outer regions of these giant galaxies. In contrast, the $\sigma_0$ values used to calibrate the baseline relation herein are centrally concentrated measurements. Because the velocity dispersion profiles of massive ETGs typically decline with radius, integrating over a larger spatial aperture systematically depresses the measured $\sigma$ relative to the central $\sigma_0$, artificially shifting these data points to the left in the $M_{\rm bh}$--$\sigma_0$ diagram.

Beyond methodological aperture effects, this behaviour is explained by a well-established dynamical mechanism, demonstrated analytically for core-S\'ersic galaxies by \citet{2005MNRAS.362..197T} and \citet{2007MNRAS.377..855T}: a fundamental physical process that intrinsically deflates the central velocity dispersions of these extreme systems.  Massive E galaxies and most BCGs are predominantly built via dry major mergers, which inevitably lead to the formation of binary supermassive black holes. The orbital coalescence of these binaries scours the central stellar phase space, preferentially ejecting stars on radial orbits \citep{2005MNRAS.362..197T}. This excavation creates a phase-space `loss cone' that manifests photometrically as a flattened, depleted central core  \citep[e.g.,][]{1966ApJ...143.1002K, 1972IAUS...44...87K, 1991A&A...244L..37N, 1994AJ....108.1598F, 1996AJ....111.1889B}, best described by a core-S\'ersic light profile \citep{2003AJ....125.2951G, 2004ApJ...613L..33G}.  As demonstrated analytically for both spherical and triaxial \citep{2007MNRAS.377..855T} core-S\'ersic models, the removal of these inner, high-radial-velocity  stars fundamentally alters the dynamical structure of the galaxy's centre, resulting in a suppressed central velocity dispersion. Consequently, the very core-scouring process that accommodates the growth of UMBHs simultaneously deflates the host galaxy's $\sigma_0$. This physical reduction operates in tandem with the virial mechanics of dry merging, further steepening the high-mass end of the $M_{\rm bh}$--$\sigma_0$ relation.

\subsection{A central \texorpdfstring{$M_{\rm bh}$--$\sigma_0$}{Mbh--sigma0} relation}
\label{Sec_central}

Having addressed the E, ES,e, and BCGs, attention is now turned to the disc galaxies, including S and S0 galaxies. 
It is apparent from Figure~\ref{Fig-M-sigma-2} that the S galaxies, at least in the current sample, do not define a tight correlation between black hole mass and central velocity dispersion. This will be addressed in Section~\ref{Sec_spiral}, and may be due to the recurrence of bars. Here, the focus is initially on spiral-less, dust-rich galaxies (predominantly) formed by wet major mergers.

\begin{figure}
\begin{center}
\includegraphics[width=1.0\columnwidth]{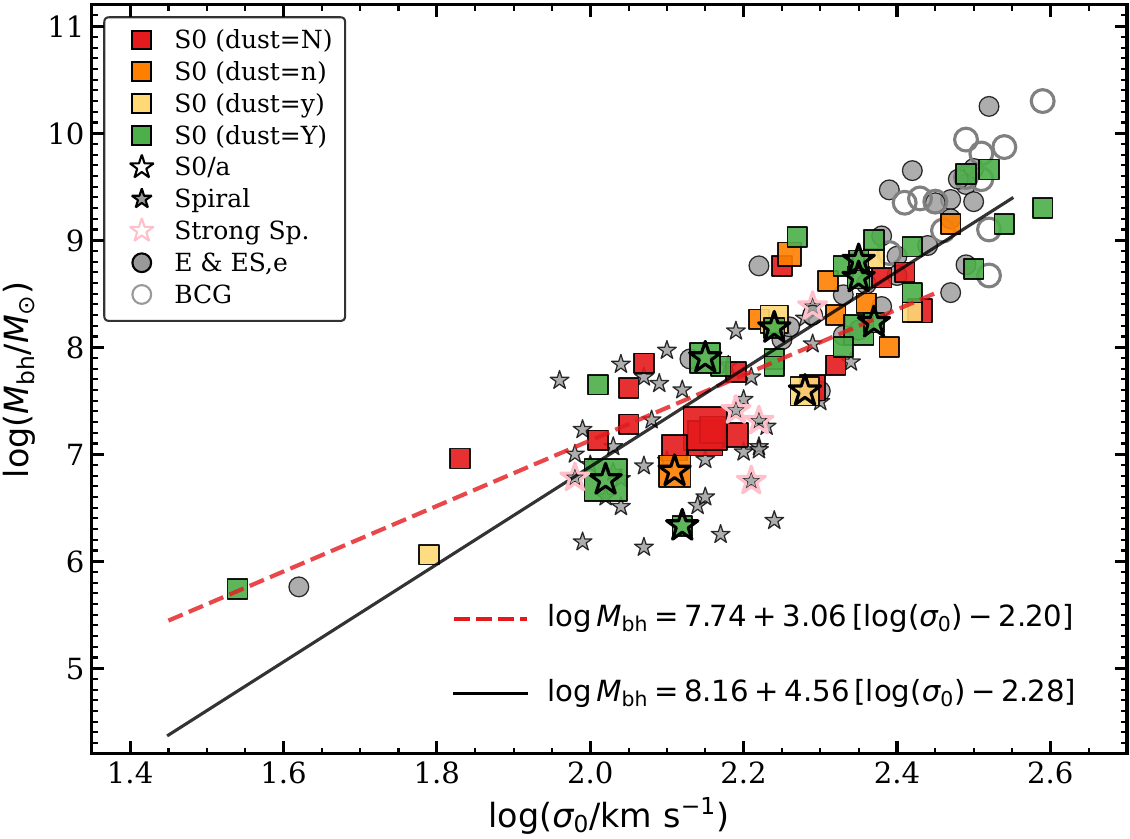}
\caption{
Variant of Figure~\ref{Fig-M-sigma-2} to better display the S0 galaxies.   
The symbol size still reflects the bar strength, but it now 
includes the contribution from barlenses and inner discs for five galaxies, as noted in Section~\ref{Sec_flat}. 
There is a tendency at $M_{\rm bh} \lesssim 10^7$ M$_\odot$ for strongly barred S0 and S0/a galaxies to reside to the right of the $M_{\rm bh}$--$\sigma_0$ relation defined by the ensemble of S0 galaxies (Equation~\ref{Eq_S0_all}) and shown by the solid line.  This is not too dissimilar to the relation defined by the dust-rich S0 galaxies (Equation~\ref{Eq_wet}). 
The dashed red line is the fit to the primeval dust $=$ N S0 galaxies (Equation~\ref{Eq_dust-N}). 
The four galaxies with the lowest velocity dispersions were not used to derive the relations shown here. 
}
\label{Fig-M-sigma-S0}
\end{center}
\end{figure}

These dust-rich (dust $=$ Y) S0 galaxies define a scaling relation given by 
\begin{equation}
\log M_{\rm bh} = (8.38 \pm 0.17) + (4.43 \pm 0.77) \left(\log \sigma_0 - 2.32 \right),
\label{Eq_wet}
\end{equation}
with the scatter reported in Table~\ref{Table_rel}. 
Including the ES,e galaxies, with their intermediate-scale discs, one obtains the slightly steeper relation 
\begin{equation}
\log M_{\rm bh} = (8.36 \pm 0.13) + (4.56 \pm 0.67) \left(\log \sigma_0 - 2.33 \right). 
\label{Eq_dusty-ESe}
\end{equation}
Although ES,e galaxies are photometrically more akin to E galaxies than S0 galaxies, their abutment with dust $=$ Y S0 galaxies in Figures~\ref{Fig-M-sigma-2} and \ref{Fig-M-sigma-S0}, and their link in the galactic speciation chain, makes it reasonable to consider a dust $=$ Y S0 galaxy plus ES,e galaxy sequence. Such a relation also removes the need to separate ES,e galaxies from dust $=$ Y S0 galaxies in integral field spectroscopy (IFS) samples of galaxies that contain substantial rotation over the inner IFS-sampled region, but with uncertainty as to whether an intermediate- or a large-scale disc exists due to the limited field-of-view or limited signal-to-noise at large radii.

Removing the ES,e galaxies and combining all the S0 galaxies together (dust $=$ Y, y, n, N) gives: 
\begin{equation}
\log M_{\rm bh} = (8.16 \pm 0.11 ) + (4.56 \pm 0.49) \left(\log \sigma_0 - 2.28 \right). 
\label{Eq_S0_all}
\end{equation}
This relation is shown in Figures~\ref{Fig-M-sigma-2} and \ref{Fig-M-sigma-S0}.   
From Figure~\ref{Fig-M-sigma-2}, it is apparent how the inclusion of E galaxies (built from dry mergers) 
would act to steepen the slope of the $M_{\rm bh}$--$\sigma_0$ relation.

For the convenience of theoretical modellers and simulators requiring physically motivated scaling relations, Table~\ref{Table_rel} provides a comprehensive summary of the optimal, morphology-specific regression parameters (slopes, intercepts, and intrinsic scatters) to be utilised for population synthesis and galaxy mock catalogues.

The somewhat stochastic nature of accretion and mergers means that the separation, i.e., the different location of different  galaxy morphologies ---  itself a reflection of a galaxy's accretion and merger history --- is not perfectly clean in the $M_{\rm bh}$--$\sigma_0$ diagram.  Nonetheless, a mass-dependent pattern exists, with `wet' mergers, then possibly `damp' mergers, and finally `dry' mergers, sequentially defining the higher-mass ETGs \citep{2026arXiv260424084G}.\footnote{S0 galaxies in the dust $=$ n bin may have formed from a `damp' merger, in which the little available cold gas formed a nuclear disc. In parallel, some of these systems might have  small nuclear discs today because their once larger gas/dust discs have been heavily pared-back if embedded in a group- or cluster-sized halo of hot X-ray-emitting gas.} 
Low-mass dust-poor S0 galaxies can also merge, bypassing the `wet road' to galaxy formation, although the probability of this occurring in clusters (where many such gas-poor galaxies reside) would have diminished over time as the clusters grew in mass and the faster speeds of these low-mass galaxies inhibited them from falling back in on each other rather than zipping past each other. 
Despite the stochasticity, 
it can be seen in Figure~\ref{Fig-M-sigma-S0} that dust $=$ n S0 galaxies somewhat bridge the dust $=$ N and dust $=$ Y S0 sample of galaxies.
This behaviour was first seen in the $M_{\rm bh}$--$M_{\rm \star,gal}$ diagram shown in \citet{2026arXiv260424084G}, where it was dubbed the `Dust Attrition' or `Dust Retention' sequence.

\subsection{The flats, the bottom end of town, and a possibly `hidden' bar signal in dust-poor S0 galaxies}
\label{Sec_flat}

The separation of galaxy types in Figure~\ref{Fig-M-sigma-2} reveals distinct substructures often averaged over in previous studies.
Here, the focus is turned to the dust-poor (dust $=$ N) S0 galaxies. Their scaling relation is consistent with  
\begin{equation}
\log M_{\rm bh} = (7.74 \pm 0.16) + (3.1 \pm 1.1) \left(\log \sigma_0 - 2.20 \right)
\label{Eq_dust-N}
\end{equation}
and shown in Figure~\ref{Fig-M-sigma-S0}. 
Including the three S0 galaxies (NGC~3091, NGC~3640, and NGC~4649) that had been excluded because they were (appropriately) reassigned from the dust $=$ N bin to the (appropriate evolutionary stage) dust $=$ Y bin, 
one obtains the (probably inappropriately steep) relation 
$\log M_{\rm bh} = (7.95 \pm 0.19) + (4.35 \pm 0.92) \left(\log \sigma_0 - 2.23 \right)$, 
with $\Delta_{\rm rms} = 0.46 \pm 0.04$~dex   
and $\sigma_{y|x} = 0.46\pm0.11$~dex.         
%

While large-scale bars do not appear to dictate the position of galaxies in the $M_{\rm bh}$--$M_\star$ diagrams \citep{2026arXiv260424084G}, their dynamical influence on the $M_{\rm bh}$--$\sigma_0$ relation remains an open question.
As noted in the Introduction, dynamical studies indicate that bars increase the vertical velocity dispersion and drive radial motions. However, this offset was not clearly evident in the analysis of \citet{2019ApJ...887...10S}. 

A key insight emerges when comparing the sub-samples in Figure~\ref{Fig-M-sigma-2}: the long-sought `bar offset' --- where barred galaxies reside at higher $\sigma_0$ than unbarred galaxies --- may be hidden within the noise of the full galaxy population.
In the S galaxy sample (stars), the dynamical signal of the bar may be drowned out by competing effects (discussed in Section~\ref{Sec_scat}). An S galaxy might have a bar heating it (pushing $\sigma_0$ up), spiral arms, or dynamically-cold disc contamination (pushing $\sigma_0$ down), resulting in no net offset, just scatter.
However, the dust-poor S0 galaxies (dust $=$ N, red squares) represent a `cleaner' dynamical laboratory. They lack the cold gas, dust, and young stars that complicate the measurement in S galaxies. In this more pristine environment, the bar's dynamical heating is the \textit{dominant} secondary effect. 
Consequently, when the dust-poor S0 galaxies are isolated, this is interpreted here as possibly revealing the dynamical signature of bars: a systematic offset to higher $\sigma_0$ compared to their unbarred counterparts.
The failure of previous studies to confirm this offset likely stemmed from diluting this clean S0 sample with noise-dominated S galaxies.

\citet{2026arXiv260424084G} reported the presence of inner lenses and inner discs in six galaxies in the current sample.  If these features are included with the bar flux, then four of these galaxies (NGC~2787, NGC~4371, NGC~4596, and NGC~6926) would be considered to have a strong bar rather than a weak bar.  NGC~6926 appears to be bulgeless and is not included here, reducing the sample to five.\footnote{The unbarred galaxy NGC~2748 and the weakly-barred galaxy NGC~4395 were the only other bulgeless galaxies from the initial sample of 137.}  Of the two other galaxies, IC~2560 retains its weak bar classification, and NGC~4762 (which was the only dust $=$ N S0 of this initial set of six) was already considered strongly barred.  Figure~\ref{Fig-M-sigma-S0} shows the results of applying these adjustments, with an emphasis on the S0 galaxies.

Removing the three dust $=$ N S0 galaxies with strong bars (NGC~3384, NGC~4762, and NGC~2549), equation~\ref{Eq_dust-N} adjusts to become\footnote{As with footnote~\ref{foot_prior}, 
replacing the $\eta=2$ LKJ prior with a uniform $\eta=1$ prior, and then additionally broadening the scale of the intrinsic scatter and population mean priors by a factor of 5, changes the slope at less than the $1\sigma$ level, yielding $2.95\pm1.08$ and then $3.03\pm1.10$.}
\begin{equation}
\log M_{\rm bh} = (7.86 \pm 0.18) + (2.6 \pm 1.1) \left(\log \sigma_0 - 2.21 \right).
\label{Eq_dust-bar}
\end{equation}
With those three removed, and now including NGC~7457 --- the dust $=$ N S0 with the lowest velocity dispersion --- the result even more strongly supports a decreased slope with a fit given by 
\begin{equation}
\log M_{\rm bh} = (7.79 \pm 0.17) + (2.54 \pm 0.89) \left(\log \sigma_0 - 2.18 \right). 
\label{Eq_S0_N_edit}
\end{equation}
However, larger numbers of galaxies will be required to better probe the tentative offset of strongly barred, dust-poor S0 galaxies and solidify the trend in which the $M_{\rm bh}$--$\sigma_0$ relation becomes shallower as one progresses to more primeval S0 galaxies, potentially unaffected by the development of bars.  Equation~\ref{Eq_dust-N} is therefore regarded as the primary relation for low-mass, dust-poor S0 galaxies. 
Given the small sample size and large uncertainties, these slopes should be regarded as indicative rather than definitive.

In passing, it is noted that the dust $=$ Y S0 galaxy with a strong bar and the highest $\sigma_0$ value of the three such galaxies in Figure~\ref{Fig-M-sigma-S0} is NGC~1316 (Fornax~A). Its bar-like feature may not be a true bar but rather an unrelaxed feature of its merger history.

\begin{table*}
\centering
\caption{Galaxies with low velocity dispersions and the few excluded galaxies noted in Section~\ref{Sec_Data}.  Systems above the line are from the primary sample described at the start of Section~\ref{Sec_Data} and 
shown in Figure~\ref{Fig-M-sigma-2}, while systems below the line are new inclusions.  
The uncertainties on the black hole masses are the reported 1$\sigma$ values from their respective papers. 
The full galaxy sample is provided through GitHUB as a machine readable table accompanying the code \textsc{scope}. 
Reference: 
[1] = HyperLeda; 
[2] \citet{2006MNRAS.369.1321D}; 
[3] = I.\ Chilingarian 2011, private communication; 
[4] = \citet{2009ApJS..183....1H}. 
$^{\dag}$ = bulgeless. 
}
\label{Table_IMBH}
\renewcommand{\arraystretch}{1.3} 
\begin{tabular}{lcclcll}
\hline
\hline
Galaxy   & $\log(M_{\rm bh}/M_\odot)$ & $\log(\sigma_0/{\rm km\,s^{-1}})$ &  Type  & Dust & Dist  & $M_{\rm bh}$ reference \\
         &         (dex)              &            (dex)                  &        & Bin  & (Mpc) &   \\
\hline
NGC 0404  & 5.74$^{+0.11}_{-0.07}$    &        1.54$^{+0.06}_{-0.06}$ [1] &  SA0   &  Y   &  3.06 & \citet{2020MNRAS.496.4061D} \\
NGC 3377 &  7.88$^{+0.06}_{-0.06}$  &        2.13$^{+0.06}_{-0.06}$ [1] & ES,e   &  y   & 10.8  & \citet{2004AandA...415..889C} \\
NGC 5102 & 6.06$^{+0.02}_{-0.03}$     &        1.79$^{+0.04}_{-0.05}$ [1] &  SA0   &  y   &  4.00 & \citet{2019ApJ...872..104N} \\ 
NGC 5206 & 5.76$^{+0.05}_{-0.13}$     &        1.62$^{+0.06}_{-0.06}$ [1] &  dE    &  N   &  3.20 & \citet{2019ApJ...872..104N} \\ 
NGC 7457 & 6.96$^{+0.20}_{-0.40}$     &        1.83$^{+0.06}_{-0.06}$ [1] &  SA0   &  N   & 12.70 & \citet{2011ApJ...729...21S} \\
\hline
NGC 0205  &  3.83$^{+0.43}_{-0.60}$   &        1.52$^{+0.06}_{-0.06}$ [2] &  dS0   &  y   & 0.82 & \citet{2019ApJ...872..104N}  \\ 
NGC 0221 &  6.38$^{+0.03}_{-0.06}$    &        1.74$^{+0.06}_{-0.06}$ [3] &  dS0   &  N   & 0.75 & \citet{2018ApJ...858..118N} \\ 
NGC 2748$^{\dag}$ &  7.68$^{+0.14}_{-0.23}$ &  1.98$^{+0.04}_{-0.04}$ [4] & SAbc   &  Y   & 25.1 & \citet{2005MNRAS.359..504A} \\ 
NGC 4395$^{\dag}$ & 5.62$^{+0.48}_{-0.27}$ &   1.42$^{+0.06}_{-0.06}$ [1] & SABm   &  Y   &  4.56 & \citet{2015ApJ...809..101D} \\ 

\hline
\end{tabular}
\end{table*}

\subsubsection{Towards intermediate-mass black holes}
\label{Sec_flatter}

While a few galaxies with AGN have been inferred to contain IMBHs, their black hole masses are based on calibrations using the black hole scaling relations for galaxies with directly measured black hole masses.  As such, it is circular logic and not appropriate to include those systems in these current scaling diagrams.

\begin{figure}
\begin{center}
\includegraphics[width=1.0\columnwidth]{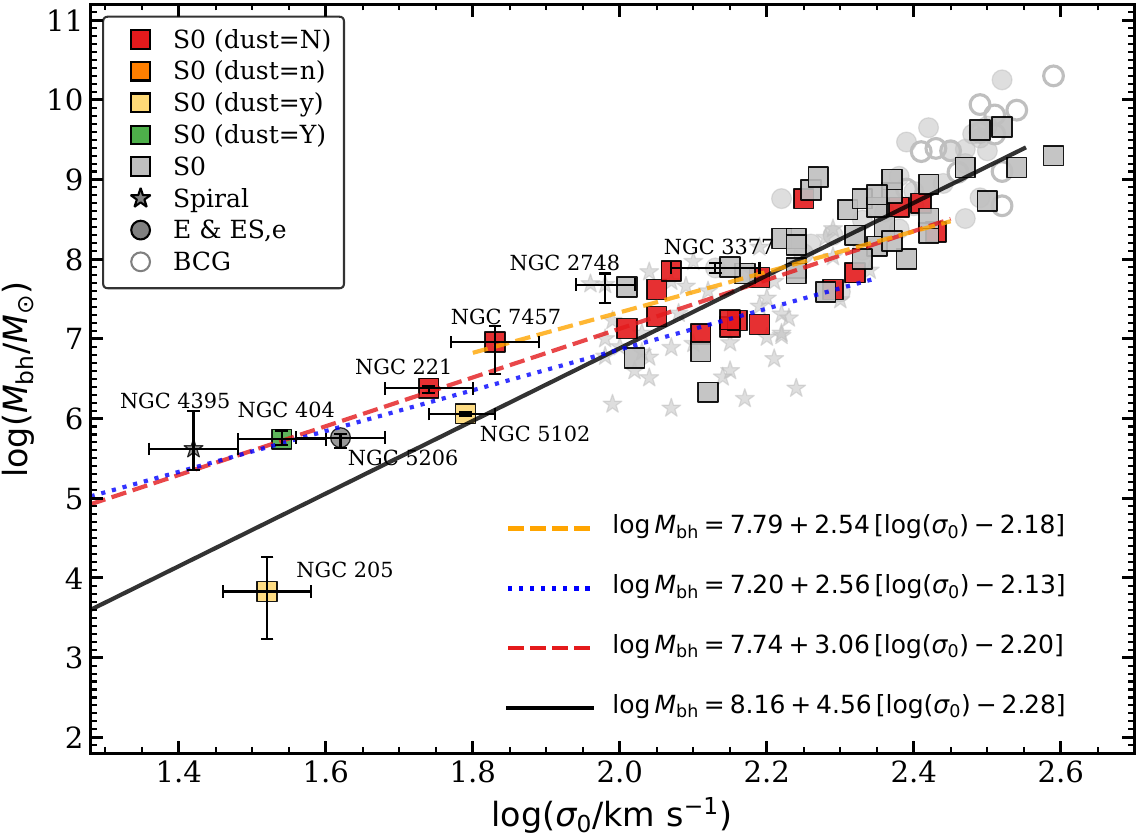}
\caption{
Variant of Figure~\ref{Fig-M-sigma-2} to better display the departure at low velocity dispersions.
The thick solid line (Equation~\ref{Eq_S0_all}) represents the ensemble of S0 galaxies, and 
the dashed red line is (still) the fit to the primeval dust $=$ N S0 galaxies (Equation~\ref{Eq_dust-N}). 
The dashed orange line with the shallower slope is Equation~\ref{Eq_S0_N_edit}, which excludes all three strongly barred dust $=$ N S0 galaxies but includes NGC~7457 (possibly biasing the slope), and the dotted blue line is the relation for the S galaxies (Equation~\ref{Eq_S}).
With the exception of NGC~7457 for Equation~\ref{Eq_S0_N_edit}, the galaxies labelled here and listed in Table~\ref{Table_IMBH} were not used to derive the relations shown here. }
\label{Fig-M-sigma-IMBH}
\end{center}
\end{figure}

To date, there are very few direct (and robust) measurements of IMBHs.
However, there are a few in the $10^5$--$10^6$ M$_\odot$ mass range.
Table~\ref{Table_IMBH} lists those galaxies from \citet{2026arXiv260424084G} without bulges or with the lowest velocity dispersion for their morphological type and dust bin, that were 
noted in Section~\ref{Sec_Data} and excluded from the linear regressions.
Two additional galaxies are noted in Table~\ref{Table_IMBH} and shown in Figure~\ref{Fig-M-sigma-IMBH}.  They are NGC~205, which has a rather insecure black hole mass measurement when considering the 3$\sigma$ uncertainty ($M_{\rm bh}=6.8^{+95.6}_{-6.7}\times10^3$ M$_\odot$), and NGC~221 (M32), which has a reliable black hole mass, but the galaxy has likely been heavily stripped of stars, placing a question mark over the applicability of its (altered?) velocity dispersion.

Comprehensive reviews of the IMBH regime \citep[e.g.,][]{2020ARA&A..58..257G} have frequently relied on the extrapolation of monolithic, single-power-law scaling relations derived from higher-mass, mixed-morphology samples. This practice overlooks the morphology-dependent relations established by, for example, \citet{2019ApJ...876..155S} and \citet{2019ApJ...887...10S}, which demonstrated that different galaxy types follow distinct evolutionary tracks. As established herein, the $M_{\rm bh}$--$\sigma_0$ relation for the primeval, dust-poor S0 galaxies --- which occupy the low-mass regime where many IMBH candidates are sought --- may possess a notably shallow slope of $\sim$2.5 to 3.0. In contrast, extrapolating a steeper monolithic relation (with slopes of $\sim$4.5--5.0) down to velocity dispersions below $100\ \rm{km\ s^{-1}}$ would systematically underpredict the expected black hole mass. Consequently, when IMBHs are detected in dETGs, they risk being erroneously classified as ``over-massive'' outliers. This apparent tension is largely an artefact of comparing these low-mass systems against a baseline that may be inappropriately steep, underscoring the necessity of applying morphology-aware scaling relations when probing the IMBH domain.  Section~\ref{sec:feedback_and_mergers} explains why a shallow slope of $\sim$2.5 (Equations~\ref{Eq_dust-bar}--\ref{Eq_S0_N_edit}) may be expected for the dETGs in the $M_{\rm bh}$--$\sigma_0$ diagram, giving rise to the set of three relations advocated in Section~\ref{Sec_unknown}.

\begin{figure*}
\begin{center}
\includegraphics[width=1.0\columnwidth]{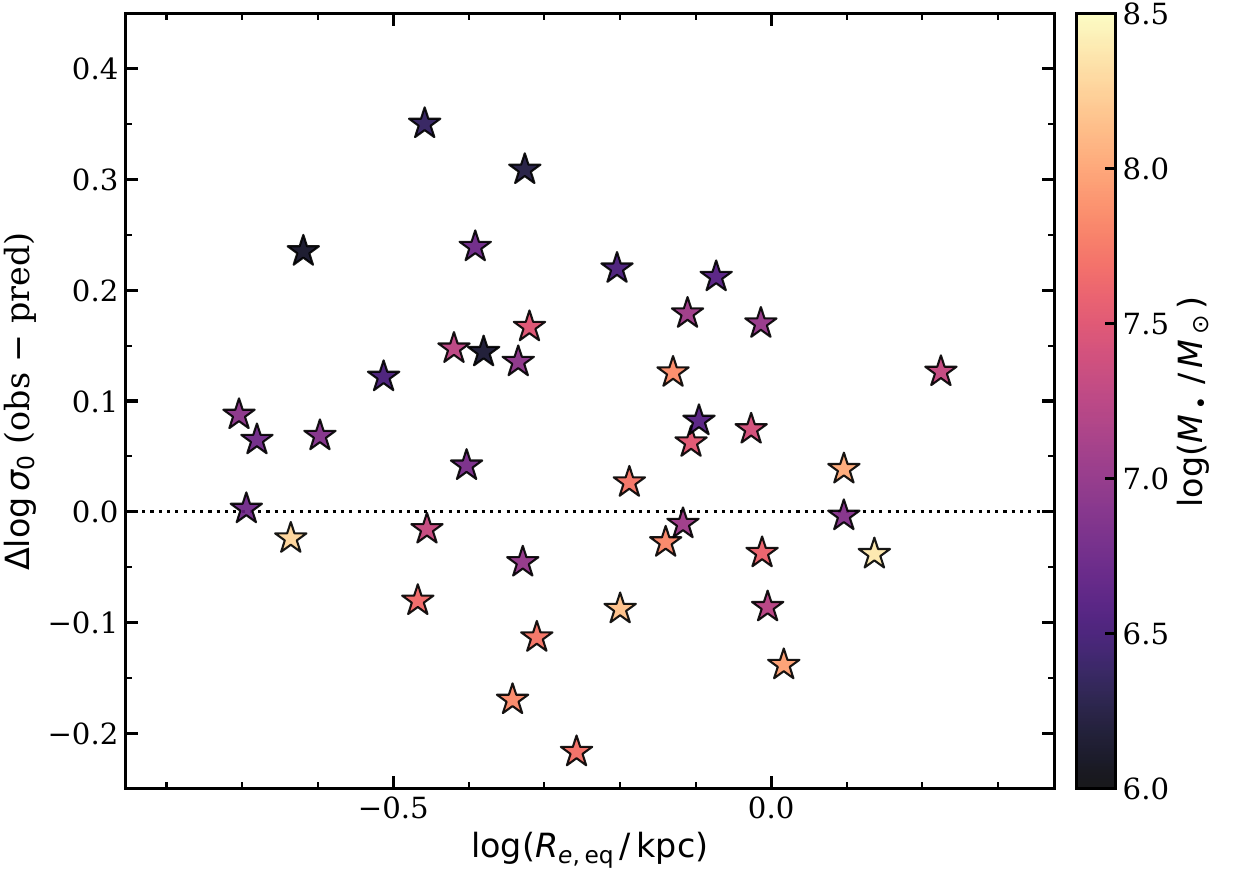}
\includegraphics[width=0.98\columnwidth]{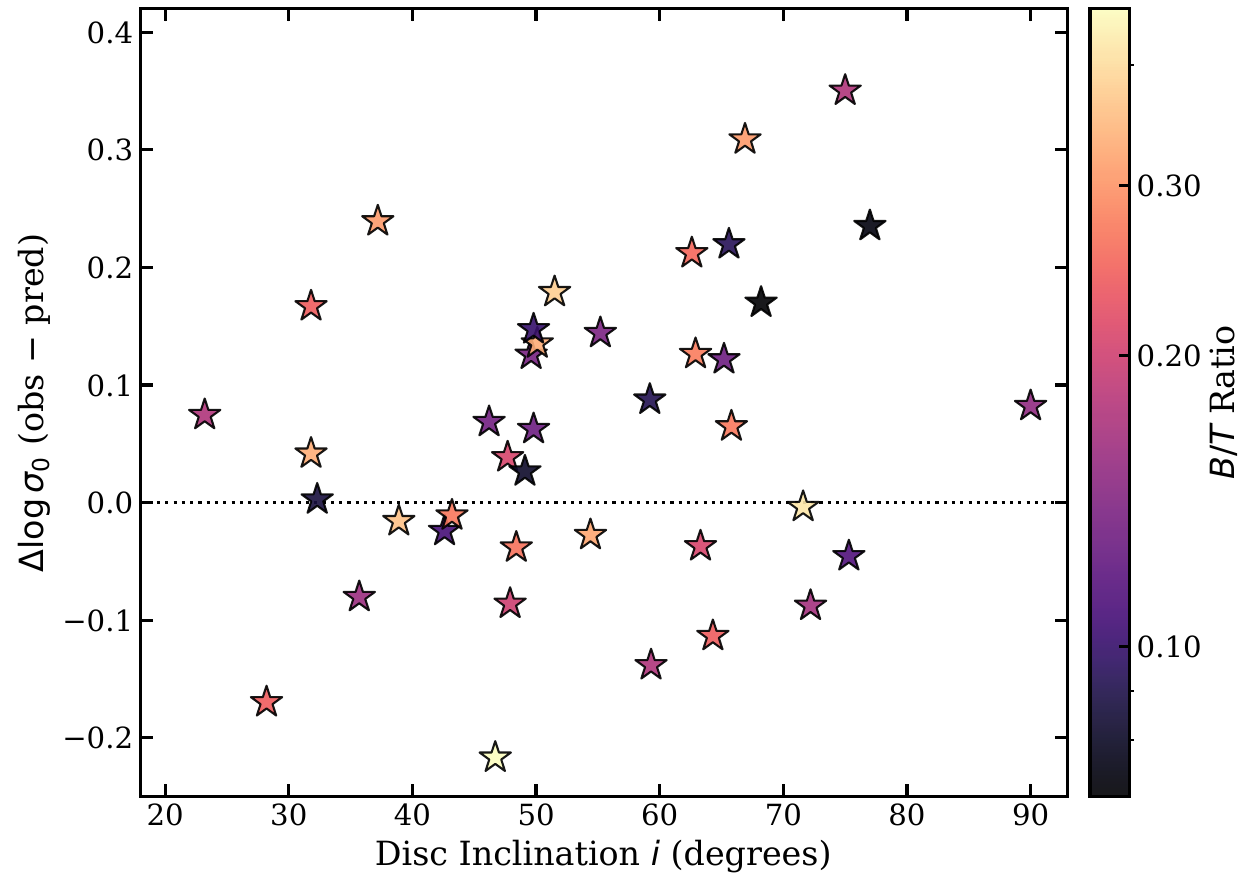}
\caption{Left: horizontal kinematic offset, $\Delta \log \sigma_0$ (see Section~\ref{Sec_spiral}), plotted as a function of the geometric-mean axis' half-light radii of the S galaxy bulges.  
The data points are colour-coded by the black hole mass.
Right: $\Delta \log \sigma_0$ of the S galaxy sample versus their disc inclination. Here, the data points are colour-coded by their bulge-to-total ($B/T$) stellar mass ratio on a logarithmic scale.
A weak positive correlation (Pearson $r = 0.26$, $p=0.15$) illustrates the expected kinematic inflation from line-of-sight disc rotation, though the highly scattered distribution of $B/T$ ratios confirms that inclination alone is insufficient to correct for the competing kinematic biases present in S galaxies.
}
\label{Fig-M-sigma-size}
\end{center}
\end{figure*}

\subsection{Weak and strong spiral patterns}
\label{Sec_spiral}

\citet{2026arXiv260424084G} identified seven likely (and two possible) S0/a galaxies, i.e., systems exhibiting weak spiral structure (Figure~\ref{Fig-M-sigma-S0}). Of these, four with the lowest black hole masses reside in the region occupied by many S galaxies, offset to the right of the $M_{\rm bh}$--$\sigma_0$ relation defined by the dust $=$ N S0 galaxies. Notably, three of these four host strong bars. This subset of four includes the suspected faded spiral NGC~4371 (dust $=$ n) and  
NGC~2787 (dust $=$ y), which may be a primeval S0 galaxy accreting gas and dust. The two lower-(black hole mass) dust $=$ Y S0/a galaxies are possibly more advanced in terms of accretion and minor mergers. The five higher-black-hole-mass (dust $=$ Y) S0/a galaxies are wet major merger products.  

The two more uncertain S0/a classifications are NGC~4594 (Sombrero), which has the highest black hole mass in this group of nine, and Mrk~1029, which has the lowest and exhibits shells and ripples indicative of past merger activity.

The S galaxies themselves display substantial scatter in the $M_{\rm bh}$--$\sigma_0$ diagram. The relation obtained by \textsc{scope} is
\begin{equation}
\log M_{\rm bh} = (7.20 \pm 0.09) + (2.56 \pm 0.98) \left(\log \sigma_0 - 2.13 \right),
\label{Eq_S}
\end{equation}
with an observed rms vertical scatter of 0.54~dex\footnote{The observed RMS (posterior mean), which marginalises over uncertainty in the fitted relation parameters and is therefore slightly larger, is $0.55 \pm 0.02$~dex.} and shown in Figure~\ref{Fig-M-sigma-IMBH}. This scatter is larger than that about the relations for the other morphological subsamples, with the exception of Equation~\ref{Eq_ESeE_tentative} that included the tentative UMBH sample; however, that relation has a much steeper slope, which amplifies the vertical scatter.  
The intrinsic population correlation for the S galaxies is comparatively weak, with $\rho = 0.42\pm0.14$.
The relatively shallow conditional slope reflects this weak intrinsic correlation. 
Within the adopted bivariate normal framework (Appendix~\ref{Sec_Appdx}), the regression slope depends not only on the ratio of the marginal dispersions, but also on the intrinsic correlation coefficient, such that $\beta = \rho(\sigma_y/\sigma_x)$. 
Although the S galaxies span a broad range in $M_{\rm bh}$ relative to their range in $\sigma_0$, the weak correlation implies that $\sigma_0$ provides only limited predictive leverage on $M_{\rm bh}$. Consequently, the inferred conditional relation is substantially flatter, and the posterior uncertainty on the slope correspondingly broad ($\pm0.98$). 
This large uncertainty is itself informative, indicating that the current S galaxy sample provides only weak constraints on the conditional $M_{\rm bh}$--$\sigma_0$ relation. 
This contrasts with the S0 galaxy populations, where stronger intrinsic correlations yield more stable and visually intuitive regression slopes. 
Despite this, Equation~\ref{Eq_S} is consistent with the location of the S galaxy NGC~4395, even though that galaxy was excluded from the regression owing to its position at the extremity of the S galaxy distribution in the $M_{\rm bh}$--$\sigma_0$ diagram.

The subset of galaxies with particularly strong spiral patterns \citep[identified in][]{2026arXiv260424084G} does not occupy a distinct region of the $M_{\rm bh}$--$\sigma_0$ diagram (see Figure~\ref{Fig-M-sigma-S0}).

\begin{table*}
\centering
\caption{Posterior mean parameters for the symmetric covariance fits to various galaxy sub-samples of size $N$.  The scaling relations take the pivot-centred form: $\log(M_{\rm bh}) = \mu_{\rm bh} + \beta[\log(\sigma_0) - \mu_\sigma]$, 
where $\mu_{\mathrm{bh}} \equiv \mu_y$ and $\mu_\sigma \equiv \mu_x$ in the notation of Appendix~\ref{Sec_Appdx}. 
All uncertainties represent the $1\sigma$ posterior standard deviations. 
Because the regression slope $\beta = \rho\,\sigma_y/\sigma_x$ is derived 
from the intrinsic population covariance, it is mathematically equivalent 
to the conditional expectation $\mathrm{E}[Y|X]$ while remaining 
directionally invariant. The reported slope ($\beta$) and intrinsic scatter 
($\sigma_{y|x}$) therefore provide the statistically optimal parameters for 
predicting black hole masses from observed velocity dispersions, as well as 
for theoretical population modelling. 
The observed root mean square (posterior mean) is also provided.
The relation defined by Sample~1, and tentatively by Sample~2, is applicable to massive E and ES,e galaxies formed by multiple mergers, and for probing UMBHs.
The relation defined by Sample~5 applies to typical high-surface-brightness S0 galaxies.
The relation defined by Sample~6 is applicable for primeval dETGs --- many of which will be dS0 galaxies with thick discs or rotating oblate-to-triaxial structures --- and probing IMBHs.
As revealed by the intrinsic population correlation, $\rho$, there is no strong $M_{\rm bh}$--$\sigma_0$ relation for the S galaxies.}
\label{Table_rel}
\renewcommand{\arraystretch}{1.3} 
\begin{tabular}{l l c c c c c c c}
\hline
\hline
\# & Sample & $N$ & $\mu_{\rm bh}$         & $\mu_\sigma$              & $\beta$ (Slope) & $\rho$ & $\sigma_{y|x}$ & $\Delta_{\rm rms}$ \\
   &        &     & $(\log {\rm M_\odot})$ & $(\log {\rm km\,s^{-1}})$ &                 &        & (dex) & (dex) \\
\hline
1. & {\bf (13 BCG) + E + ES,e}                      & 38 & $9.09 \pm 0.10$ & $2.43 \pm 0.01$ & $7.8 \pm 1.3$ & $0.87 \pm 0.08$ & $0.28 \pm 0.08$ & $0.42 \pm 0.04$ \\
2. & (13 + 16 BCG) + E + ES,e                       & 54 & $9.36 \pm 0.09$ & $2.44 \pm 0.01$ & $8.6 \pm 1.5$ & $0.80 \pm 0.09$ & $0.39 \pm 0.08$ & $0.53 \pm 0.04$ \\
3. & S0 (dust = Y)                                  & 23 & $8.38 \pm 0.17$ & $2.32 \pm 0.03$ & $4.43 \pm 0.77$ & $0.82 \pm 0.09$ & $0.44 \pm 0.10$ & $0.47 \pm 0.03$ \\
4. & S0 (dust = Y) + ES,e                           & 32 & $8.36 \pm 0.13$ & $2.33 \pm 0.02$ & $4.56 \pm 0.67$ & $0.83 \pm 0.08$ & $0.38 \pm 0.08$ & $0.43 \pm 0.02$ \\
5. & {\bf S0 (dust = Y, y, n, N)}                   & 51 & $8.16 \pm 0.11$ & $2.28 \pm 0.02$ & $4.56 \pm 0.49$ & $0.85 \pm 0.05$ & $0.38 \pm 0.06$ & $0.45 \pm 0.01$ \\
6. & {\bf S0 (dust = N)}                            & 16 & $7.74 \pm 0.16$ & $2.20 \pm 0.03$ & $3.1 \pm 1.1$ & $0.64 \pm 0.17$ & $0.44 \pm 0.11$ & $0.44 \pm 0.04$ \\
7. & S0 (dust = N, ex.\ 3 strong bars)              & 13 & $7.86 \pm 0.18$ & $2.21 \pm 0.04$ & $2.6 \pm 1.1$ & $0.60 \pm 0.20$ & $0.46 \pm 0.12$ & $0.45 \pm 0.05$ \\
8. & S0 (dust = N, ex.\ 3 strong bars + NGC 7457)   & 14 & $7.79 \pm 0.17$ & $2.18 \pm 0.05$ & $2.54 \pm 0.89$ & $0.66 \pm 0.17$ & $0.45 \pm 0.11$ & $0.43 \pm 0.05$ \\
9. & Spiral                                         & 41 & $7.20 \pm 0.09$ & $2.13 \pm 0.02$ & $2.56 \pm 0.98$ & $0.42 \pm 0.14$ & $0.52 \pm 0.07$ & $0.55 \pm 0.02$ \\
\hline
\end{tabular}
\end{table*}

Additional insight is provided by the left-hand panel of Figure~\ref{Fig-M-sigma-size}, which shows the horizontal kinematic offsets/residuals (observed minus predicted $\sigma_0$) relative to the S0-defined baseline relation (Equation~\ref{Eq_S0_all}). A weak trend is present, with smaller bulges ($R_{\rm e,eq} < 0.4$~kpc) tending toward positive offsets, although the correlation is marginal and visually biased by just three or four points in the upper-left of the diagram. Folding in the black hole mass, as visualised by the colour bar, reveals the offset to higher velocity dispersions may not simply be associated with small bulge sizes, but S galaxies with lower black hole masses and thus S galaxies of lower mass in general.  Such galaxies have less prominent bulges and thus greater contributions from their disc \citep{2001AJ....121..820G} --- and thus greater disc kinematic contamination of the measured $\sigma_0$ --- and they have higher ratios of dark matter \citep[$M_{\rm \star,gal}/M_{\rm DM}\propto M_{\rm DM}^{0.3}$: ][]{2019ApJ...877...64D}.

The (geometric-mean axis) bulge sizes used here are primarily drawn from \citet{2020ApJ...903...97S}, with updates for a small number of galaxies from \citet{2015ApJS..219....4S}, \citet{Graham:Sahu:22b}, and \citet{2026arXiv260424084G}. They have been rescaled to the adopted distances in \citet{2026arXiv260424084G}. 
In passing, it is noted that the bulge sizes of NGC~4501 (the largest seen in Figure~\ref{Fig-M-sigma-size}) and NGC~5495 (not shown due to bright-star contamination) may have been overestimated in their decomposition \citep{2019ApJ...873...85D}. 

The right-hand panel of Figure~\ref{Fig-M-sigma-size} displays the same `horizontal kinematic residuals' as a function of disc inclination, with bulge-to-total stellar mass ratio indicated as a third parameter. No clear dependence on disc inclination or dominance is evident. This is discussed further in Section~\ref{Sec_scat}.

\subsection{Slope comparison}

While the reported 1$\sigma$ uncertainties on the slopes, collated in 
Table~\ref{Table_rel}, 
already suggest distinct distributions, it is possible to quantify this significance using the metric described in 
Section~\ref{sec:slope_comparisons}. 
This allows the posterior probability that one slope is steeper than another ($P(\beta_A > \beta_B)$) to be calculated directly from the MCMC samples, which is subsequently mapped to an equivalent standard normal $Z$-score ($\sigma$ significance):

\begin{itemize}
    \item 
    Comparing the dry-merger-built elliptical and ellicular galaxies against the dust-rich 
    lenticular galaxies 
    (Equation~\ref{Eq_ESeE} versus Equation~\ref{Eq_wet}) 
    yields $P(\beta_{\rm E+ES,e} > \beta_{\rm S0, dust=Y}) \approx 0.99$, corresponding to a $2.3\sigma$ significance, providing moderate evidence that the relation steepens at the high-mass end.
    \item 
    Comparing the dust-rich S0 galaxy population against the primeval, dust-poor S0 galaxies 
    (Equation~\ref{Eq_wet} versus Equation~\ref{Eq_dust-N}) 
    yields $P(\beta_{\rm S0, dust=Y} > \beta_{\rm S0, dust=N}) \approx 0.85$, corresponding to only $1.0\sigma$; this difference does not reach conventional statistical significance and is consistent with these two S0 sub-populations sharing a common underlying slope, given the current sample sizes.
    \item 
    Comparing the dry-merger-built elliptical and ellicular galaxies against the primeval dust-poor S0 galaxies 
    (Equation~\ref{Eq_ESeE} versus Equation~\ref{Eq_dust-N}) 
    yields $P(\beta_{\rm E+ES,e} > \beta_{\rm S0, dust=N}) \approx 0.997$, corresponding to a $2.8\sigma$ significance, providing reasonably strong evidence that the primeval S0 sequence is shallower than the dry-merger sequence.
\end{itemize}
These quantitative probabilities provide a robust, regression-direction-independent 
method of assessing whether these morphological classes follow statistically distinct 
evolutionary scaling relations, bypassing the necessity of qualitative, visual comparisons.

Throughout, the quoted significances denote a \emph{one-tailed} $Z$-score,
$Z = \Phi^{-1}(P)$, appropriate because the sense of each comparison --- i.e.,
which population is expected \emph{a priori} to be steeper --- was specified
in advance, motivated by the `Virial Mirror' framework of
Section~\ref{sec:feedback_and_mergers}.
It is also cautioned that, given the modest sample sizes underlying some of these
subsamples 
(Table~\ref{Table_rel}), 
particularly the dust\,$=$\,N S0
population ($N=13$--$16$), these significances should be regarded as
indicative rather than definitive, consistent with the caveats raised in
Section~\ref{Sec_flat}.

Because the marginal posterior distributions of the regression slopes are nearly Gaussian, the MCMC-derived probability $P(\beta_A > \beta_B)$ and its equivalent standard normal $Z$-score ($\sigma$ significance) align closely with the analytical approximation obtained from standard error propagation:
\begin{equation}
Z \approx \frac{\bar{\beta}_A - \bar{\beta}_B}{\sqrt{\sigma_{\beta_A}^2 + \sigma_{\beta_B}^2}},
\label{eq:analytical_sigma}
\end{equation}
where $\bar{\beta}_A$ and $\bar{\beta}_B$ represent the posterior mean slopes of the respective sub-populations, and $\sigma_{\beta_A}$ and $\sigma_{\beta_B}$ denote their corresponding posterior standard deviations. Evaluating 
Equation~\ref{eq:analytical_sigma} 
using the values collated in 
Table~\ref{Table_rel} 
yields $\approx2.2\sigma$ for the E/ES,e versus dust-rich S0 comparison, $\approx1.0\sigma$ for the dust-rich versus dust-poor S0 comparison, and $\approx2.8\sigma$ for the E/ES,e versus dust-poor S0 comparison, confirming excellent agreement with the exact MCMC-derived $Z$-scores ($\approx2.3\sigma$, $1.0\sigma$, and $2.8\sigma$, respectively).

\subsection{Exclusions} 
\label{sec:exclusions_results}

As described in 
Section~\ref{sec:sample_pruning}, 
five low-dispersion, high-leverage galaxies were excluded from the primary regression fits. No other data points in the sample were found to exert a comparable biasing influence on the derived scaling relations. 
The slopes that would have been obtained for 
equations~\ref{Eq_ESeE} to
\ref{Eq_S0_N_edit}, 
had these galaxies been included, are reported here,
along with the significance of the change calculated using
Equation~\ref{eq:analytical_sigma}.

\begin{itemize} 
\item Equation~\ref{Eq_ESeE}: 4.91$\pm$0.58 (2.0$\sigma$) $\rightarrow$ 6.56$\pm$1.03 (0.8$\sigma$), without NGC~5206.
\item Equation~\ref{Eq_ESeE_tentative}: 5.39$\pm$0.64 (1.9$\sigma$) $\rightarrow$ 7.37$\pm$1.16 (0.6$\sigma$), without NGC~5206.
\item Equation~\ref{Eq_wet}: 3.80$\pm$0.51 (0.7$\sigma$).
\item Equation~\ref{Eq_dusty-ESe}: 3.81$\pm$0.43 (0.9$\sigma$).
\item Equation~\ref{Eq_S0_all}: 3.96$\pm$0.35 (1.0$\sigma$).
\item Equation~\ref{Eq_dust-N}: 2.68$\pm$0.85 (0.3$\sigma$).
\item Equation~\ref{Eq_dust-bar}: 2.41$\pm$0.85 (0.2$\sigma$).
\item Equation~\ref{Eq_S0_N_edit}: 2.41$\pm$0.85 (0.1$\sigma$).
\end{itemize}

\begin{figure}
\begin{center}
\includegraphics[width=1.0\columnwidth]{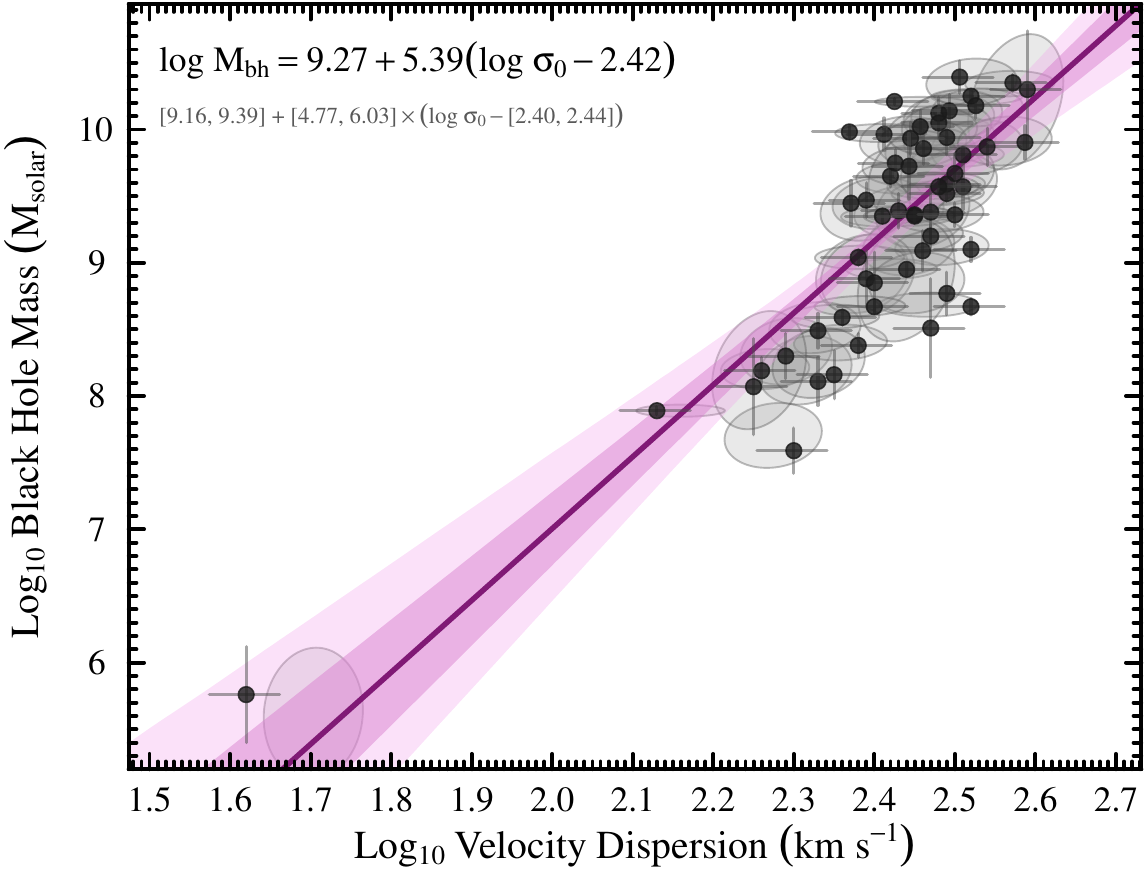}
\caption{
Illustration of low-dispersion high-leverage points --- NGC~3377 ($\rm ES,e$: $\log\sigma_0=2.13$~dex) and NGC~5206 ($\rm E$: $\log\sigma_0=1.62$~dex) --- influencing the regression involving the (13 $+$ 16 BCG) $+$ 16 E $+$ 9 ES,e galaxies. Excluding these two points yields the baseline relation shown in Figure~\ref{Fig-M-sigma-UMBH}.
}
\label{Fig-biased-fit}
\end{center}
\end{figure}

As illustrated in 
Figure~\ref{Fig-biased-fit}, 
the inclusion of these high-leverage points can introduce substantial bias into the regressions, with the resulting slope not representing the typical scaling behavior of the bulk population. 
The substantial shift in slope for the first two samples is primarily driven by the inclusion of NGC~5206 with the E galaxies. If NGC~5206 (sometimes classified as a transition $\rm dE/S0$ galaxy; 
e.g., \citealt{2018ApJ...858..118N}) 
is instead categorised with the dust-poor S0 population, the shift from the original 
Equations~\ref{Eq_ESeE} and \ref{Eq_ESeE_tentative} is largely resolved 
(see Figure~\ref{Fig-M-sigma-IMBH}).

\subsection{If the galaxy morphology is unknown}
\label{Sec_unknown}

The following prescription is suggested if one does not know the morphological type (and `dust bin') of an ETG.
\begin{itemize}
\item 
From Figure~\ref{Fig-M-sigma-2}, if $\sigma_0 > 230$~km~s$^{-1}$, use Equation~\ref{Eq_ESeE}.
\item 
From Figure~\ref{Fig-M-sigma-S0}, if $150 \le \sigma_0 \le 230$~km~s$^{-1}$, use Equation~\ref{Eq_S0_all}.
\item 
From Figures~\ref{Fig-M-sigma-S0} and \ref{Fig-M-sigma-IMBH}, if $\sim(25\pm5) < \sigma_0 < 150$~km~s$^{-1}$, use Equation~\ref{Eq_dust-N}.
\end{itemize} 
These three prescriptions are the observational manifestation of the 'Virial Mirror' framework developed in Section~\ref{sec:feedback_and_mergers}. 
Expectations for ETGs, including ultra-diffuse galaxies (UDGs), with $\sigma_0 \lesssim 25$ km s$^{-1}$ and $M_{\rm bh} < 10^5$~M$\odot$ are discussed further in Section~\ref{Sec_IMBH}.

Finally, if the galaxy under investigation is an S galaxy, there is no strong $M_{\rm bh}$--$\sigma_0$ relation (see Equation~\ref{Eq_S}).  The scatter about the best-fit relation for the S galaxies is considerable, as reported in Table~\ref{Table_rel}, and the intrinsic population correlation, $\rho$, is low.  For the S galaxies, the $M_{\rm bh}$--$M_{\rm \star}$ relations \citep{2023MNRAS.522.3588G} are only slightly preferable to the $M_{\rm bh}$--$\sigma_0$ relations, with a standard rms scatter of 0.52~dex.

\section{Discussion}
\label{Sec_disc}

\subsection{Off the straight and narrow: are massive black holes passengers or drivers?}
\label{sec:feedback_and_mergers}

The morphology-dependent $M_{\rm bh}$--$\sigma_0$ relations derived here offer a natural bridge between theoretical models of AGN feedback and the dynamical mechanics of hierarchical galaxy assembly. 
This context sets the stage for examining how theory aligns with observation.  

Early theoretical frameworks posited a single, universal slope for the $M_{\rm bh}$--$\sigma_0$ relation based on the physics of gas expulsion. \citet{1998MNRAS.300..817H} predicted a slope of 4 in the momentum-conserving feedback limit, while \citet{1998A&A...331L...1S} predicted a slope of 5 in the energy-conserving limit. Recently, \citet{2024ApJ...961L..39S} proposed a transitional evolutionary model, suggesting that early ($z \gtrsim 6$) compact galaxies are regulated by momentum-driven outflows (slope $\approx 4$), before transitioning to energy-conserving feedback (slope $\approx 5$) as the gas reservoir expands. 

The empirical results for the dust-rich (wet-merger-built) S0 galaxies align remarkably well with these theoretical expectations: the derived slope of $4.43 \pm 0.77$ for the dust $=$ Y S0 galaxies (Equation~\ref{Eq_wet}) sits perfectly within the expected theoretical window of $4$ to $5$. As these systems were built from gas-rich major mergers, they provide the physical environment in which AGN feedback is expected to  strongly couple with the interstellar medium, thereby regulating the growth of the central black hole and establishing this baseline scaling law.

However, the observational data demand a significant addition to these feedback theories to account for the most (and the least) massive systems. While early feedback models cast massive black holes as the primary drivers regulating the galaxy's stellar mass --- particularly during gas-rich growth phases --- the steepening of the $M_{\rm bh}$--$\sigma_0$ relation at high masses reveals that during dry mergers, supermassive black holes act more as passengers, responding to the hierarchical assembly of the host galaxy rather than dictating its mass. In these massive elliptical galaxies, AGN feedback likely transitions into a ``care-taker'' or ``maintenance'' mode --- periodically heating the circumgalactic medium to prevent gas cooling and subsequent star formation --- rather than acting as the foundational driver of the scaling relations. Instead, these systems' masses simply sum together while their velocity dispersion remains anchored by virial mechanics. Recognition that the $M_{\rm bh}$--$\sigma_0$ relation  steepens for massive, dry-merger-built galaxies was first quantified in \citet{2013ApJ...764..151G}, who showed that core-S\'ersic E galaxies follow a significantly steeper relation than lower-mass S\'ersic systems. 
As shown by Equation~\ref{Eq_ESeE}, massive E and ES,e galaxies follow a steep $M_{\rm bh}$--$\sigma_0$ relation with a slope of $7.8 \pm 1.3$. This steepening, relative to the relation for dust$=$Y S0 galaxies, 
is quantitatively predicted by the dry-merger virial framework of 
\citet{2023MNRAS.518.6293G}, 
which derived the expected steepening directly from the scalar virial theorem and the ETGs' size--mass relation, rather than inferring it qualitatively from the observed trend alone.
After the gas-rich phase ends and galaxies quench, subsequent growth  primarily occurs via relatively cold-gas-poor hierarchical assembly. During such a dry major (equal mass) merger, the mass doubles, and the effective radius, $R_{\rm e}$, of the spheroid/galaxy nearly doubles. However, as dictated by the virial Theorem ($M_{\rm dyn} \propto \sigma_0^2 R_{\rm e}$), the velocity dispersion $\sigma_0$ remains relatively constant 
--- consistent with dissipationless N-body simulations of dry spheroid re-mergers, which similarly find little accompanying increase in velocity dispersion 
\citep[e.g.,][]{2006ApJ...650..791C, 2006ApJ...641...21R, 2009ApJ...703.1531N} --- while the central black hole mass roughly doubles following the coalescence of the binary black holes. Consequently, dry mergers drive galaxies almost vertically upward in the $M_{\rm bh}$--$\sigma_0$ diagram, steepening the initially feedback-regulated slope of $\sim 4$--$5$ into the observed $\sim 7$--$9$ regime \citep{2013ApJ...764..151G, 2018ApJ...852..131B, 2019ApJ...887...10S}.
This is not merely a qualitative expectation: using the observed $M_{\star,\rm sph}$--$R_{\rm e,sph}$ and $M_{\rm dyn,sph}$--$M_{\star,\rm sph}$ relations for real galaxies, coupled with the virial theorem, 
\citet{2023MNRAS.518.6293G} 
explicitly predicted the expected $M_{\rm bh}$--$\sigma_0$ slope for specific merger histories: an S0+S0 merger building a massive E galaxy is predicted to move systems along a slope of $\sim$5, a subsequent E+E merger along a slope of $\sim$6--9 (depending on the adopted virial mass estimator), and mergers of lower-mass S0 galaxies along a slope of $\sim$3. These predictions, derived independently of the present $M_{\rm bh}$--$\sigma_0$ fits, bracket the observed slope of $7.8\pm1.3$ for the E and ES,e galaxies here, demonstrating that the steepening is quantitatively, not merely qualitatively, reproduced by realistic merger histories.

This virial framework may extend to the opposite end of the mass spectrum, offering a physical basis for the shallow slope ($\sim$2.5-3.1) observed for the primeval, dust-poor S0 galaxies. At the high-mass end, dry mergers drive the steep slope because the merger remnants expand spatially (nearly doubling $R_{\rm e}$ with a doubling of stellar mass) while the velocity dispersion remains relatively static \citep{2023MNRAS.518.6293G}. 
Conversely, at the low-mass end, dETGs (encompassing primeval S0 galaxies) 
are independently known, from structural studies predating and unrelated to the present $M_{\rm bh}$--$\sigma_0$ analysis, to 
exhibit a rather constant $R_{\rm e,gal} \approx 1$~kpc for stellar masses ranging from $\sim$$10^8$ to $\sim$$10^{10}$ \citep{2006AJ....132.2711G, 2008MNRAS.389.1924F}.\footnote{Importantly, one should be aware that when descending further into the realm of UDGs, the curved size-mass relation for ETGs \citep{2008MNRAS.388.1708G, 2019PASA...36...35G} reveals how their effective half light radii, $R_{\rm e}$, actually increase with decreasing stellar mass --- as explained by \citet{2025PASA...42..155G}. The connection of these low-mass ETGs with dwarf and ordinary ETGs, and with low surface brightness (LSB) late-type galaxies, which have large stellar disc sizes, is also presented there. Descending to lower {\it bulge} stellar masses, the size-mass relation appears rather linear down to $\sim$0.1--0.2~kpc \citep[e.g.,][]{2010MNRAS.405.1089L, 2023MNRAS.519.4651H}.}  Applying the simplified scalar virial approximation ($M_{\rm dyn} \propto \sigma_0^2 R_{\rm e,gal}$) and assuming $M_{\star} \propto M_{\rm dyn}$ (roughly) 
--- consistent with the near-constant dynamical-to-stellar mass ratio independently reported for low-mass early-type systems over this mass range
\citep[e.g.,][figures~12--14]{2008MNRAS.389.1924F} --- any increase in stellar mass (for a fixed $R_{\rm e}$) demands a commensurate increase in the velocity dispersion ($\sigma_0 \propto M_{\rm dyn}^{0.5}$).
If $M_{\rm bh}$ scales approximately linearly with $M_{\rm \star,gal}$ for this population, this relatively rapid evolution in $\sigma_0$ compared to stellar mass growth can give rise to a shallower $M_{\rm bh}$--$\sigma_0$ relation, providing a mathematical mirror to the high-mass steepening.
This near-linear scaling between $M_{\rm \star,gal}$  and $M_{\rm dyn,gal}$ is not an assumption introduced here: it was reported by 
\citet{1998AJ....115.2285M}, 
and persists in modern, expanded datasets 
\citep[e.g.,][]{2026arXiv260424084G} 
whenever ETGs of differing formation histories are combined into a single fitted relation. As shown by \citet{2026arXiv260424084G}, 
this apparent near-linearity arises from averaging over several parallel, near-quadratic sub-relations for individual morphological populations (S0, E, and BCG), offset from one another; the lumped fit is a blend of sub-relations rather than the true underlying scaling for any one population.

Herein, the term `Virial Mirror' is introduced as a shorthand to encapsulate this virial bounding in which coupling of the curved ETG size-mass relation \citep{2019PASA...36...35G} with the scalar virial theorem dictates the current kinematic extremities of the $M_{\rm bh}$--$\sigma_0$ relations. Just as dry mergers lead to a virial anchoring of the velocity dispersion to drive a steep vertical bound at the high-mass end, the near-static radii of primeval dwarfs anchor the spatial extent, forcing $\sigma_0$ to absorb the associated mass growth \citep[prior to ETG size growth in lower-mass UDGs:][]{2025PASA...42..155G}. 
Consequently, the Virial Mirror might elegantly frame the distinct $M_{\rm bh}$--$\sigma_0$ relations: a shallow track at low masses, an intermediate feedback-regulated sequence for gas-rich systems, and a steep dry-merger ascent at the high-mass end. Although, it is noted that the present data does not yield a statistically significant difference in slope between the low- and intermediate regime, and as such the `mirror' formally has just two faces.

If the steeper, near-quadratic scaling between black hole mass and stellar mass reported for other morphological types \citep[e.g.,][]{2025PASA...42...68G} were to extend the low-mass regime below the masses probed here, then the corresponding $M_{\rm bh}$--$\sigma_0$ relation would not flatten. In this sense, the observed shallow primeval S0 track in the $M_{\rm bh}$--$\sigma_0$ diagram implicitly favours a near-linear $M_{\rm bh}$--$M_{\rm \star,gal}$ scaling over the observed data range.

It is noted that while reliance on $R_{\rm e,gal}$ introduces some structural biases, switching to a radius containing, for example, 10 or 90 per cent of the light does {\it not} alter the existence of a near-constancy of size among the dETGs (see figure~4 of \citealt{2019PASA...36...35G}) and thus the need for greater changes in $\sigma_0$ with changing $M_{\rm bh}$, i.e., the flattening of the $M_{\rm bh}$--$\sigma_0$ relation at low $\sigma_0$ for the ETGs in Figures~\ref{Fig-M-sigma-S0} and \ref{Fig-M-sigma-IMBH}, remains.

However, interpreting this shallow primeval track through a simplified virial argument is not without well-known physical complexities. 
Unlike massive, pressure-supported E galaxies, low-mass ETGs frequently exhibit significant rotational support 
\citep[e.g.,][]{2002MNRAS.332L..59P}, 
as do the more massive S0 galaxies 
\citep[e.g.,][]{1998A&AS..133..325G, 2011MNRAS.416.1680C}. 
The scalar virial theorem, as invoked here, equates kinetic energy solely with random motions ($\propto\sigma_0^2$), whereas the full virial theorem requires that ordered rotational kinetic energy also be included. Relying on $\sigma_0$ alone therefore underestimates the true kinetic-energy budget --- and hence the dynamical mass --- in systems with dynamically important rotation. Because no clear trend in the rotation-to-dispersion ratio with stellar mass is established for the present sample of S0 galaxies, the net effect of this simplification on the slope cannot be confidently predicted at this time.  
Resolving the rotational contribution would require a dedicated program investigating ordered versus random motions across the full sample.  

Given the non-homology, i.e., the range of S\'ersic indices, across the ETG
sequence, there is also likely to be a range of velocity dispersion profiles \citep{1991A&A...249...99C, 2002A&A...386..149B}, modulo the shape of the dark matter profile. Conversion from $\sigma_0$ to a velocity dispersion within a larger, and ideally infinite, aperture that is more representative of the galaxy will also have a hand to play.  

Additional complications arise from the choice of size (the `bulge' component may be more appropriate than the full stellar body of S0 galaxies) and from possible variations in the baryon-to-dark-matter fraction at lower masses than probed here (e.g., in UDGs). 
However, this fraction is expected to remain roughly constant over the mass ranges used in the present study 
\citep{2008MNRAS.389.1924F}, 
and the other complexities may be subdominant to the near-constant size of the dETGs’ stellar component, thereby producing the shallower $M_{\rm bh}$--$\sigma_0$ relation relative to the dry-merger-built E galaxies (whose velocity dispersions remain nearly constant). This point is debated further in 
Section~\ref{Sec_IMBH}.

\subsection{Revisiting the notion of offset barred galaxies}
  \label{Sec_bar_offset}

The origin of the debate over offset barred galaxies in the $M_{\rm bh}$--$\sigma_0$ diagram can be clarified by examining the morphology of the sample used in \citet{2008ApJ...680..143G}, who first flagged this alleged offset. That study, which reported a systematic offset of barred galaxies to higher velocity dispersions (or lower black hole masses), relied on a sample dominated by ETGs. Of the six barred galaxies presented there, only two were S galaxies (including the Milky Way and NGC~4258). Crucially, three of the four barred ETGs driving the offset were NGC~1023, NGC~2778, and NGC~3384. In the `Triangal' classification, these are all dust-poor (dust $=$ N) S0 galaxies.
The fourth barred ETG was NGC~2787 (SAB0/a, dust $=$ y).

As discussed in Section~\ref{Sec_flat}, the dust-poor S0 sequence represents a dynamically `clean' environment where the kinematic heating of the bar is not masked by the effects of cold gas (nuclear and extended young discs) and spiral arms.
Excluding the few massive merger-built S0 galaxies whose dust has been removed by ram-pressure-stripping \citep{1972ApJ...176....1G} or fragmented into metals by X-rays coming from a hot halo of gas \citep{1979ApJ...231...77D, 1979ApJ...231..438D}, these predominantly low-mass dust-poor S0 galaxies are a laboratory that has long been frozen in time, and therefore likely underwent fewer generations of bar growth/destruction than the S galaxies.
Some simulations have suggested that bars in ETGs can be long-lived structures
\citep{2017MNRAS.469.1054A, 2022MNRAS.512.5339R, 2026arXiv260321279D}. 
Therefore, it is speculated that \citet{2008ApJ...680..143G} may have detected a bar offset precisely because the sample was weighted toward these clean S0 systems. 
In contrast, larger, more diverse samples \citep[e.g.,][]{2019ApJ...887...10S} mix these S0 galaxies with gas-rich S galaxies, thereby drowning out any potential signal.

Now armed with a quantitative measure of bar strength and a slightly larger sample, six S0 galaxies with {\rm strong} bars have been recognised \citep{2026arXiv260424084G}. 
They tend to reside in low-mass S0 galaxies (e.g., NGC~2549, NGC~3384, NGC~3489, and NGC~4762) and are notably rare in high-mass, dust-rich S0 galaxies. The only such dust $=$ Y S0 galaxy with a (debatable) strong bar is Fornax~A (NGC~1316). NGC~4026 (dust $=$ y) rounds out the set of six S0 galaxies with strong bars; it displays no offset to the lower-right of the $M_{\rm bh}$--$\sigma_0$ diagram.  The dozen 
S0 galaxies in the upper-right-most portion of the S0 distribution in the $M_{\rm bh}$--$\sigma_0$ diagram do not contain even a weak bar.   The distribution can be seen in Figure~\ref{Fig-M-sigma-2}.  

Recent cosmological-scale hydrodynamical simulations offer theoretical support for the persistence and kinematic impact of bars in these ETGs. Utilising the TNG50 simulation, \citet{2026arXiv260321279D} demonstrated that genuine, long-lived bar structures are common in dispersion-dominated ETGs. By tracing the progenitor histories of these galaxies, they revealed that these structures originate as fast bars in gas-rich discs that survive quenching. As the host galaxy `dries out', the bar acts as a hardy fossil, undergoing secular braking via dynamical friction and transferring angular momentum to the surrounding stellar spheroid and dark matter halo \citep[e.g.,][]{2003MNRAS.341.1179A, 2023MNRAS.518.1002R}. 
This evolution dynamically heats the host system, providing an additional physical mechanism for elevated velocity dispersions --- the `bar offset' --- observed here in the dust-poor S0 galaxies. Furthermore, \citet{2026arXiv260321279D} highlight that many morphological classification pipelines inherently suffer from an observational blind spot, frequently misclassifying barred ETGs as featureless spheroids, which may explain why this kinematic offset has been historically diluted and overlooked in mixed-morphology censuses.

\subsection{Dissecting the Scatter in Spiral Galaxies}
\label{Sec_scat}

Although the sample size remains modest, Figure~\ref{Fig-M-sigma-IMBH} suggests a connection between the primeval (dust $=$ N at $z=0$) S0 galaxies as their discs evolve, developing bars, and weak spiral patterns to become S0/a galaxies, and transition toward the locus occupied by S galaxies in the $M_{\rm bh}$--$\sigma_0$ diagram. 
While the (barless) dust-poor S0 galaxies define a relatively tight relation (Section~\ref{Sec_flat}), the S galaxies exhibit substantial scatter, often offset toward higher velocity dispersions relative to the mean relation defined by the dust-poor (and the combined) S0 sample. 
This offset is also observed among LSB late-type galaxies \citep{2009RAA.....9..269M, 2011MNRAS.418..789R, 2016MNRAS.455.3148S}.  The gas-rich nature of these galaxies and, in general, low-mass LTGs in contrast to low-mass ETGs, reveals their central, massive black holes through their AGN \citep{1998AJ....116.1650S, 2021ApJ...923..246G}. 
Refined virial $f$-factors (Graham et al.\ 2026, in preparation) will enable refined placement of AGN-hosting LSB galaxies and low-mass S galaxies in the $M_{\rm bh}$--$\sigma_0$ diagram, in turn facilitating exploration of what may drive the scatter among the LTGs.

The heightened scatter among the S galaxies in the $M_{\rm bh}$--$\sigma_0$ diagram is unlikely to be purely stochastic; instead, it may reflect the influence of disc-related processes that are largely absent in the dynamically simpler S0 population.
In particular, the central kinematics of S galaxies can be coupled to (optically-bright `thin discs' with young populations, and bulge) rotation, spiral structure, and bar-driven evolution, such that the measured $\sigma_0$ may not cleanly trace the depth of the bulge potential. Despite exploring below several candidate drivers, no single mechanism cleanly accounts for the observed dispersion. 
The heightened scatter in this galaxy population does not mean that black hole mass does not correlate with S galaxies, as a strong trend is seen in the $M_{\rm bh}$--(arm pitch angle) diagram \citep{2008ApJ...678L..93S, 2017MNRAS.471.2187D} and a trend exists in the $M_{\rm bh}$--$M_{\rm DM}$ diagram \citep[][and references therein]{2019ApJ...877...64D}.

\subsubsection{Drivers of Elevated Velocity Dispersion}

The tendency for some S galaxies to scatter toward higher $\sigma_0$ may arise from a combination of observational contamination and physical effects, including disc kinematics and structural scaling relations such as those implied by the virial theorem. 
\begin{enumerate}
    \item Inclination and Disc Contamination: 
    S galaxies possess dynamically cold, thin discs where rotational velocity, $V_{\rm rot}$, can dominate over $\sigma_0$ when the bulge is small. When a spiral galaxy disc is viewed at high inclination (edge-on), the spectroscopic aperture integrates along the line of sight through the rotating disc. This can introduce a rotational (shear) component into the line broadening, 
    thereby broadening the line-of-sight velocity distribution and increasing the measured dispersion.

    To test the impact of rotational contamination, the horizontal residuals of the S galaxies relative to the central $M_{\rm bh}$--$\sigma_0$ relation defined by the combined sample of S0 galaxies (Equation~\ref{Eq_S0_all}) were explored. Using disc inclinations taken from \citet{2017MNRAS.471.2187D, 2019ApJ...877...64D}, a weak positive trend is observed between the inclination angle and the above mentioned kinematic residual (Pearson $r = 0.26$; Figure~\ref{Fig-M-sigma-size}, right panel). 
    While this slope is qualitatively consistent with expectations that more edge-on discs contribute greater line-of-sight rotational broadening, the correlation lacks strong statistical significance.  The measured probability value ($p = 0.15$) indicates a 15~per~cent chance that this distribution could arise by chance, falling short of the standard $p < 0.05$ ($\approx 2\sigma$) threshold typically required to claim a robust correlation. 

A promising avenue to mitigate potential inclination and disc contamination is the spectral decomposition of bulge and disc kinematics using integral-field spectroscopy. For instance, \citet{2020MNRAS.495.4638O} utilised data from the Sydney-AAO Multi-object Integral-field spectrograph (SAMI) Galaxy Survey \citep{2015MNRAS.447.2857B} to simultaneously estimate the velocity dispersions of bulge and disc components. They demonstrated that isolated bulges follow a more well-defined luminosity--(velocity dispersion) relation than the integrated galaxy light, confirming that the underlying bulge kinematics are more stable than the composite measurement. 
However, applying such spectral decomposition to large samples remains challenging due to the structural complexity of disc galaxies. The approach of \citet{2020MNRAS.495.4638O}, which fixes spectral weights using simple bulge+disc decompositions, can struggle in systems hosting strong bars, spiral arms, or nuclear rings, potentially leading to biased bulge estimates. Consequently, extracting a truly isolated classical spheroid velocity dispersion --- particularly in barred S galaxies --- remains difficult.

    \item Bar-Driven Heating:
Bars can induce non-circular streaming motions and vertical (out of disc plane) heating, which may act to increase the measured $\sigma_0$. However, this mechanism does not appear to dominate in the present sample, as strongly barred S galaxies do not occupy a distinct region of the $M_{\rm bh}$--$\sigma_0$ diagram (Figure~\ref{Fig-M-sigma-2}). 
 Plausibly, one could argue that the spiral arms themselves shift some S galaxies to the notably higher $\sigma_0$ values seen in the lower-right of Figure~\ref{Fig-M-sigma-2}, although those galaxies do not contain the strongest, most visually prominent arms (Figure~\ref{Fig-M-sigma-S0}).  It may, however, be related to whether the arms extend in to the galaxy centre and what their orientation is relative to our line-of-sight. 

    \item The Virial Theorem:
    For a given  mass (and S\'ersic concentration), pressure-supported bulges with smaller sizes should have larger velocity dispersions. Addressing scatter in the $M_{\rm bh}$--$\sigma_0$ diagram, \citet[][figure~2]{2008ApJ...680..143G} explored the potential for a `fundamental plane' for black holes involving the spheroids' effective radii, $R_{\rm e,sph}$. It was noted that the few barred galaxies, which were offset to higher $\sigma_0$ for a given $M_{\rm bh}$,  appeared to have smaller effective radii than non-barred galaxies of similar mass, consistent with expectations from the virial theorem. Checking for this trend with the current sample of S galaxies, and as noted in Section~\ref{Sec_spiral}, the left-hand panel of Figure~\ref{Fig-M-sigma-size} displays the `horizontal kinematic residuals' 
plotted against the half-light radii of the S galaxy bulges, and colour-coded by the black hole mass to check for offsets at fixed black hole mass.  
A trend of larger positive offsets with smaller $R_{\rm e,sph}$ would (naively) be expected if a strong $M_{\rm bh}$--$M_{\rm \star,sph}$--$M_{\rm dyn}$ relation exists for the S galaxies. 
However, the behaviour seen here does not provide support for such a simple virial interpretation in which smaller bulges systematically shift S galaxies to higher $\sigma_0$ at fixed black hole mass. In fact, the opposite may be occurring, as evidenced by the broad diagonal (lower-left to upper-right) distributions for galaxies of similar black hole mass. 
For a constant $R_{\rm e,bulge}/h_{\rm disc}$ size ratio \citep[][and references therein]{1996ApJ...457L..73C, 2008MNRAS.388.1708G}, these larger kinematic residuals (higher $\sigma_0$) at fixed $M_{\rm bh}$ are associated with large discs, possibly connecting the disc growth shown in \citet[][figure~12]{2025PASA...42..155G}. 
Removing potential outliers such as NGC~4501 does not significantly strengthen the correlation (Pearson $r = -0.18$, $p=0.29$; Spearman $r = -0.21$, $p=0.21$, where $p$ is the probability that the correlation could arise by chance).

\end{enumerate}

\subsubsection{Drivers of Depressed Velocity Dispersion}

Counteracting these effects are processes that can artificially reduce the measured velocity dispersion. 
As cautioned by \citet{2011MNRAS.412.2211G}, the presence of cold nuclear discs or nuclear bars can create central `$\sigma_0$-drops' in the kinematics --- an issue that is exacerbated by two specific observational biases in the optical regime:
\begin{enumerate}
    \item The `Shining' of Young Populations: 
    Spectroscopy measures a luminosity-weighted velocity dispersion. S galaxies routinely contain populations of young, massive stars formed in cold, dense gas. These populations are dynamically cold (low $\sigma_0$) and  luminous at rest-frame optical wavelengths. They mix with the older, dynamically hotter bulge stars within the spectroscopic aperture, biasing the measurement toward the lower dispersion of the young disc stars.
    \item Dust Extinction:
    In dust-rich S galaxies, the central regions (where $\sigma_0$ is highest) are often partly obscured. An optical spectrum may preferentially sample light from the `surface' of the bulge or layers above the dusty mid-plane, where the intrinsic velocity dispersion is lower than in the true centre.
\end{enumerate}

To circumvent these dual issues, observational campaigns have increasingly shifted to measuring velocity dispersions in the near-infrared. Following early work by \citet{2001A&A...368...52E} that utilised the CO stellar absorption bandheads at $\sim$2.3 $\mu\mathrm{m}$ to peer through dust and avoid nuclear $\sigma_0$-drops, this regime offers benefits for improving black hole scaling relations  \citep[e.g.,][]{2013ApJ...767...72R, 2015MNRAS.446.2823R, 2017ApJ...835..271B}. Because near-infrared light penetrates the obscuring dust and preferentially traces the older, dynamically relaxed red giant and supergiant populations, it should be superior at recovering the kinematics of the underlying classical spheroid. Notably, \citet{2013ApJ...767...72R} demonstrated this effect explicitly for dusty, star-forming galaxies, revealing a  ``$\sigma_0$ discrepancy'': while optical measurements yielded artificially depressed velocity dispersions, near-infrared measurements recovered the deeper potential well.
Such treatment, however, is beyond the scope of the present investigation but worth being mindful of.  

Returning to the primary sample, the relative tightness of the dust-poor S0 galaxy $M_{\rm bh}$--$\sigma_0$ relation may arise because these systems are less impacted by the competing effects described above.

\subsection{AGN virial \texorpdfstring{$f$}{f}-factors and early-Universe black holes}
\label{sec:f_factors}

Determining black hole masses in distant AGN via `reverberation mapping' relies on the dimensionless virial factor, $f$, to scale the measured virial product ($\mathrm{VP} = R_{\rm BLR} \sigma_{\rm BLR}^2 / G$) into a  mass \citep[e.g.,][]{2004ApJ...615..645O, 2004ApJ...613..682P, 2011MNRAS.412.2211G}. The value of $f$ has historically been calibrated by forcing the AGN virial products to align with a single log-linear $M_{\rm bh}$--$\sigma_0$ or $M_{\rm bh}$--(luminosity or stellar mass) relation fit to local ($z=0$) galaxies with predominantly inactive galactic nuclei (IGN) and directly measured black hole masses.  However, this derivation overlooks the fundamentally bent nature of these distributions and their strict morphological dependencies. 

As demonstrated herein, S galaxies exhibit extensive scatter in the $M_{\rm bh}$--$\sigma_0$ diagram. Because this scatter is likely driven by complex disc kinematics and is not alleviated simply by isolating the barred systems, the $M_{\rm bh}$--$\sigma_0$ relation is considered here to be an unreliable tool for calibrating the $f$-factor of AGN in S galaxies. 
For deriving the $f$-factor in S galaxies, the morphology-dependent $M_{\rm bh}$--$M_{\rm \star,sph}$ and $M_{\rm bh}$--$M_{\rm \star,gal}$ relations \citep{2026arXiv260424084G} offer a superior framework. 
For early-type host galaxies, the $M_{\rm bh}$--$\sigma_0$ relation remains a viable calibration tool, but only if the changing behaviour of the relation with galaxy morphology --- and thus with increasing $M_{\rm bh}$ and $\sigma_0$ --- is  accounted for. 
This requires that the local calibrator galaxies and the distant AGN hosts be strictly segregated by morphological type and dust properties. This task will be performed in future work, bringing the `Triangal' framework to bare on the derivation of refined virial factors, or factor, for deriving the masses of the many distant AGN that the Vera C.\ Rubin Observatory's 
Legacy Survey of Space and Time 
\citep[LSST:][]{2019ApJ...873..111I}\footnote{\url{https://www.lsst.org/}}, conducted with the Simonyi Survey Telescope, 
will uncover, and revising masses from studies using the Dark Energy Spectroscopic Instrument  \citep[DESI:][]{2016arXiv161100036D}. 
Furthermore, dynamically heated sub-populations, such as the strongly barred, dust-poor (dust $=$ N) S0 galaxies, should be explicitly removed from the regression to avoid artificially flattening the calibration baseline.

Failing to separate galaxies with different evolutionary pathways when calibrating the virial factor can introduce systematic, order-of-magnitude errors into the predicted black hole masses.  Rectifying this morphological bias is therefore critical, as an overestimated $f$-factor artificially inflates derived AGN masses, impacting assumptions about `seed black hole' masses \citep[e.g.,][]{2023ApJ...957L...3P}. Consequently, properly accounting for galaxy morphology in $f$-factor derivations may naturally resolve some of the current astrophysical tension surrounding the unexpectedly massive black holes recently reported in the early Universe by the \textit{James Webb Space Telescope} \citep[e.g.,][]{2021ApJ...907L...1W, 2024A&A...691A.145M}.

\subsection{IMBHs: scaling relations and observational limits}
\label{Sec_IMBH}

This subsection examines both the expected behaviour of scaling relations in the IMBH regime and the observational limitations that currently obscure it.  
While extrapolation into the IMBH mass range in the $M_{\rm bh}$--$\sigma_0$ diagram remains uncertain, there is reason to expect that the relations may exhibit slope changes, possibly rendering them more tightly constrained than the linear extensions shown in Figures~\ref{Fig-M-sigma-S0} and \ref{Fig-M-sigma-IMBH}.

First, consider the scenario in which the shallow trend $M_{\rm bh} \propto \sigma_0^{2.5}$ (Equations~\ref{Eq_dust-bar} and~\ref{Eq_S0_N_edit}) continues to lower velocity dispersions. This relation would define an upper envelope for the primeval S0 galaxies in Figure~\ref{Fig-M-sigma-IMBH}. The plausibility of this extrapolation can be assessed by linking it to the corresponding stellar-mass scaling relations. For the dust-poor dETGs from which this relation is derived, combining $M_{\rm bh} \propto \sigma_0^{2.5}$ with the established stellar mass--velocity dispersion relation ($M_{\rm \star,gal} \propto \sigma_0^{2\text{--}2.5}$: \citealt{1983ApJ...266...41D, 1992AJ....103..851H, 2005MNRAS.362..289M, 2014ApJS..215...17T}) implies a near-linear scaling between black hole mass and galaxy stellar mass ($M_{\rm bh} \propto M_{\rm \star,gal}^{1\text{--}1.25}$), though this inference inherits any bias present in the underlying $M_{\rm \star,gal}$--$\sigma_0$ relation.\footnote{Biases may arise when samples mix primeval dETGs with S0 galaxies that are faded S galaxies or dust-stripped (dust $=$ Y $\rightarrow$ N) post-major-merger S0 galaxies.}

\begin{figure}
\begin{center}
\includegraphics[width=1.0\columnwidth]{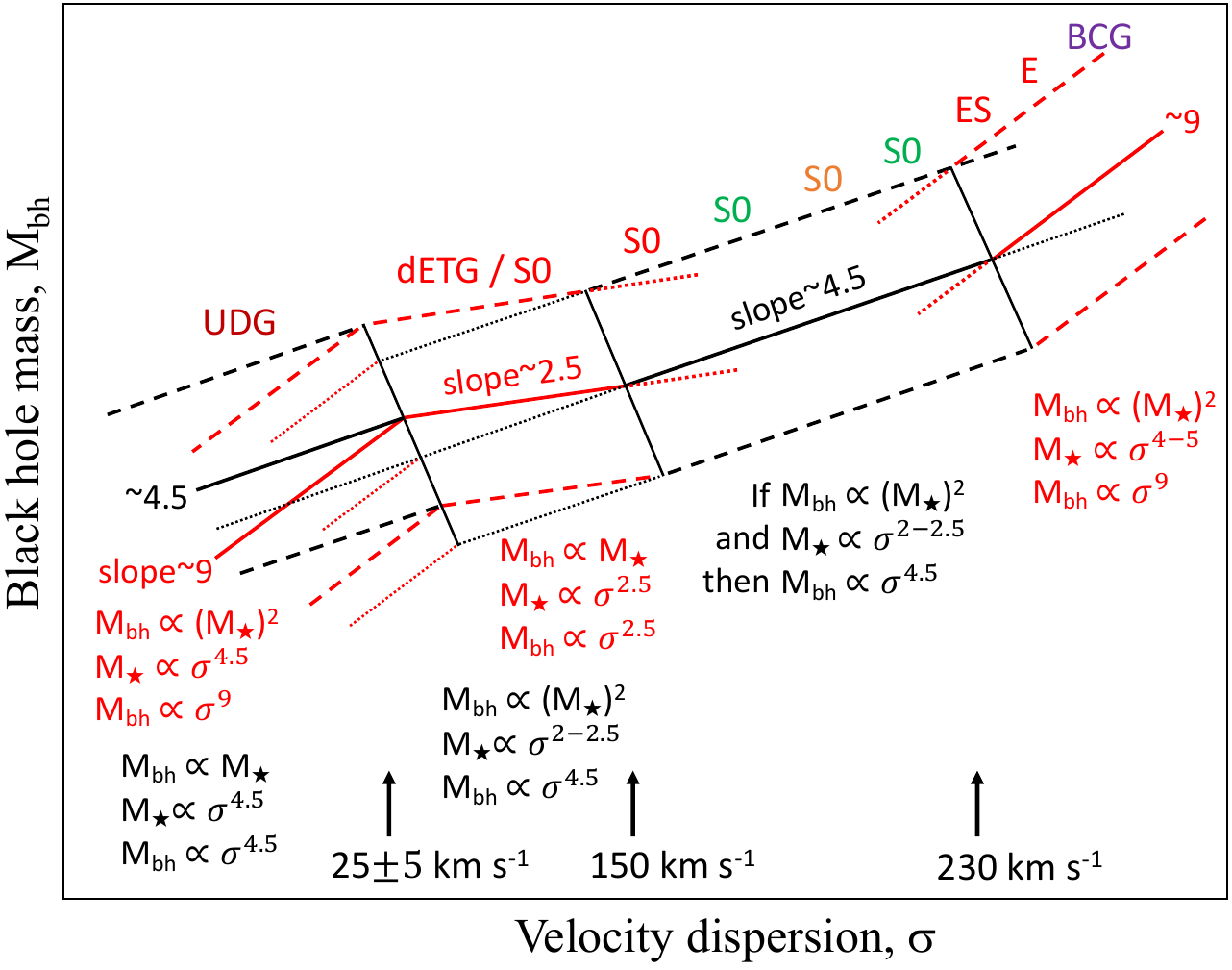}
\caption{Consistency checks for ETGs and reframing the IMBH regime.  Each nested set of three equations has a slope for the $M_{\rm bh}$--$\sigma$ relation that is consistent with the other two relations. For simplicity of expression, the subscript `0' is omitted here from the symbol $\sigma$.
The $\sigma$-values running along the bottom correspond to the suggested transition points for the  $M_{\rm bh}$--$\sigma_0$ relations (Section~\ref{Sec_unknown}). 
While the stellar and black hole mass double via individual, dry, equal-mass mergers, collectively, such E $\rightarrow$ BCG mergers shift the ensemble of E points in such a way as to maintain that distribution's slope, which follows an $M_{\rm bh}\propto M_{\rm \star,gal}^2$ scaling \citep[Figure~\ref{bh-ratio-sph} and][figure~A3]{2023MNRAS.522.3588G} that has been used here. Uncertainty in the slope of the $M_{\rm bh}$--$M_{\rm \star,gal}$ relation for primeval S0 galaxies (including dETGs and UDGs) results in two pairs of relations at the low-mass end of the diagram.  Data to date (Figure~\ref{Fig-M-sigma-IMBH}) suggests $M_{\rm bh}$--$\sigma^{2.5}$ for the dETGs. The shallower slope over the dETG regime aligns well with the near-constant half-light radii for such galaxies \citep{2019PASA...36...35G} and the simple scalar virial theorem.
}
\label{M-sigma-virial}
\end{center}
\end{figure}

At very low velocity dispersions ($\sigma_0 \lesssim 20~\mathrm{km\,s^{-1}}$), galaxies appear to follow a steeper stellar mass--velocity dispersion relation, $M_{\rm \star,gal} \propto \sigma_0^{4.5}$ \citep[see figure~17 of][]{2014ApJS..215...17T}. If a near-linear black hole-to-stellar mass scaling approximately persists into this regime --- an extrapolation that remains empirically untested at such low masses --- then consistency requires that the $M_{\rm bh}$--$\sigma_0$ relation steepens accordingly, approaching $M_{\rm bh} \propto \sigma_0^{4.5}$.
If, however, the steeper near-quadratic scaling between black hole mass and stellar mass reported for other morphological types were to apply in this low-mass regime, then the corresponding decline in the $M_{\rm bh}$--$\sigma_0$ plane would be even more pronounced.  This is illustrated in Figure~\ref{M-sigma-virial}.

There are other approaches to get at IMBH masses. 
Reviews of the IMBH regime in star clusters and dwarf galaxies \citep[e.g.,][]{2024bheg.book..149A} highlight the intense observational and theoretical efforts to locate IMBHs. A fundamental predictive tool in this domain is the direct scaling relation between the mass of the central black hole and the mass of its host nuclear star cluster. This $M_{\rm bh}$--$M_{\rm nsc}$ relation was originally established by mathematically uniting the low-mass 
$M_{\rm bh}$--$\sigma_0$ and $M_{\rm nsc}$--$\sigma_0$ scaling relations to eliminate the host velocity dispersion \citep{2016IAUS..312..269G}. 
Performing the same trick with the 
$M_{\rm bh}$--$M_{\rm bulge}$ and $M_{\rm nsc}$--$M_{\rm bulge}$ scaling relations to eliminate the host bulge mass yielded a consistent result: $M_{\rm bh} \propto M_{\rm nsc}^{2.7\pm0.7}$ \citep{2016IAUS..312..269G}. 
The physical validity of this relation was subsequently demonstrated when it successfully predicted the black hole masses within ultra-compact dwarf (UCD) galaxies \citep{2020MNRAS.492.3263G}. That consistency test provided quantitative evidence that UCDs are the tidally threshed remnant nuclei of larger galaxies \citep{1988IAUS..126..603Z, 1993ASPC...48..608F, 2001ApJ...552L.105B}. Recognising this connection between black holes and nuclear star clusters is important for modern IMBH searches. However, if the velocity dispersion of a nuclear star cluster is erroneously conflated with, or used as a proxy for, the broader spheroidal or galaxy velocity dispersion, one will invariably miscalculate the expected black hole mass and misconstrue the underlying baseline scaling relations at the low-mass end.

Equally important is an awareness of when and how sample selection 
boundaries skew low-mass distributions.
\citet{2003AJ....125.2936G} first quantified the above-mentioned relation between nuclear star cluster mass and host galaxy stellar mass for dETGs, establishing 
the baseline scaling subsequently extended to a broad range of environments 
by \citet{2013ApJ...763...76S}.
Recent empirical work involving nuclear star cluster masses 
\citep[][their figure~4]{2020A&ARv..28....4N, 2025A&A...704A.113P} exhibits a flattening of the running 
mean mass ratio with host galaxy stellar mass that deviates from this underlying baseline relation.
However, this apparent departure appears to be a direct consequence of an 
observational floor: nuclear star clusters less massive than 
$\sim\!10^{4.5}\,M_\odot$ are  
difficult to detect. 
When this censored distribution is mapped into the $M_{\rm NSC}/M_{\rm gal}$ versus $M_{\rm gal}$ plane, the running mean is pulled systematically upward at the low-mass end, creating a significant observational --- rather than physical --- discrepancy with the  baseline relation. 

This is mentioned because 
this limitation is not unique to suspected IMBH-hosting nuclear star clusters, but reflects a more general observational constraint on low-mass black hole detections. In this context, the same level of scepticism must also be applied when interpreting the $M_{\rm bh}$--$\sigma_0$ relations derived herein.
Dynamical measurements of black holes become profoundly challenging 
below $\sim\!10^6\,M_\odot$, because the black hole's sphere of influence typically 
shrinks below current spatial resolution limits, imposing an analogous observational floor on the $M_{\rm bh}$ distribution at the low-mass end.
It is therefore prudent to acknowledge that the notably shallow slope 
of $\sim$2.5--3.0 reported here for the primeval, dust-poor S0 galaxies (Section~\ref{Sec_flat}) 
may be, at least in part, a manifestation of this sample selection bias 
rather than a purely physical flattening of the evolutionary track.
A more complete census of black holes in the $10^5$--$10^6\,M_\odot$ and lower range 
--- currently inaccessible with existing facilities --- may ultimately 
be required to confirm whether the shallow slope reflects an intrinsic 
physical feature of the primeval S0 evolutionary track, or whether it 
is an artefact of a censored low-mass boundary of the directly measured 
black hole sample. Nonetheless, this observed shallowing of the slope among dETGs (excluding UDGs and systems with $\sigma_0 \lesssim 20$ km s$^{-1}$) in the $M_{\rm bh}$--$\sigma_0$ diagram is 
anticipated by the `Virial Mirror' framework presented for the first time in Section~\ref{sec:feedback_and_mergers}, wherein the static spatial radii of dETGs enforce a kinematic flattening.

\subsubsection{Outlook}

A notably shallower slope 
carries significant implications for future observational campaigns targeting the IMBH regime. The longstanding assumption of a relatively steep, monolithic scaling relation may have led to systematic underestimates of black hole masses in low-dispersion environments. Because the ability to dynamically confirm a black hole critically depends on resolving its sphere of influence \citep[$r_h = G M_{\rm bh} / \sigma^2$;][]{2001ASPC..249..335M}, such underestimates naturally translate into more pessimistic expectations regarding the feasibility of IMBH detections.  The revised, shallower scaling indicates that black holes in the $10^5$--$10^6 \, M_\odot$ range, and lower, may reside in galaxies with lower velocity dispersions than previously assumed, implying larger spheres of influence and, consequently, improved prospects for spatially resolving their kinematics. 

Such measurements became feasible with facilities capable of delivering spatial resolutions of $\sim\!50$--$100$ milliarcseconds (mas) (e.g., Keck/OSIRIS, VLT/MUSE, and JWST/NIRSpec), and that situation will further improve with the next generation of adaptive-optics-assisted instruments approaching $\sim\!10$--$20$~mas resolution (e.g., GMT/GMTIFS, and ELT/HARMONI).
Moreover, the identification of the shallow, dust-poor S0 sequence substantially improves the observational outlook for dynamical IMBH searches at {\it optical} wavelengths. This is particularly relevant for the MAVIS instrument \citep{Agapito_2022}, which is expected to provide visible-light adaptive-optics imaging and spectroscopy at resolutions approaching $\sim\!20$--$30$~mas on the VLT.

\subsection{UMBHs in core-S\'ersic galaxies}
\label{Sec_UMBH}

The steep $M_{\rm bh}$--$\sigma_0$ relation for core-S\'ersic galaxies \citep{2013ApJ...764..151G, 2019ApJ...887...10S}, and, in particular, for BCGs and BGGs, as shown by \citep{2018ApJ...852..131B}, 
implies the existence of much larger black holes than predicted by the older $M_{\rm bh}$--$\sigma_0$ relations with their shallower slopes around 4 to 5 \citep{2000ApJ...539L...9F, 2000ApJ...539L..13G}.  The relation for E$+$ES,e galaxies presented in Equation~\ref{Eq_ESeE} 
is somewhat consistent with the $10^{10}$--$10^{11}$~M$_\odot$ UMBH masses predicted by \citet{1969Natur.223..690L}.  It has a steep slope of 7.8$\pm$1.3 and predicts black hole masses exceeding $10^{10}$~M$_\odot$ at $\sigma_0 > 352$ km s$^{-1}$.  
The BGG$+$BGG $M_{\rm bh}$--$\sigma_0$ relation from \citet{2018ApJ...852..131B} predicted this could occur at $\sigma_0 > 337$ km s$^{-1}$, while Equation~\ref{Eq_ESeE_tentative} suggests it occurs at $\sigma_0 > 327$ km s$^{-1}$.  
Given the dry-merger-induced steepening of the $M_{\rm bh}$--$\sigma_0$ relation explained in \citet{2023MNRAS.518.6293G}, 
it seems likely that the inclusion of the lower-mass E and ES,e galaxies (and BGGs) is resulting in a shallower slope than is observed for samples of giant core-S\'ersic galaxies with UMBHs, as studied by \citet{2019ApJ...886...80D}.  Unfortunately, the narrow range in velocity dispersion for the current BCG-only sample prevents the derivation of a reliable $M_{\rm bh}$--$\sigma_0$ relation using \textsc{scope}.

It has frequently been asserted in the literature that the $M_{\rm bh}$--$\sigma_0$ relation `saturates' at the high-mass end and therefore fails to predict the existence of UMBHs \citep{2007ApJ...662..808L, 2019ApJ...886...80D, 2025arXiv251204178D}. For instance, \citet{2019ApJ...886...80D} noted that classical $M_{\rm bh}$--$\sigma_0$ relations restrict predicted black hole masses to $\lesssim 5 \times 10^9 M_\odot$ for velocity dispersions of $300$--$390\ \rm{km\ s^{-1}}$, creating an apparent discrepancy with the predictions from $M_{\rm bh}$--$L$ relations. This perceived tension has often been used to argue that $\sigma_0$ becomes a poor predictor of black hole mass for the most luminous spheroids built by dry mergers.
However, this discrepancy is partly an artifact of applying shallow, monolithic $M_{\rm bh}$--$\sigma_0$ scaling relations --- historically with slopes of $\sim$4--5 --- that improperly blend diverse morphological types. Such analyses overlook the steeper, morphology-dependent relations already established for massive ETGs.  When a single regression is forced through a heterogeneous sample --- for example, the slope of 5.8 reported by \citet{2025ApJ...978...48S} --- the most massive BCGs appear to lie above the fit and are often described as hosting ``over-massive'' black holes; indeed, \citet{2025ApJ...978...48S} themselves conclude that massive galaxies must be treated independently. This offset, however, is a mathematical consequence of averaging the shallower, accretion-driven tracks of disc galaxies with the steeper, dry-merger-driven evolution of massive E galaxies. When E and ES,e galaxies are considered separately, they define a much steeper relation (slope $\sim$8--9), consistent with collisionless virial growth. Relative to this physically appropriate baseline, the $>10^{10}\,M_\odot$ black holes in BCGs are not anomalous, but represent the natural endpoints of hierarchical dry assembly.

When appropriate, morphology-specific scaling relations are applied, the false dichotomy between the predictive power of luminosity and velocity dispersion vanishes. As quantified by the updated relation for E and ES,e galaxies presented here (Equation~\ref{Eq_ESeE_tentative}, featuring a slope of $8.6$), velocity dispersions in the range of $326$--$426\ \rm{km\ s^{-1}}$ yield $10^{10}$--$10^{11} M_\odot$ UMBHs. Allowing for scatter about the $M_{\rm bh}$--$\sigma_0$ relation, lower velocity dispersions can reach these black hole masses. Consequently, there is no failure of the $M_{\rm bh}$--$\sigma_0$ relation at the high-mass end; rather, there simply is a transition to the steeper scaling track dictated by dry merger dynamics \citep{2023MNRAS.518.6293G}.

\subsection{Implications for gravitational wave predictions}
\label{sec:gravitational_waves}

The sensitivity of gravitational wave event rate predictions to the precise calibration of black hole scaling relations cannot be overstated. A striking precedent for this was demonstrated by \citet{2012A&A...542A.102M} in their cosmological forecasts for the targeting of Extreme Mass-Ratio Inspirals (EMRIs) by LISA \citep{2017arXiv170200786A}. Because EMRIs frequently involve stellar-mass black holes within dense nuclear star clusters inspiraling into a central massive black hole, the predicted event rates are highly dependent on the low-to-intermediate mass end of the scaling relations. \citet{2012A&A...542A.102M} showed that replacing a traditional, monolithic near-linear scaling baseline with a steeper, morphology-aware $M_{\rm bh}$--$M_{\rm \star,sph}$ relation applicable to S\'ersic galaxies reduced the expected LISA EMRI detection rate by a full order of magnitude. Because the steeper $M_{\rm bh}$--$M_{\rm \star,sph}$ relation predicts significantly lower black hole masses for spheroids in the $10^8$--$10^{10}\,M_\odot$ range, the target demographics were fundamentally altered. This extreme sensitivity demonstrates that failing to account for morphology-specific black-hole scaling relations can systematically and dramatically skew predicted demographics across the gravitational-wave spectrum.

Just as the low-mass end of the black hole mass scaling relations dictate EMRI rates, the high-mass end governs the nanohertz stochastic gravitational-wave background targeted by Pulsar Timing Arrays (PTAs). 
Theoretical models of
this background typically begin with simulated dark matter halos or empirical galaxy stellar-mass and velocity-dispersion functions, which are then `populated' with supermassive black holes by drawing from an observed scaling relation. 
Historically, the community relied on monolithic scaling relations that blended diverse morphological types, thereby diluting the steepness of the high-mass end.  To reconcile predictions with early, stringent PTA upper limits \citep[e.g.,][]{2015Sci...349.1522S}, some models therefore advocated for the use of shallow monolithic fits \citep[see the discussion in][]{2016MNRAS.463L...6S}. However, the recent detection of an unexpectedly loud nanohertz gravitational-wave background by global PTAs \citep[e.g.,][]{2023ApJ...951L...9A} strongly implies the existence of a robust population of UMBHs that shallower, monolithic relations struggle to produce.

The morphology-dependent $M_{\rm bh}$--$\sigma_0$ relations derived here, together with the steep morphology-dependent $M_{\rm bh}$--$M_{\star}$ relations of 
\citet{2019ApJ...876..155S} and 
\citet{2023MNRAS.522.3588G}, 
supply the natural inputs for such population models. In particular, dry-merger-built E and ES,e galaxies follow an $M_{\rm bh}$--$\sigma_0$ relation 
(Equations~\ref{Eq_ESeE} and \ref{Eq_ESeE_tentative}) 
that is steeper than that defined by disc galaxies. 
Yet current PTA host-identification pipelines and many population-synthesis models implicitly assume that a single black-hole–galaxy scaling relation applies across all morphological types.  The present results indicate that this assumption should be revised.

Population- synthesis studies generally predict that individually resolvable continuous gravitational-wave sources reside predominantly in massive ($M_\star \gtrsim 10^{11}\, M_\odot$), bulge-dominated early-type galaxies at low redshift 
\citep[e.g.,][]{2025A&A...694A.282T,2026A&A...706A.115T}. 
This population overlaps strongly with the galaxies that define the steep $M_{\rm bh}$--$\sigma_0$ relation reported here for E and ES,e systems, including BCGs (slope $7.8\pm1.3$) and the near-quadratic $M_{\rm bh}$--$M_{\star,\rm gal}$ relation for major-merger-built galaxies (slope $\beta\approx2.06$--$2.27$; 
\citealt[][table~1]{2023MNRAS.522.3588G}).

Because the characteristic strain depends strongly on chirp mass\footnote{The chirp mass of a binary is defined as 
$\mathcal{M}_{\rm c}=(m_1 m_2)^{3/5}(m_1+m_2)^{-1/5}$, 
where $m_1$ and $m_2$ are the individual black-hole masses.} ($h_{\rm c}\propto\mathcal{M}_{\rm c}^{5/3}$), systematic revisions to the black hole masses of the most massive hosts can affect both the ranking of candidate hosts for individually resolvable sources and, potentially, predictions for the stochastic gravitational-wave background. 
Existing Bayesian host-identification pipelines assign candidate galaxies a prior probability using an empirical, morphology-independent scaling relation 
\citep[e.g.,][]{2019MNRAS.485..248G,2026A&A...706A.115T}.
Adopting a near-linear relation 
\citep[e.g.,][]{2019MNRAS.488..401C}\footnote{\citet{2025A&A...694A.282T} quote a normalisation $\alpha=8.40$ from \citep{2019MNRAS.488..401C}. The actual median of the prior reported by \citep[][their table~1 and figure~5]{2019MNRAS.488..401C} is 
$\log_{10}M_*=8.17^{+0.35}_{-0.32}$. The difference of $+0.23$\,dex corresponds to black-hole 
masses $\sim$70\,per\,cent higher at fixed bulge mass. Because the characteristic strain 
scales as $h_{\rm c}\propto\mathcal{M}_{\rm c}^{5/3}$, this would raise the predicted 
gravitational-wave background amplitude by a factor of roughly $1.7^{5/3}\approx2.3$.}
rather than the steeper major-merger relation underestimates the black hole mass of a galaxy with $M_\star=5\times10^{11}\,M_\odot$ by a factor of $\sim4$, and by $\sim9$ at $M_\star = 10^{12}\,M_\odot$, or $\sim5$ and $\sim10$ if the spheroid-mass relation is used instead. 
Brightest-cluster galaxies, which follow a comparably steep sequence but with a lower normalisation \citep{2023MNRAS.522.3588G}, still reach the highest absolute black-hole masses and would therefore be affected still more strongly.
On another practical note, it is noted that 
when converting an observed galaxy luminosity function into a stellar-mass function for use with these $M_{\rm bh}$--$M_{\star}$ relations, the same initial mass function adopted in the construction of the relations must be employed.

Substituting morphology-specific relations for the single, morphology-blind prior currently in use would therefore supply more physically motivated black hole mass priors for the galaxies expected to dominate the PTA source population. A dedicated population-synthesis study that adopts the morphology-dependent $M_{\rm bh}$--$\sigma_0$ and $M_{\rm bh}$--$M_\star$ relations presented here (and in 
\citealt{2023MNRAS.522.3588G}) 
is required to quantify the resulting changes in the stochastic background amplitude, the expected number of individually resolvable continuous-wave sources, and the efficacy of Bayesian host-galaxy ranking. 
While a heavier high-mass end naturally raises the predicted strain amplitude, any impact on the detailed spectral shape of the background remains to be determined by such modelling.

\subsection{Connections to the Triangal}
\label{Sec_connect}

The morphology-dependent $M_{\rm bh}$--$\sigma_0$ relations derived herein provide a kinematic validation of the `Triangal' galaxy evolution schema \citep{2023MNRAS.522.3588G}. While traditional morphological schemata --- such as the Tuning Fork \citep{1926ApJ....64..321H, 1928asco.book.....J, 1936rene.book.....H} or the Trident \citep{1976ApJ...206..883V} --- do not encode the distinct formation histories that govern black hole growth, the results presented in Section~\ref{Sec_RandI} demonstrate that environmental quenching, accretion, and hierarchical merging define kinematically distinct tracks in the $M_{\rm bh}$--$\sigma_0$ plane. By contrast, with the exception of bars in dust-poor S0 galaxies, bars and spiral strength do not cleanly dictate a galaxy's location in this diagram.

The low-mass realm explored here captures the evolutionary bifurcation driven by early environmental quenching versus  major mergers or sustained disc growth. Recent high-redshift observations confirm that strong, bar-driven gas inflows were already underway by $z \approx 2.5$ \citep{2017ApJ...850...61Y, 2025Natur.641..861H, 2025A&A...696A.156X}, consistent with the `Triangal' expectation that primordial gas-rich discs successfully evolve into secularly complex S galaxies provided their evolution is not prematurely truncated. 
As shown in Section~\ref{Sec_scat}, this ongoing secular complexity manifests as the substantial scatter observed in the $M_{\rm bh}$--$\sigma_0$ diagram for S galaxies, a picture ultimately completed by the major mergers depicted in Figure~\ref{bh-ratio-sph}. 

Conversely, systems stripped and quenched early by dense environments were denied this extended spiral future. While these systems would have possessed significantly higher gas fractions at high redshift, they are observed today as the $z \approx 0$ dust-poor S0 population, anchored to the notably shallower, kinematically clean track of the $M_{\rm bh}$--$\sigma_0$ diagram (Section~\ref{Sec_flat}).

At the opposite extreme, the `Disc Down-sizing' sequence \citep{2026arXiv260424084G} traces the merger-dominated pathway of the `Triangal'. As gas-rich discs are destroyed in wet mergers to build massive dust-rich S0 galaxies, and subsequently scoured by repeated dry mergers to build ES,e and E galaxies, the host kinematics transition into the steep, collisionless virial track (Section~\ref{sec:feedback_and_mergers}). 
A concrete nearby example is the anticipated future merger of the Milky Way and Andromeda galaxies. As two gas-rich S galaxies, their collision should produce a dust-rich (dust = Y) S0 galaxy. Rather than appearing milky-white, the remnant will likely have a dusty, chocolate-brown appearance due to the stirred-up interstellar dust, prompting the informal suggestion of ‘Chocomeda’ as a more apt name than the occasionally used ‘Milkomeda’.
This galaxy-speciation-by-merger scenario is structurally corroborated by the emergence of anti-truncated discs and the progressive S\'ersicification of the ETG light profiles \citep{2014A&A...570A.103B, 2024MNRAS.535..299G}. 
Ultimately, when armed with precise morphological classifications, the $M_{\rm bh}$--$\sigma_0$ diagram ceases to be a monolithic scatter plot and emerges as a kinematic map of galaxy speciation.
The 'Virial Mirror' provides the physical backbone of this map.

\subsection{Benchmarking cosmological simulations and semi-analytic models}
\label{sec:simulations}

The morphology-dependent scaling relations and the `Triangal' framework offer a refined observational benchmark for cosmological hydrodynamical simulations and semi-analytic models \citep[e.g.,][]{2006MNRAS.365...11C, 2006MNRAS.370..645B, 2008MNRAS.391..481S, 2018MNRAS.481.3573L}. 
Historically, theoretical models have calibrated their sub-grid AGN feedback efficiencies by requiring their $z=0$ simulated galaxy populations to reproduce a single, monolithic $M_{\rm bh}$--$\sigma_0$ relation, or monolithic $M_{\rm bh}$--$M_\star$ relation. However, tuning simulations to a universal slope artificially blurs distinct physical mechanisms and risks improper calibration of the feedback parameters. 
Specifically, reproducing a single blended slope is a necessary but not a
sufficient test of a feedback prescription: numerically similar ensemble
slopes can emerge from different underlying seeding, accretion, and feedback
physics once averaged over a heterogeneous mix of formation histories. 
In practice, this morphology-dependent benchmark can be implemented by tracking the simulated galaxies' individual merger and star-formation histories, and classifying them at $z=0$ into morphological sub-populations based on their stellar mass fractions built via dry merging versus dissipative accretion. This allows for a direct, category-by-category comparison with the empirical relations acquired here (Table~\ref{Table_rel}). A simulation's sub-grid parameters are only successfully validated if they can simultaneously reproduce (i) the steep, dry-merger-driven slope ($\beta \approx 7.8$) for galaxies dominated by dissipationless merging, (ii) the intermediate slope ($\beta \approx 4.6$) for systems with gas-rich major merger histories, and (iii) the shallow slope ($\beta \approx 3.1$) for quiescent, accretion-starved systems. This multi-track benchmarking provides a highly sensitive diagnostic to constrain the relative strengths of momentum-driven versus energy-driven AGN feedback and the transition from gas-rich to gas-poor coevolution, granting a far more discriminating benchmark than matching a single blended relation alone.

Recent simulations \citep[e.g., Horizon-AGN, EAGLE, Illustris~TNG and TNG50, SIMBA, Magneticum, ASTRID;][]{2014MNRAS.444.1453D, 2015MNRAS.446..521S, 2018MNRAS.473.4077P, 2019MNRAS.486.2827D, 2019MNRAS.490.3234N, 2021ApJ...910...87K, 2022MNRAS.513..670N} now routinely track detailed merger histories and kinematic morphologies. Recent theoretical studies exploring black hole scaling relations have capitalised on this, typically separating simulated galaxies by general star formation activity or broad kinematic disc fractions \citep[e.g.,][]{2020MNRAS.493.1888T, 2021MNRAS.503.1940H, 2022MNRAS.509.3015H}. However, empirical evidence now reveals that a galaxy's specific morphological type --- which encodes its unique accretion and merger history --- is a more fundamental determinant of its position in the $M_{\rm bh}$--$M_*$ diagrams than its current star formation rate \citep{2024MNRAS.52710059G}. Therefore, to fully capture the observational realities of black hole--galaxy coevolution, there is a compelling opportunity to move beyond monolithic or SFR-binned scaling relations. By leveraging existing merger trees to explicitly distinguish between primeval galaxies, S galaxies, wet-major-merger-built S0 galaxies, and dry-major-merger-built E galaxies, future theoretical studies can more accurately trace the distinct physical mechanisms driving black hole growth across different evolutionary pathways.

For instance, within the `Triangal' framework, the apparent diversity of the black hole scaling relations reflects the relative importance of three distinct physical processes that simulations must independently capture: 
\begin{enumerate}
    \item[(i)] Gas-rich (wet) mergers, which promote rapid, feedback-regulated black hole growth resulting in the $\sim$4.4--4.6 slope of dust-rich S0 galaxies;
    \item[(ii)] Accretion, minor mergers, and secular evolution, where complex disc kinematics and transient bar-heating, coupled with variations in global disc structure (such as scale-length) and central dark matter fractions that contribute to the depth of the composite gravitational potential, can introduce significant scatter in the measured $\sigma_0$ for S galaxies; and
    \item[(iii)] Gas-poor (dry) mergers, which build the most massive ES,e and E galaxies, which is expected to steepen the $M_{\rm bh}$--$\sigma_0$ relation to a slope of $\sim$8--9 by preserving the virial equilibrium of the spheroids while summing the black hole masses \citep{2023MNRAS.518.6293G}.
\end{enumerate}

The dynamical findings, presented herein, also complement the `Disc Down-sizing' sequence identified in \citet{2026arXiv260424084G}. The evolutionary transition from S galaxy $\to$ Dust-Rich S0 $\to$ ES,e $\to$ E with a nuclear disc $\to$ pure E represents not just a reduction of disc mass, but a transformation of galaxy kinematics. As portions of colliding spiral discs are violently redistributed during wet major mergers (creating a dust $=$ Y S0 galaxy), the $M_{\rm bh}$--$\sigma_0$ relation tightens. This tightening suggests an underlying connection between the growing spheroid and the central black hole as the kinematic noise of cold rotating discs, spiral arms, and bar-driven heating is erased.

Enforcing limited calibrations may severely limit and bias a simulation's insight into these processes, as well as  its predictive power at the mass extremes. As discussed in Section~\ref{sec:gravitational_waves}, at the high-mass end, failing to capture the dry-merger-driven steepening can systematically under-predict the mass and abundance of UMBHs hosted by E and ES,e galaxies, thereby skewing forecasts for the nanohertz gravitational wave background. Conversely, at the low-mass end, imposing a monolithic slope overlooks what may be a notably shallower relation (slope $\sim$3.0) in primeval gas-stripped ETGs, artificially altering the expected demographics of IMBHs and misrepresenting early-Universe seed growth.

To accurately reflect the hierarchical assembly of the Universe, simulations must move beyond monolithic targets and verify that their simulated galaxies follow distinct morphological tracks. The relations presented in Table~\ref{Table_rel} provide the exact conditional slopes ($\beta$) and intrinsic scatters ($\sigma_{y|x}$) required for this validation. Ultimately, the morphology-dependent scaling relations presented here effectively disentangle the regimes dominated by AGN feedback from those dominated by collisionless merger dynamics, providing updated prescriptions for incorporating accurate black hole--galaxy co-evolution into the next generation of semi-analytic and cosmological models.

\section{Conclusions}
\label{Sec_Conclusions}

Morphology- and dust-bin-dependent $M_{\rm bh}$--$\sigma_0$ relations have been
derived for a sample of 137 galaxies using the  hierarchical
Bayesian regression code \textsc{scope}, introduced in 
Section~\ref{Sec_Methodology} 
and detailed in 
Appendix~\ref{Sec_Appdx}. 
The data are divide into four broad regimes, yielding: a steep relation (slope $7.8\pm1.3$) for
E and ES,e galaxies 
(Equation~\ref{Eq_ESeE}); 
an intermediate relation (slope $4.56\pm0.49$) for the ensemble of S0 galaxies
(Equation~\ref{Eq_S0_all}); 
a shallow relation (slope $\sim$2.5--3.1) for dust-poor S0 galaxies 
(Equations~\ref{Eq_dust-N}--\ref{Eq_S0_N_edit});
and a poorly constrained, low-significance relation for S galaxies
(Equation~\ref{Eq_S}). 
These, and other, divisions are summarised in 
Table~\ref{Table_rel}.
The statistical significance of the differences between the slopes of the $M_{\rm bh}$--$\sigma_0$ relations has been
quantified directly from their posterior distributions 
(Section~\ref{sec:slope_comparisons}). 
The comparison of E/ES,e
versus dust-poor S0 reveals a difference at $\sim2.8\sigma$; the comparison of dry-merger versus
wet-merger reveals a $\sim2.3\sigma$ difference; while the two sub-populations of S0 are 
not significantly distinguished ($\sim1.0\sigma$). 
As discussed in Section~\ref{sec:sample_bias}, 
while the dynamically measured black hole sample is unavoidably bounded by the spatial resolution limits of modern instrumentation, the compiled host galaxies are consistent with being representative in terms of their global scaling properties. Consequently, these relations provide a well-calibrated empirical baseline that can be applied to other systems and used as benchmarks in cosmological simulations, subject to standard observational caveats.

A distinction should be maintained between these primary empirical results and the evolutionary frameworks proposed to interpret them. While the morphological classifications and dust-content categories are observationally grounded 
(Section~\ref{Sec_Data}), 
they may not guarantee a unique evolutionary pathway for any individual galaxy.  Nonetheless, 
generalisations can be made or inferred.  
Spiral arms in S galaxies require a pre-existing disc from which to emerge, and the `Triangal' framework postulates past gas accretion and minor mergers onto primeval S0 galaxies as the origin of the S galaxies.  
Within this framework, those S0 galaxies that did not form substantial discs and did not produce a spiral pattern are interpreted as the surviving primeval population, which is considered to be today’s low-mass, dust-poor S0 galaxies.

Regarding today's dust-rich S0 galaxies, the majority of those in the present sample are known major-merger remnants 
\citep{2023MNRAS.521.1023G}, 
thought to involve at least one S galaxy given their dust-rich nature. Their subsequent merger is believed to produce the ES,e and E galaxies 
\citep[e.g.,][]{2006ApJ...636L..81N}.  
Indeed, 
E and ES,e galaxies without large-scale discs or significant rotation are widely understood to have been assembled through multiple major mergers that erased the ordered motions of their progenitor galaxies.  Collectively, this provides a scaffolding for understanding evolutionary pathways through the $M_{\rm bh}$--$\sigma_0$ diagram, in tandem with those illustrated through the $M_{\rm bh}$--$M_\star$ diagram 
\citep[see][figure~1]{2023MNRAS.522.3588G}, 
and the (central surface brightness)-size diagram 
\citep[see][figure~12]{2025PASA...42..155G}, 
and for explaining the S\'ersicification of ETGs 
\citep{2024MNRAS.535..299G}.
In this context, strongly barred dust-poor S0 galaxies show a potential systematic offset to higher velocity dispersions 
(Equations~\ref{Eq_dust-bar}--\ref{Eq_S0_N_edit}). 
Although this signature is qualitatively consistent with theoretical expectations of kinematic heating by stellar bars, it remains tentative given the modest sample sizes currently available at the low-mass end.

Similarly, the `Virial Mirror' 
(Section~\ref{sec:feedback_and_mergers}) 
is presented as an analytical framework derived directly from fundamental scaling laws and physical principles. Its key ingredients --- namely, the classical scalar virial theorem, the near-constant spatial sizes of dwarf ETGs ($R_{\rm e} \approx 1\ \rm{kpc}$), and their near-constant dynamical-to-stellar mass ratios --- are independently established empirical realities of galaxy structure 
\citep[e.g.,][]{2008MNRAS.389.1924F, 2019PASA...36...35G}. 
Under these physical constraints, the resulting shallow $M_{\rm bh}$--$\sigma_0$ slope at low masses is an analytical requirement. To further test this framework, future work is required. Larger samples are needed to reduce uncertainties in fitted relations; velocity dispersions measured within larger apertures are needed to mitigate the effects of non-homologous dispersion profiles; and a dedicated kinematic decomposition of ordered versus random motions is required to quantify the impact of rotational support. Additionally, as explored in 
Section~\ref{Sec_IMBH}, 
searches for IMBHs at the lowest masses must account for observational censoring and detection floors for both low-mass black holes and nuclear star clusters, which can strongly shape the observed scaling relations.

Beyond these coevolutionary considerations, the morphology-dependent scaling relations carry immediate practical implications for three major astrophysical domains. First, they supply physically motivated, multi-track calibration targets for the next generation of cosmological simulations, offering a more discriminating benchmark than a single blended relation. Second, they demonstrate that combining all early-type host galaxies into one relation can systematically bias the scaling laws used to calibrate AGN virial $f$-factors. Finally, the steepening observed for massive elliptical and ellicular galaxies ($\beta \approx 7.8$--$8.6$) indicates that monolithic, single-power-law scaling relations under-predict the UMBH population ($M_{\rm bh}>10^{10}\,M_\odot$) thought to contribute to the nanohertz stochastic gravitational-wave background detected by pulsar timing arrays.

Finally, to facilitate broader application, the \textsc{scope} regression package is released as a public, open-source tool designed to perform robust, coordinate-invariant regressions that natively handle asymmetric measurement errors on any bivariate scaling relation whose intrinsic scatter is Gaussian.

\section*{Acknowledgements}

Publication costs were funded through the Australia and New Zealand
Institutions (Council of Australian University Librarians affiliated) Open
Access Agreement.
This work has used the 
NASA/IPAC Extragalactic Database (NED),
funded by NASA and operated by the California Institute of Technology, and the 
Hyperleda database (http://atlas.obs-hp.fr/hyperleda/). 
This research has also used the SAO/NASA Astrophysics Data System (ADS)
bibliographic services 
and the {\sc Rstan} package available at \url{https://mc-stan.org/}.
The Stan Development Team is thanked for making  the 
computational framework available that enabled the hierarchical modelling 
implemented in \textsc{scope}.
Ewan Cameron is thanked for providing the baseline routine for mapping asymmetric uncertainties to skew-normal probability distributions.


\section{Data Availability}

%
The measured and derived galaxy parameters utilised herein are available in the online supplementary material of \citet{2026arXiv260424084G} and \citet{2019ApJ...876..155S}. To facilitate reproduction of the analyses, the specific data subsets used to derive the scaling relations have also been formatted as machine-readable `.csv` files and are provided alongside the software repository linked below.

\section*{Code Availability}

The \texttt{R} implementation of the Symmetric COvariance Population Estimator (\textsc{scope}) developed for this work is publicly available at \url{https://github.com/A-Graham/SCOPE}. The repository includes the Bayesian hierarchical regression code, the \textsc{stan} model file, the input data catalogues, and the execution script required to reproduce the results presented in this paper. 
The \textsc{scope} code repository is archived on Zenodo (\href{https://doi.org/10.5281/zenodo.20536917}{DOI: 10.5281/zenodo.20536917}).

\bibliographystyle{mnras}
\bibliography{Paper_M-sigma-2}{}

\appendix

\section{A symmetric Bayesian hierarchical regression: \textsc{scope}}
\label{Sec_Appdx}

The challenge of fitting a straight line to astronomical data with measurement uncertainties in both axes and intrinsic population scatter has a rich history, with early, foundational Bayesian solutions provided by \citet{Gull1989} and \citet{DAgostini2005}. Over the subsequent decades, Bayesian hierarchical modelling has been increasingly adopted across astrophysics for its ability to rigorously handle latent variables and complex selection effects \citep[e.g.,][]{2018MNRAS.475.1203C, 2019ApJ...874..150B, 2026A&A...708A..53M, 2026arXiv260119424H}. The analysis developed and used here builds on this legacy by using modern Bayesian hierarchical regression \citep{Barnard2000, Gelman_Hill_2006, Gelman2013}, and the resulting code is referred to as the Symmetric COvariance Population Estimator (SCOPE).

The joint distribution of $x = \log \sigma_0$ and 
$y = \log M_{\rm bh}$ is modelled as a bivariate normal distribution 
with mean vector $(\mu_x, \mu_y)$ and covariance matrix $\Sigma_0$,
\begin{equation}
\begin{pmatrix}
x \\
y
\end{pmatrix}
\sim
\mathcal{N}
\left(
\begin{pmatrix}
\mu_x \\
\mu_y
\end{pmatrix},
\Sigma
\right),
\end{equation}
where the covariance matrix is parameterised in terms of marginal 
standard deviations $\sigma_x$, $\sigma_y$, and correlation coefficient $\rho$,
\begin{equation}
\Sigma =
\begin{pmatrix}
\sigma_x^2 & \rho \sigma_x \sigma_y \\
\rho \sigma_x \sigma_y & \sigma_y^2
\end{pmatrix}.
\end{equation}

A Lewandowski-Kurowicka-Joe (LKJ) prior \citep{2009JMA...100.1989L} is adopted for the correlation matrix. The LKJ prior is governed by a single shape parameter, $\eta$. Setting $\eta = 1$ yields a uniform prior distribution over all valid correlation matrices. In contrast, values of $\eta > 1$ concentrate the prior probability mass toward zero correlation. Here, a default weakly informative prior of $\eta = 2$ is adopted. This gently suppresses implicit mathematical preference for extreme correlations ($\rho \to \pm 1$), improving numerical stability while ensuring the posterior remains overwhelmingly data-driven. The
measurement uncertainties in both variables are explicitly modelled, 
with asymmetric uncertainties in $y$ represented using a skew-normal 
distribution.
The regression of $y$ on $x$ follows from the conditional expectation 
of the bivariate normal distribution,
\begin{equation}
E[y \mid x]
=
\mu_y
+
\beta (x - \mu_x),
\end{equation}
where the slope is
\begin{equation}
\beta = \rho \frac{\sigma_y}{\sigma_x}.
\end{equation}
The underlying bivariate normal model treats both variables symmetrically, 
accounting for their intrinsic covariance and measurement uncertainties. The regression slope reported here is the conditional slope 
$\beta = \rho \sigma_y / \sigma_x$, corresponding to $E[y \mid x]$, 
rather than the major-axis slope\footnote{
The slope $\beta = \rho\,\sigma_y/\sigma_x$ is derived from the symmetric 
intrinsic covariance matrix and is mathematically identical to the 
conditional expectation $\mathrm{E}[y|x]$, while remaining invariant to 
the choice of dependent variable. The major-axis slope $\sigma_y/\sigma_x$ 
characterises the orientation of the intrinsic covariance ellipse and 
differs from $\beta$ by the factor $\rho$.}
$\sigma_y / \sigma_x$.
This distinction between the orientation of the covariance ellipse and the conditional regression slope is particularly relevant for the S galaxy sample examined herein. 
Although the S galaxies span a broad range in $M_{\rm bh}$ relative to $\sigma_0$, their intrinsic correlation is comparatively weak ($\rho \approx 0.4$), yielding a shallow conditional slope with substantial uncertainty. 
In this regime, the conditional expectation $\mathbb{E}(y|x)$ is naturally much flatter than the major-axis orientation of the covariance ellipse, reflecting the limited predictive power of $\sigma_0$ for determining $M_{\rm bh}$ in S galaxies.

Relative to the earlier implementation \citep[][their Appendix~A]{2019ApJ...873...85D}, this formulation adopts a 
Cholesky parameterisation of the covariance matrix with an LKJ prior, 
yielding improved numerical stability and a statistically well-behaved 
treatment of the intrinsic correlation. The slope is derived directly 
from the covariance structure rather than imposed asymmetrically, 
ensuring consistent propagation of parameter uncertainties within the 
hierarchical framework.

Given that the default priors are weakly informative and data-driven, the likelihood dominates the posterior geometry. To explicitly confirm this, an `ultra-flat' model configuration was tested on the smallest $N=13$ sample (Equation~\ref{Eq_dust-bar}). The $\eta=2$ LKJ prior was replaced with a uniform $\eta=1$ prior, and the scale of the intrinsic scatter and population mean priors were broadened by a factor of 5. The resulting conditional slope changed by $\Delta \beta = 0.41$, less than the $1\sigma$ statistical uncertainty ($\pm 1.08$).

\subsection{Derivation of the regression slope}

For a bivariate normal distribution with covariance matrix given above, 
the conditional mean of $y$ at fixed $x$ is
\begin{equation}
E[y \mid x]
=
\mu_y
+
\frac{\mathrm{Cov}(x,y)}{\mathrm{Var}(x)}
(x - \mu_x).
\end{equation}
From the covariance matrix,
\begin{equation}
\mathrm{Cov}(x,y) = \rho \sigma_x \sigma_y,
\qquad
\mathrm{Var}(x) = \sigma_x^2.
\end{equation}
Substituting these expressions gives
\begin{equation}
E[y \mid x]
=
\mu_y
+
\rho \frac{\sigma_y}{\sigma_x}
(x - \mu_x),
\end{equation}
so that the regression slope is
\begin{equation}
\beta
=
\rho \frac{\sigma_y}{\sigma_x}.
\end{equation}
The slope is therefore determined jointly by the intrinsic correlation 
coefficient and the ratio of marginal dispersions.

\subsection{A word on the uncertainties}

The posterior uncertainties on the slope and intercept obtained from 
the present hierarchical Bayesian model are generally larger than 
those derived using classical regression estimators such as BCES 
\citep{1996ApJ...470..706A} or the modified FITEXY routine 
\citep{2002ApJ...574..740T}. This difference reflects the distinct 
inferential frameworks underlying these methods rather than an 
inconsistency in the data.

Classical estimators such as BCES and modified FITEXY determine the 
best-fitting regression line by correcting for measurement errors and, 
in the case of modified FITEXY, introducing an additional intrinsic 
scatter term to achieve $\chi^2_\nu \approx 1$. The slope is treated 
as a directly fitted parameter, and uncertainties are typically derived 
from asymptotic approximations or curvature of the likelihood surface 
near its maximum. While these approaches account for measurement 
uncertainties in both variables, they do not explicitly model the full 
intrinsic population distribution from which the data are drawn.

In contrast, the present hierarchical model assumes that 
$(x, y) = (\log \sigma_0, \log M_{\rm bh})$ arise from an intrinsic 
bivariate normal distribution characterised by the population means 
$(\mu_x, \mu_y)$, the marginal dispersions $(\sigma_x, \sigma_y)$, and 
the correlation coefficient $\rho$. The regression slope is not an 
independent free parameter but is derived from the intrinsic covariance 
structure as $\beta = \rho \sigma_y / \sigma_x$. Consequently, 
uncertainty in the slope reflects joint posterior uncertainty in the 
intrinsic dispersions and correlation coefficient, in addition to 
sampling variance and measurement errors.

This hierarchical treatment, therefore, propagates uncertainty in the 
underlying population covariance structure into the inferred scaling 
relation. In finite samples, particularly when intrinsic scatter is 
substantial or the dynamic range in $x$ is limited, this leads to 
broader posterior credible intervals than those returned by classical 
estimators. The larger uncertainties do not indicate a weaker relation; 
rather, they represent a more complete accounting of uncertainty in 
the intrinsic population parameters.

\subsection{Intrinsic scatter}

The intrinsic scatter about the regression is
\begin{equation}
\sigma_{y|x}
=
\sigma_y \sqrt{1 - \rho^2},
\end{equation}
which corresponds to the standard deviation of $y$ at fixed $x$ implied 
by the assumed bivariate normal distribution.

The intrinsic scatter $\sigma_{y|x}$ differs from the empirical rms 
vertical scatter about the fitted relation, as it represents the 
population-level conditional dispersion implied by the intrinsic 
bivariate normal model, excluding measurement errors and sampling noise.
Historically, many astronomical regression analyses estimated the 
intrinsic scatter by computing the rms vertical dispersion about a 
best-fitting line and subtracting measurement errors in quadrature. 
Such approaches implicitly assume a conditional regression model and 
do not distinguish between the population-level conditional variance 
implied by an underlying joint distribution and the sample-dependent 
rms scatter of noisy data. In a fully hierarchical framework, the 
intrinsic scatter $\sigma_{y|x}$ is instead derived directly from the 
intrinsic covariance structure and represents the true conditional 
population dispersion, independent of sampling fluctuations.

\begin{figure}
\begin{center}
\includegraphics[width=1.0\columnwidth]{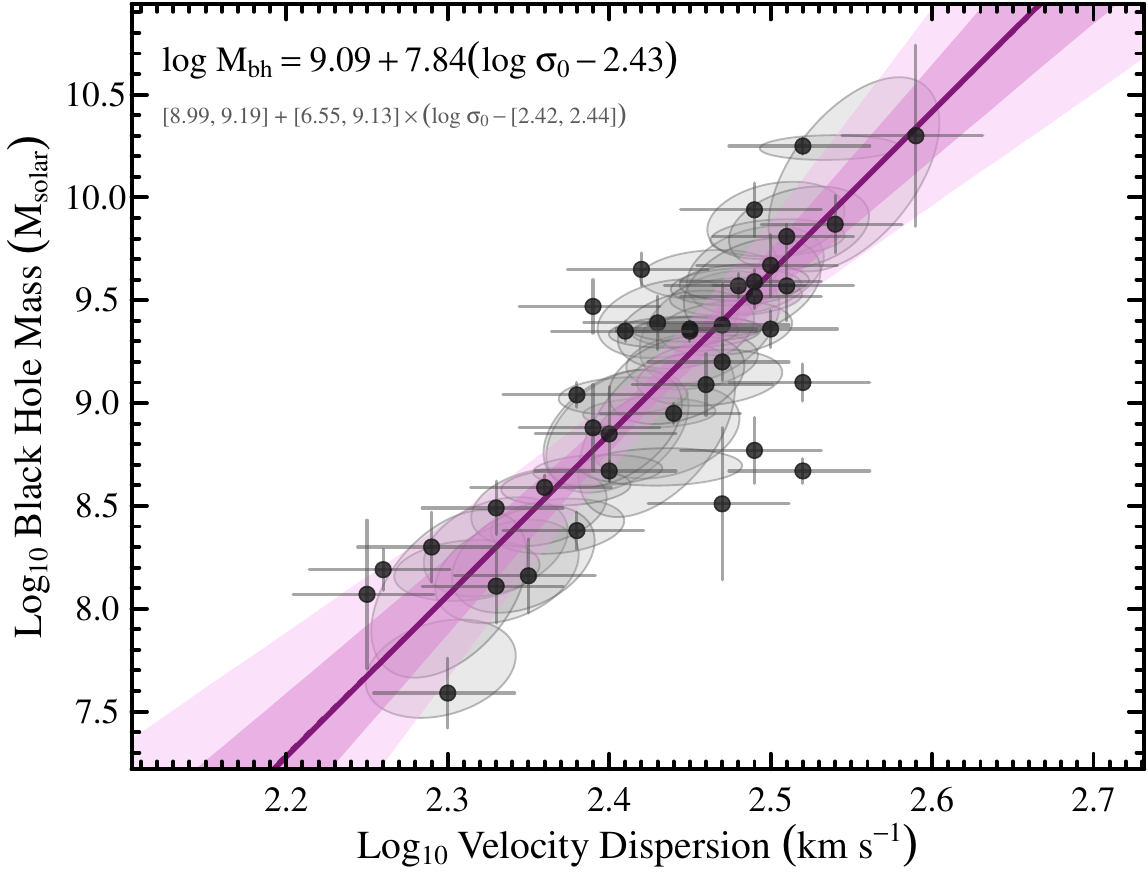}
\caption{
The \textsc{scope} fit to the sample of 38 BCGs, Es, and ES,e galaxies (Equation~\ref{Eq_ESeE}). 
The (overlapping) solid and dashed lines show 
the posterior mean and median relations respectively; for 
well-constrained posteriors these are essentially coincident. 
The darker and lighter shaded regions show the $\approx$68.3 and $\approx$95.4 per 
cent pointwise credible bands for the posterior {\em median} relation, 
evaluated at the $\Phi(\pm1)$ and $\Phi(\pm2)$ quantiles of the 
posterior predictive distribution at each abscissa value, where 
$\Phi$ denotes the standard normal cumulative distribution function. The annotated credible 
intervals (see the second inset equation) are the quantile-based 
$[\Phi(-1),\Phi(1)]$ intervals centred on the posterior {\em median}, 
and differ in general from the posterior standard deviations 
reported in the text, which are symmetric summaries centred on 
the posterior {\em mean}; the two are identical for symmetric posteriors. 
Grey ellipses show $1\sigma$ posterior credible regions for each 
galaxy's true position (see Section~\ref{Sec_fin_report}).
}
\label{Fig-SCOPE}
\end{center}
\end{figure}

\subsection{Final reported relation}
\label{Sec_fin_report}

The fitted scaling relation is written as 
\begin{equation}
\log M_{\rm bh}
=
\mu_y
+
\beta
\left(
\log \sigma_0
-
\mu_x
\right),
\label{eq:mbh_relation}
\end{equation}
where $\mu_x$ and $\mu_y$ are the population means of 
$\log \sigma_0$ and $\log M_{\rm bh}$, respectively, and 
$\beta$ is the regression slope.
Posterior means and standard deviations are reported for each parameter,
\begin{align}
\beta &= \bar{\beta} \pm \sigma_\beta, \\
\mu_x &= \bar{\mu}_x \pm \sigma_{\mu_x}, \\
\mu_y &= \bar{\mu}_y \pm \sigma_{\mu_y},
\end{align}
where the quoted uncertainties represent marginal posterior standard 
deviations and incorporate measurement errors and uncertainty in the 
intrinsic covariance structure.
Because \textsc{scope} samples the full posterior parameter space, it fundamentally does not assume that the resulting parameter uncertainties are symmetric. While the symmetric posterior standard deviations ($\sigma$) are reported here as a convenient summary, the code natively evaluates and outputs the exact quantile-based credible intervals (e.g., roughly the 16th and 84th percentiles). This allows users of the code to fully preserve and utilise asymmetric parameter posteriors when applying these relations in theoretical models or population synthesis.

For reference, the equivalent intercept at $\log \sigma_0 = 0$ is
\begin{equation}
\alpha = \mu_y - \beta \mu_x,
\end{equation}
although this quantity is derived and not fundamental to the adopted 
mean-centred formulation. The pivot $\mu_x$ is determined by the data 
and may vary for different galaxy samples.

Figure~\ref{Fig-SCOPE} displays \textsc{scope} applied to the sample of 
38 BCGs, Es, and ES,e galaxies. The grey ellipses show the 
$1\sigma$ posterior credible regions for the true positions of 
each galaxy in the $(X, Y)$ plane, derived from the joint 
posterior of the latent true values. These ellipses are not 
centred on the observed data points (shown with error bars) 
because the hierarchical model partially pools information 
across the full sample: each galaxy's inferred true position 
is drawn toward the population mean relation in proportion to 
its measurement uncertainty, a consequence of shrinkage 
inherent to hierarchical Bayesian inference. Galaxies with 
larger observational uncertainties exhibit more pronounced 
offsets between their observed and inferred positions.

\subsection{Comparison with \texorpdfstring{\textsc{linmix}}{linmix} (Kelly 2007)}

The fully Bayesian regression formalism and code of \citet{2007ApJ...665.1489K}, known as \textsc{linmix}, also provides a hierarchical treatment of measurement errors and intrinsic scatter, and merits comparison with the present approach. That approach is formulated as a conditional regression model, $p(y \mid x)$, in which the slope and intercept are treated as primary parameters. To ensure flexibility, the intrinsic marginal distribution of the independent variable is represented using a mixture-of-Gaussians model, accommodating non-Gaussian distributions of $x$. The original \textsc{linmix} implementation was written in IDL \citep{2007ApJ...665.1489K}; a Python port is also widely available \citep{linmix_python}, whereas the present covariance-based model is implemented using \textsc{Stan} via \texttt{R}.

The present analysis adopts a complementary perspective. Rather than modelling the relation conditionally, it assumes that $(x,y) = (\log \sigma_0, \log M_{\rm bh})$ arise from an intrinsic bivariate normal population characterised by its mean vector and covariance matrix. In this formulation, the intrinsic covariance structure is fundamental, and the regression slope emerges as a derived quantity,
\[
\beta = \rho \frac{\sigma_y}{\sigma_x}.
\]
This parameterisation treats both variables symmetrically at the population level and makes explicit that the scaling relation reflects the underlying covariance of the galaxy population rather than a purely predictive regression of one variable on the other.

This symmetric formulation offers several practical advantages. First, it yields direct inference on the intrinsic dispersions and correlation coefficient, quantities of astrophysical interest in their own right. Second, the regression line is constrained to pass through the intrinsic population mean, clarifying the role of the pivot point and avoiding the implicit assumption of  a fixed reference value. Third, the Cholesky parameterisation with an LKJ prior provides a statistically well-behaved, weakly informative prior over the space of admissible covariance matrices, improving numerical stability and avoiding unintended bias toward extreme correlations.

Unlike modified \textsc{fitexy} \citep{2002ApJ...574..740T}, which assumes intrinsic scatter solely in the vertical direction and thereby breaks symmetry between the variables \citep{2006ApJ...637...96N}, the present model describes the intrinsic population with a fully symmetric covariance matrix. However, while this Bivariate Normal framework elegantly preserves symmetry and treats both variables equally, it implicitly assumes that the marginal distribution of the independent variable is a single Gaussian. In contrast, \citet{2007ApJ...665.1489K} models the marginal distribution of the independent variable as a mixture of $K$ Gaussians, providing greater flexibility when the distribution of $x$ is non-Gaussian (e.g., bimodal or strongly skewed). This flexibility in modelling the marginal distribution of $x$ comes at the cost of treating the two variables asymmetrically: the independent variable receives a rich mixture model while the dependent variable does not.

This directional asymmetry in the regression mechanics has practical consequences. When \textsc{linmix} is applied to the sample of 38 ETGs shown in Figure~\ref{Fig-SCOPE}, it returns a conditional slope of $\beta_1 = 8.11 \pm 1.37$ when fitting $\log M_{\rm bh}$ as a function of $\log\sigma_0$. However, swapping the axes to perform the inverse regression yields a slope whose reciprocal is $1/\beta_2 = 9.41$. This discrepancy of $\Delta\beta = 1.30$ is comparable in magnitude to the $1\sigma$ posterior uncertainty on $\beta_1$ itself.\footnote{The asymmetry is amplified here by the substantial 
difference in measurement precision between the two variables. 
The $\log\sigma_0$ uncertainties are small and nearly uniform 
($\sim$0.04--0.05\,dex), whereas the $\log M_{\rm bh}$ 
uncertainties are larger and heterogeneous (0.04--0.44\,dex, 
with several galaxies exceeding 0.15\,dex). In a conditional 
regression, measurement error in the independent variable causes 
attenuation bias, pulling the slope toward zero. When 
$\log\sigma_0$ is the independent variable, its small errors 
produce little attenuation; however, when $\log M_{\rm bh}$ plays that 
role, its larger errors produce stronger attenuation, depressing 
$\beta_2$ and inflating $1/\beta_2$ relative to $\beta_1$. 
The \textsc{SCOPE} slope is immune to this directional 
inconsistency because it is derived from the intrinsic covariance 
matrix, which treats both variables symmetrically regardless of 
their individual measurement precisions.} 
In contrast, \textsc{SCOPE} returns a unified scaling with $\beta = 7.8 \pm 1.3$. Because the \textsc{SCOPE} slope is derived directly from the invariant intrinsic covariance matrix, it mathematically guarantees $\beta_1 \equiv 1/\beta_2$ across every posterior draw.

A second practical distinction lies in the treatment of the observational data. Standard implementations of \textsc{linmix} accept only a single, symmetric measurement uncertainty per variable. In observational astronomy, however, quoted uncertainties can be asymmetric. \textsc{SCOPE} natively accommodates these asymmetric measurement errors on both variables by employing a skew-normal likelihood approximation.
The requisite location, scale, and shape parameters are derived by mapping the reported asymmetric error bounds via a least-squares quantile-matching procedure 
\citep[adapted from the methodology of][who applied this approximation to the $\log M_{\rm bh}$ ($y$) variable only, and which is extended here to both variables] {2019ApJ...873...85D}. This circumvents the need to artificially symmetrise or average disparate upper and lower error bounds before fitting, thereby preserving the precise informational content of the data.

Crucially, it must be emphasised that modelling the covariate distribution is distinct from modelling the intrinsic scatter about the regression line. Both the covariance-based approach presented here and the conditional approach of \citet{2007ApJ...665.1489K} assume that the intrinsic scatter (the residuals in $y$ given $x$) is strictly Gaussian. Neither approach natively provides robustness to extreme outliers in the dependent variable. Extending either framework to handle heavy-tailed residual distributions would require replacing the Gaussian intrinsic scatter model with, for example, a Student-$t$ distribution. In the absence of compelling evidence for non-Gaussian intrinsic structure in the residuals, a Gaussian description provides a parsimonious and physically interpretable representation.

Ultimately, the key distinction is interpretive. The conditional formulation of \citet{2007ApJ...665.1489K} estimates the expected black hole mass at a fixed velocity dispersion, $E[y \mid x]$, and is therefore naturally suited to predictive applications. In contrast, the present analysis models the intrinsic joint distribution of galaxies in $(\log \sigma_0, \log M_{\rm bh})$ space. This perspective is particularly relevant for comparison with theoretical models and numerical simulations, which generate populations of galaxies rather than conditional predictions. The regression slope in the present framework is therefore a derived property of the intrinsic covariance structure. As long as the underlying population is reasonably well described by a Bivariate Normal distribution, the covariance-based parameterisation provides a natural, physically motivated representation of the population-level structure that simulations are designed to reproduce.

\bsp
\label{lastpage}
\end{document}